\documentclass[aps,twocolumn,notitlepage,superscriptaddress]{revtex4-1}
\usepackage[colorlinks=true, citecolor=magenta, linkcolor=blue, urlcolor=blue]{hyperref}
\usepackage{graphicx}
\usepackage{float}
\usepackage{amsmath}
\usepackage{amssymb}
\usepackage{amsfonts}
\usepackage{hyperref}
\usepackage{mathtools}
\usepackage{xcolor}
\usepackage{tikz}
\usepackage{enumitem}
\usepackage{verbatim}
\usetikzlibrary{shapes.geometric, arrows}
\usetikzlibrary{decorations.markings}
\usepackage{braket}
\usepackage{siunitx}

\newcommand{\linedot}{\,\underline{\boldsymbol{\cdot}}\,}

\usepackage[T1]{fontenc}
\usepackage{textcomp, amssymb}
\usepackage[utf8]{inputenc}

\allowdisplaybreaks

\begin{document}
 
\title{Cavity-Enhanced Optical Manipulation of Antiferromagnetic Magnon-Pairs}

\author{Tahereh Sadat Parvini}
\affiliation{Walther-Meißner-Institut, Bayerische Akademie der Wissenschaften, 85748 Garching, Germany}
\affiliation{Munich Center for Quantum Science and Technology (MCQST), Schellingstr.4, 80799, Munich, Germany}
\email{Tahereh.Parvini@wmi.badw.de}
\author{Anna-Luisa E. R\"omling}
\affiliation{Condensed Matter Physics Center (IFIMAC) and Departamento de Física Teórica de la Materia Condensada, Universidad Autónoma de Madrid, E-28049 Madrid, Spain}
\author{Sanchar Sharma}
\affiliation{Laboratoire de Physique de l’\'{E}cole Normale Sup\'{e}rieure, ENS, Universit\'{e} PSL, CNRS, Sorbonne Universit\'{e}, Universit\'{e} de Paris, F-75005 Paris, France}
\author{Silvia Viola Kusminskiy}
\affiliation{Institut f{\"u}r Theoretische Festk{\"o}rperphysik, RWTH Aachen University, 52056 Aachen, Germany}
\affiliation{Max Planck Institute for the Science of Light, 91058 Erlangen, Germany}

\date{\today}

\begin{abstract}
The optical manipulation of magnon states in antiferromagnets (AFMs) holds significant potential for advancing AFM-based computing devices. In particular, two-magnon Raman scattering processes are known to generate entangled magnon-pairs with opposite momenta. We propose to harness the dynamical backaction of a driven optical cavity coupled to these processes, to obtain steady states of squeezed magnon-pairs, represented by squeezed Perelomov coherent states. The system's dynamics can be controlled by the strength and detuning of the optical drive and by the cavity losses. In the limit of a fast (or lossy) cavity, we obtain an effective equation of motion in the Perelomov representation, in terms of a light-induced frequency shift and a collective induced dissipation which sign can be controlled by the detuning of the drive. In the red-detuned regime, a critical power threshold defines a region where magnon-pair operators exhibit squeezing—a resource for quantum information— marked by distinct attractor points. Beyond this threshold, the system evolves to limit cycles of magnon-pairs. In contrast, for resonant and blue detuning regimes, the magnon-pair dynamics exhibit limit cycles and chaotic phases, respectively, for low and high pump powers. Observing strongly squeezed states, auto-oscillating limit cycles, and chaos in this platform presents promising opportunities for future quantum information processing, communication developments, and materials studies. 
\end{abstract}

\maketitle

\section{Introduction}

\begin{figure}[h!]
  \centering
  \includegraphics[width=0.48\textwidth]{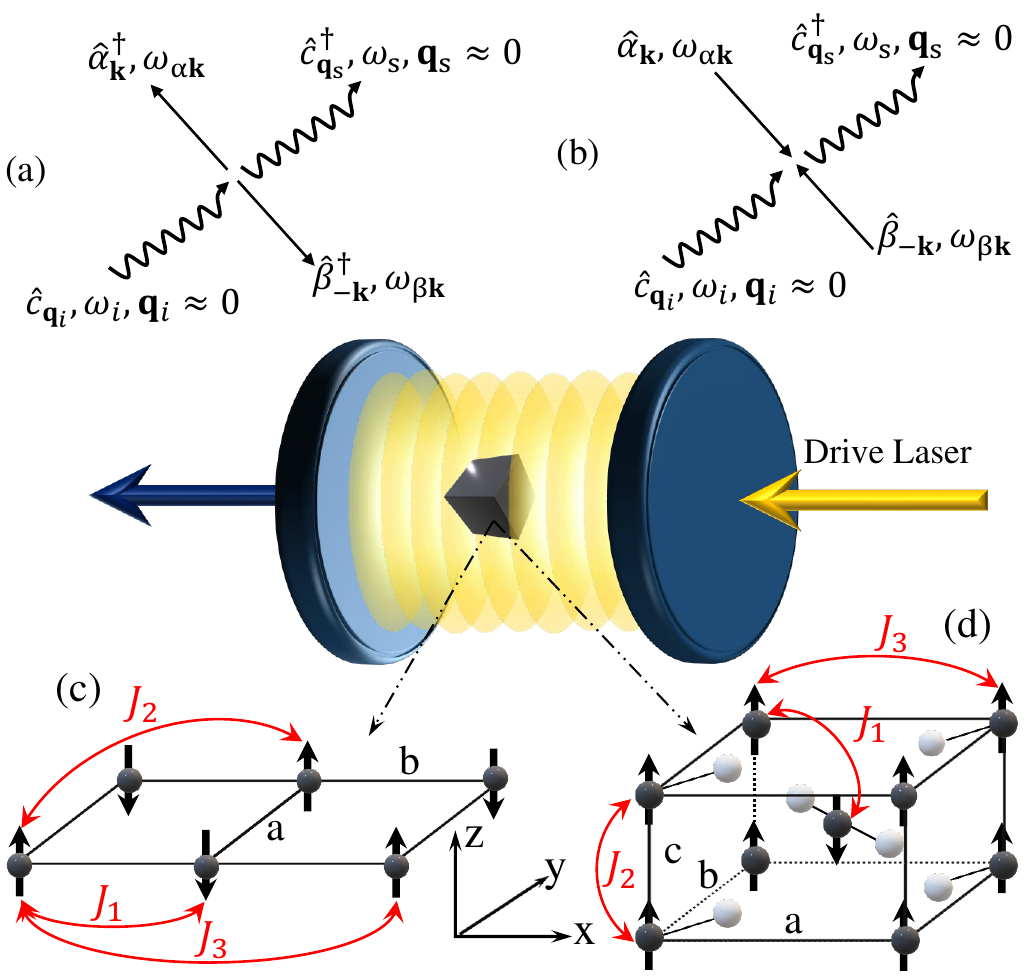}
  \caption{AFM optomagnonic cavity. Light-matter coupling within the domain of nonresonant two-magnon Raman scattering processes: (a) Stokes scattering with a redshifted frequency of $\omega_{s}=\omega_{i}-\omega_{\alpha{\bf k}}-\omega_{\beta{\bf k}}$ and (b) anti-Stokes scattering with blueshifted frequency $\omega_{as}=\omega_{i}+\omega_{\alpha{\bf k}}+\omega_{\beta{\bf k}}$. Crystal and magnetic structure of (c) the Rutile- and (d) quasi-two-dimensional AFMs. Short-range exchange constants $J_{\mathbf{R'}}$ are defined as $J_{l}$ ($l=1,2,3$).}
\label{Fig1}
\end{figure}

The rapid evolution of information technology demands devices with high storage density, energy efficiency, and fast read/write speeds. As traditional silicon-based electronics approach their physical limits, new paradigms are essential to meet the growing demands of modern technology~\cite{keyes1981limitations,vashchenko2008physical, keyes2001fundamental}. Magnonics, which explores magnons — the quanta of spin waves in magnetic materials — as alternative information carriers, has emerged as a promising direction~\cite{Flebus_2024, ChumakIEEE2022, AFM_Rev1, AFM_Rev2, spaldin2010magnetic, duine2018synthetic, han2023coherent, jungwirth2016antiferromagnetic, qaiumzadeh2018,chumak2015magnon, chumak2019magnon, wang2020magnonic, hortensius2021coherent}. Among magnetic materials, antiferromagnets (AFMs) stand out due to their reduced magnetic noise and high-frequency operation up to the terahertz (THz) range~\cite{BLS_Rev}. Effectively manipulating and controlling magnons in AFMs is currently an active area of research. Various external stimuli, such as electrical currents~\cite{vzelezny2014, wadley2016electrical, moriyama2018spin} and short laser pulses~\cite{wienholdt2012thz, bossini2019laser, olejnik2018terahertz, satoh2015writing, KIMEL2024172039, nvemec2018antiferromagnetic, surynek2024picosecond}, are being investigated for this purpose.

An alternative approach for magnon manipulation is to leverage the interaction with electromagnetic cavities, which enhances the coupling between magnons and confined photons~\cite{Silvia_Book, OMag_Book, rameshti2022cavity, lee2023cavity, sharma2022protocol}. While most efforts to date have concentrated on the ferrimagnetic material Yttrium Iron Garnet (YIG) coupled to microwave~\cite{zhang2014strongly, hisatomi2016bidirectional} and optical cavities ~\cite{kusminskiy2016coupled, liu2016optomagnonics, osada2018brillouin, liang2023all, sharma2019optimal, haigh2016triple, simic2020coherent, sharma2024tomo}, the interplay between electromagnetic cavities and antiferromagnetic magnons has begun to be explored both theoretically~\cite{parvini2020, xiao2019magnon, curtis2022cavity} and experimentally~\cite{boventer2023antiferromagnetic, bialek2021strong, zhang2021zero}. In optical cavities, AFM magnons can predominantly couple to photons through one-magnon and two-magnon Raman scattering processes, involving three- and four-particle interactions, respectively~\cite{PhysRev.Fleury, fleury1966light, loudon1968theory}.  It has been proposed that coherent, cavity-enhanced one-magnon scattering could be harnessed e.g. for magnon cooling and quantum memory protocols in Rutile-AFMs with two collinear spin sublattices (e.g., MnF$_2$, FeF$_2$, NiO)~\cite{parvini2020}. However, unlike ferromagnets, two-magnon scattering in AFMs dominates their one-magnon counterpart as the former arises from the strong exchange interactions between opposite spin sublattices. Therefore, in this study, we concentrate on the regime of cavity-enhanced two-magnon Raman scattering in AFMs.

Experimental evidence demonstrates that this scattering in MnF$_2$ and FeF$_2$ is 2-3 times stronger than one-magnon Raman scattering~\cite{fleury1967two, PhysRevB.35.1973, cottam1983raman}. Furthermore, in high-transition temperature ($T_c$) superconductors such as La$_{2}$CuO$_4$ and YBa$_2$Cu$_3$O$_6$ (with spin $S=\frac{1}{2}$), it becomes the dominant light-matter coupling mechanism within their insulating phases~\cite{Mott_Rev, Moriya_2003, erlandsen2019enhancement, canali1992theory}. In the cuprate superconductors, quantum spin fluctuations play a crucial role in the electric conductivity within the CuO$_2$ planes~\cite{kastner1998magnetic, betto2021multiple, scalapino2012common, mitrano2016possible, mankowsky2014nonlinear}. By creating magnon pairs \cite{Odagaki1971, fedianin2023selection}, two-magnon Raman scattering can be used for generating squeezed magnon states~\cite{zhao2004magnon}, where spin fluctuations can fall below the quantum vacuum noise level. This opens exciting possibilities beyond established phonon squeezing applications~\cite{hu1996quantum, hu1996squeezed}. Therefore, the interest in exploring the dynamics of an optical cavity coupled to two-magnon Raman processes is two-fold: on the one hand, this kind of interaction goes beyond the usual ``optomechanical interaction'' paradigm~\cite{aspelmeyer2014cavity} (given by an interaction term of the kind $g_0\hat{a}^\dagger\hat{a}(\hat{m}+\hat{m}^\dagger)$ where $g_0$ is the vacuum coupling strength, $\hat{a}^{(\dagger)}$ are the photonic annihilation (creation) operators, and, in the case of magnons, $\hat{m}^{(\dagger)}$ are the magnonic annihilation (creation) operators), where protocols for manipulation of magnon states, e.g. for cooling or the generation of quantum states, are well understood. On the other hand, the possibility of manipulating magnon states by cavity-enhanced two-magnon Raman scattering could provide insights into how to control fluctuations in strongly correlated systems supporting antiferromagnetism and their potential impact on the underlying electronic phase~\cite{erlandsen2019enhancement}. 

In this work, we explore the coupled dynamics of optical photons in a cavity and two-magnon processes focusing on two examples of Rutile-AFM insulators and cuprate parent compounds. The outline of the paper is as follows. In Sec.~\ref{II}, we present the theoretical model based on the magnon-photon interaction Hamiltonian. The emergence of magnon-pair operators within the total Hamiltonian leads us to use the Perelomov representation for magnon operators~\cite{perelomov1975coherent, Kastrup2007, novaes2004some, Aravind:88}, satisfying the commutation relations of the SU(1,1) Lie algebra~\cite{gerry1985dynamics}. We end the section by deriving the coupled equations of motion for the system in the presence of a drive laser and losses. In Sec.~\ref{sec:FastCavity} we obtain effective equations of motion in the fast cavity limit, where the dynamics of the two-magnon operators are slow compared to the characteristic time scales of the cavity photons. By integrating out the photons, we obtain an optically induced (anti-) damping term, reminiscent of an anisotropic Gilbert term in the SU(1,1) space, and we show that it can surpass the intrinsic magnon dissipation, even with a modest number of circulating photons, enabling dynamic control of the Perelomov coherent states through the interplay of cavity drive and dissipation. We explore the possible attractors of the resulting nonlinear system,  showing the possibility of steady-state squeezed Perelomov coherent states. In Sec.~\ref{IV}, we further characterize the system's full dynamics beyond the fast cavity limit, showing the possibility of bringing the system into a chaotic attractor in the blue-detuned regime. This regime is of interest for applications in spin-torque oscillator nano-oscillators (STNOs) and unconventional signal-processing schemes~\cite{mayergoyz2009nonlinear, elyasi2020resources, rezende1990spin, yuan2022quantum}. In App.~\ref{app:AFM_Ham} we present the derivation of the AFM Hamiltonian for Rutile AFMs and cuprate parent compounds, along with the magnon frequency spectrum and density of states along the Brillouin zone. For more details on the full dynamics of magnon-pairs, we direct readers to App.~\ref{app:DROC}. App.~\ref{app:Perelomov} provides the derivation of the squeezing formula for Perelomov operators.

\section{Theoretical Model}
\label{II}

We theoretically model a high-finesse optical cavity externally pumped by a laser, with an AFM insulator placed inside, as illustrated in Fig.~\ref{Fig1}. We focus on AFMs where the dominant coupling mechanism with optical photons arises from two-magnon Raman scattering, specifically: (i) Rutile-AFM compounds, such as XF$_2$ (X=Fe, Mn), and (ii) cuprate parent compounds exhibiting Mott insulating behavior, e.g. La$_2$CuO$_4$ and YBa$_2$Cu$_3$O$_6$ (YBCO). The spin configurations of the AFMs are shown in Fig.~\ref{Fig1}(c,d). It is important to note that the spin order in cuprates can exhibit complex structures, including in-plane, out-of-plane, and canted configurations, depending on the experimental conditions and composition~\cite{coldea2001spin, bonesteel1993theory, curtis2022cavity, chen2011angle, govind2001magnetic, dalla2016excitation}. However, the primary focus of this study is to formulate the system's dynamics in the presence of an unconventional coupling mechanism, namely, the two-magnon Raman scattering process. To this end, we have adopted a simplified two-dimensional geometry, disregarding interlayer couplings and assuming an out-of-plane spin order~\cite{wan2009calculated, sandvik1998numerical, manousakis1991spin, chakravarty1989two}. Our work paves the way for further studies considering bilayer cuprates and more complex spin order configurations.

\subsection{Microscopic Hamiltonian}

We consider a spin Hamiltonian given by
\begin{equation}
   \hat{\mathcal{H}}_m = \sum_{\mathbf{R}} \left[ \sum_{\mathbf{R}'} J_{\mathbf{R'}} \hat{\boldsymbol{S}}_{\mathbf{R}} \cdot \hat{\boldsymbol{S}}_{\mathbf{R} + \mathbf{R}'} - D \left(\hat{S}_{\mathbf{R}}^{z}\right)^{2}\right], 
   \label{Spi-Ham}
\end{equation}
where $\hat{\boldsymbol{S}}_{\mathbf{R}}$ denotes a spin-$S$ particle located at site $\mathbf{R}$, and $J_{\mathbf{R}'}$ is the exchange interaction between spins separated by $\mathbf{R}'$. In the case of two-magnon Raman scattering, momentum conservation allows the excitation of magnons with any equal and opposite momentum $\bf{k}$. In our model, we consider interactions up to the third nearest neighbor, as depicted in Fig.~\ref{Fig1}, since $J_2$ and $J_3$ are relevant when considering magnons with non-zero momentum (${\bf{k}} \neq 0$), see Eq. A2 (App.~\ref{app:AFM_Ham}). The second term in Eq.~\eqref{Spi-Ham} denotes an easy-axis anisotropy aligning the N\'eel vector along the $\boldsymbol{z}$ direction. This is associated with the anisotropy field $H_A$ through $D = g\mu_BH_A/(2S)$, where $g$ is the spectroscopic splitting factor and $\mu_B$ is the Bohr magneton. Employing a Holstein-Primakoff transformation~\cite{holstein1940field}, applying a Fourier transformation, and a 
Bogoliubov transformation (see App.~\ref{app:AFM_Ham}), the Hamiltonian in Eq.~\eqref{Spi-Ham} becomes
\begin{equation}
    \hat{\mathcal{H}}_{m} = \hbar\sum_{{\bf k}} \omega_{m\mathbf{k}} \left[\alpha_{{\bf k}}^{\dagger}\alpha_{{\bf k}} +\beta_{-{\bf k}}^{\dagger} \beta_{-{\bf k}}+1 \right],
    \label{Hm}
\end{equation} 
where $\alpha_{\bf k}$ and $\beta_{\bf -k}$ ($\alpha_{\bf k}^\dagger$ and $\beta_{\bf -k}^\dagger$) are the annihilation (creation) operators of bosonic magnon modes with eigenfrequencies $\omega_{m \mathbf{k}}$. The eigenfrequencies of the considered cuprates are higher than those of Rutile-AFMs owing to the presence of stronger exchange interactions (App.~\ref{app:AFM_Ham}).

We consider an intercavity electric field $\hat{\mathbf{E}} = \hat{\boldsymbol{\cal E}} + \hat{\boldsymbol{\cal E}^{\dagger}}$ such that
\begin{equation} 
    \hat{\boldsymbol{\cal E}}(\mathbf{R}) = i \sqrt{\frac{2\hbar \omega_c}{\varepsilon V}}\ \boldsymbol{p}  \hat{c} e^{i\bf {q}.\bf R}.
    \label{quantization}
\end{equation}
Here, $\varepsilon$ and $V$ are the permittivity and volume of the cavity, respectively. $\boldsymbol{p}$, $\mathbf{q}$, $\omega_c$, and $\hat{c}$ denote the polarization, wavevector, resonance frequency, and annihilation operator of the cavity photon, respectively. 
The empty-cavity Hamiltonian is $\hat{\mathcal{H}}_c = \hbar \omega_c \hat{c}^\dagger \hat{c}$.

The interaction between the electric field in the cavity and the AFM is described through linear and quadratic magneto-optical couplings~\cite{cottam1975temperature, cottam1986light, moriya1967theory}. The corresponding interaction Hamiltonian is given by
\begin{equation}
    \hat{\mathcal{H}}_{\rm int} = \sum_{\mu,\nu} \sum_{\bf{R}} \hat{\cal E}^{\mu}(\mathbf{R}) \hat{\cal E}^{\nu,\dagger}(\mathbf{R}) \hat{\chi}^{\mu\nu}\left(\mathbf{R}\right),
    \label{Hint:Micro1}
\end{equation}
where $\hat{\cal E}^{\mu}(\mathbf{R})$ is the $\mu$-component of $\hat{\boldsymbol{\cal E}}$ at site $\bf{R}$ in the crystal and $\hat{\chi}^{\mu\nu}({\bf R})$ represents a component of the spin-dependent polarizability tensor,
\begin{equation}
\begin{aligned}
    \hat{\chi}^{\mu\nu}({\bf R}) &= \sum_{\gamma}K_{\mu\nu\gamma}\left(\mathbf{R}\right)\hat{S}_{\mathbf{R}}^{\gamma}+\sum_{\gamma,\delta}G_{\mu\nu\gamma\delta}\left(\mathbf{R}\right)\hat{S}_{\mathbf{R}}^{\gamma}\hat{S}_{\mathbf{R}}^{\delta} \\
    &+\sum_{\gamma,\delta,\mathbf{R^\prime}}B_{\mu\nu\gamma\delta}\left(\mathbf{R},\mathbf{R^\prime}\right)\hat{S}_{\mathbf{R}}^{\gamma}\hat{S}_{\mathbf{R}+\mathbf{R^\prime}}^{\delta}+...,
    \label{suscep}
\end{aligned}
\end{equation}
where $\gamma,\delta,\mu,\nu\in \{x,y,z\}$. The magnitudes represented by $K$, $G$, and $B$ characterize magneto-optical effects~\cite{cottam1975temperature,PhysRevB.35.1973, bostrom2023direct, lockwood2012magnetooptic}. The first two terms in Eq.~\eqref{suscep} involve spin operators at a single site $\bf{R}$, whereas the third term involves a pair of operators at different sites. The first and second terms arise primarily from indirect electric dipole coupling mediated by spin-orbit interactions~\cite{PhysRev.Fleury}. The third term is associated with the excited-state exchange interaction and involves electric-dipole transitions. In most AFM insulators, the third term emerges as the predominant light-matter interaction~\cite{fleury1967two,PhysRev.Fleury}. This interaction can lead to a non-thermal excitation of magnon-pairs when the incident photon intensity exceeds a certain threshold~\cite{Odagaki1971,odagaki1973_part1, odagaki1973_part2}. For a monochromatic photonic cavity, the principle of conservation of energy gives rise to two distinct types of two-magnon Raman scattering processes: Stokes (Fig.~\ref{Fig1}(a)) and anti-Stokes scattering (Fig.~\ref{Fig1}(b)). Raman experiments typically involve infrared to visible photons with frequencies $\sim \SI{500}{\tera\hertz}$, corresponding to wavevectors $|{\bf {q}}|\sim \SI{15}{\micro\meter^{-1}}$ which is significantly smaller than typical magnon momenta throughout the Brillouin zone. Consequently, this process leads to the creation of magnon-pairs with approximately opposite wave vectors \cite{formisano2024coherent}. Group theory analysis~\cite{fleury1967two, dimmock1962symmetry, loudon1968theory, sugano2013magneto, fleury1970magnon, amer1975two, weber2000raman, devereaux2007inelastic, poppinger1977temperature, davies1971spin, vernay2007momentum, sandvik1998numerical, chubukov1995resonant} reveals that the interaction Hamiltonian for simple cubic lattices, considering only the third term in Eq.~(\ref{suscep}), simplifies to
\begin{equation}
    \hat{\mathcal{H}}_{int}=-\sum_{\mathbf{R}\mathbf{R'}} \frac{B}{|{\mathbf{R'}}|^2}  \left( \hat{\boldsymbol{\mathcal{E}}} \cdot \mathbf{R'}\right) \left( \hat{\boldsymbol{\mathcal{E}}}^\dagger \cdot \mathbf{R'}\right)\hat{\boldsymbol{S}}_{\mathbf{R}} \cdot \hat{\boldsymbol{S}}_{\mathbf{R} + \mathbf{R'}}, \label{Hint:Micro2}
\end{equation}
where the variable $\mathbf{R'}$ spans over nearest neighbors. Theoretically, the coupling coefficient $B$ in cuprates can be derived from a half-filled single-band Hubbard model, given by $B = \frac{4t^2}{|\hat{\cal E}|^2(U - \hbar\omega_c)}$, where $t$ represents the hopping amplitude between nearest neighbor sites, and $U$ is the on-site electron-electron interaction~\cite{shastry1990theory, canali1992theory}. For Rutiles, its derivation from time-dependent perturbation theory is detailed in~\cite{PhysRev.Fleury, fleury1967two}. From an experimental perspective, the exact value of $B$ in Rutiles is not known (we only have ratios between different scattering geometries~\cite{PhysRevB.73.184434, PhysRev.Fleury}) whereas, for cuprates, $U$ and $t$ have been obtained by connecting spectroscopic measurements to the microscopic model~\cite{sheshadri2023connecting}.

\subsection{Coupling constants}

\begin{figure}[t!]
  \centering
    \centering
    \includegraphics[width=0.47\textwidth]{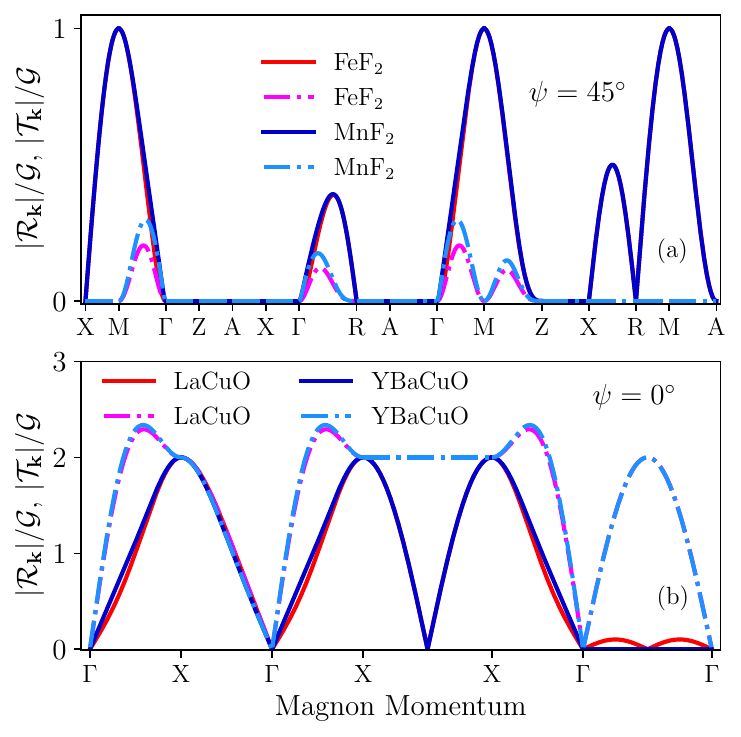}
    \caption{The magnitudes of the coefficients $\mathcal{R}_{\bf k}$ (dash-dot lines) and $\mathcal{T}_{\bf k}$ (solid lines), normalized to the coupling constant, at different wave-vectors for (a) Rutile-AFMs and (b) cuprates-AFMs. See Figs.~\ref{FigSM:Rutile}(b) and~\ref{FigSM:cuprates}(b) for the definitions of the wave-vector points on the x-axis.}
    \label{fig:micro_coup}
\end{figure}

This subsection discusses the effective interaction Hamiltonian for a given light polarization. Our analysis focuses on single-mode cavities with electric fields linearly polarized in the XY plane, described by the polarization vector
\begin{equation}
   \mathbf{p} = \cos\psi \mathbf{x} + \sin\psi \mathbf{y}.
\end{equation}

This specific choice is motivated by studies showing strong two-magnon Raman scattering in this geometry~\cite{PhysRevB.73.184434, lockwood2012magnetooptic,PhysRevB.35.1973, fleury1967two, sandvik1998numerical, cottam1983raman}. Expanding spins in the magnon basis, the interaction Hamiltonian in Eq.~\eqref{Hint:Micro2} becomes 
\begin{equation}
    \begin{aligned}
    \hat{\mathcal{H}}_{\rm int} &= \hbar\sum_{{\bf k}}\left[\mathcal{R}_{{\bf k}}\left(\hat{\alpha}_{{\bf k}}^{\dagger}\hat{\alpha}_{{\bf k}}+\hat{\beta}_{-\mathbf{k}}^{\dagger}\hat{\beta}_{-{\bf k}}+1\right)\right.\\
    &\qquad\qquad\left.+\mathcal{T}_{{\bf k}}\left(\hat{\alpha}_{{\bf k}}\hat{\beta}_{-{\bf k}}+\hat{\alpha}_{\mathbf{k}}^{\dagger}\hat{\beta}_{-\mathbf{k}}^{\dagger}\right)\right] \hat{c}^{\dagger}\hat{c},
    \end{aligned}
    \label{int}
\end{equation}
where $\mathcal{R}_{{\bf k}}=-\mathcal{G}\Pi_{\mathcal{R},\bf{k}}^{xy}\sinh\theta_{{\bf k}}$, $\mathcal{T}_{{\bf k}}=\mathcal{G}\Pi_{\cal{T},\bf{k}}^{xy}\cosh\theta_{{\bf k}}$ and $\theta_{\bf k}$ is the Bogoliubov angle defined in App.~\ref{app:AFM_Ham} quantifying the two-mode squeezing between the two sublattices. The coupling factor is given by
\begin{equation}
    \mathcal{G} = 2B \left( \frac{a}{d_{\rm neigh}} \right)^2 \frac{\omega_{c}S}{\varepsilon V},
\end{equation} 
where $d_{\rm neigh}$ is the nearest neighbor distance, given by $\sqrt{2a^2 + c^2}/2$ for rutile-AFMs and by $a$ for 2D cuprates, with $a$ and $c$ being the lattice constants, see App.~\ref{app:AFM_Ham}. To simplify calculations, we assumed equal cavity and AFM volumes as $V$. In reality, the filling factor (ratio of AFM to cavity volume) is crucial, as it significantly affects the light-matter interaction strength~\cite{weichselbaumer, graf2021design}. 

The factor $\Pi^{xy}$ encodes the dependence on the lattice structure and the optical polarization, which will be discussed in the following subsections. Eq.~\eqref{int} shows that the two-photon two-magnon light-matter interaction not only leads to the creation and annihilation of magnon-pairs (governed by the $\mathcal{T}_{{\bf k}}$ coefficient) but also induces frequency shifts (governed by the $\mathcal{R}_{{\bf k}}$ coefficient). For Rutile-AFMs, we have
\begin{equation}
    \Pi_{{\cal T},\mathbf{k}}^{xy}=\Pi_{{\cal R},\mathbf{k}}^{xy} = \sin2\psi \sin\left(\frac{ak_x}{2}\right) \sin\left(\frac{bk_y}{2}\right) \cos\left(\frac{ck_z}{2}\right),\nonumber
\end{equation}
and for 2D cuprates
\begin{equation}
\begin{aligned}
    \Pi_{{\cal T},\mathbf{k}}^{xy} &= 2\left[\cos^{2}\psi \cos\left(ak_{x}\right) + \sin^{2}\psi \cos\left(ak_{y}\right) - \tanh(\theta_{{\bf k}})\right]\nonumber\\
    \Pi_{{\cal R},\mathbf{k}}^{xy}&=2\left[\cos^{2}\psi \cos\left(ak_{x}\right) + \sin^{2}\psi \cos\left(ak_{y}\right) - \coth(\theta_{{\bf k}})\right].
\end{aligned}
\end{equation}

These expressions show that Rutile-AFMs exhibit maximum coupling at a polarization angle of $\psi=\pi/4$, whereas cuprates achieve maximum coupling at $\psi=0^\circ$ or $\psi=\pi/2$. Fig.~\ref{fig:micro_coup} illustrates the variation of $\mathcal{R}_{\bf k}$ and $\mathcal{T}_{\bf k}$ in wavevector space, revealing that $\mathcal{T}$ peaks at the M-point for Rutile-AFMs and at the X-point for cuprates. Notably, at these points, the magnon density of states (DOS) exhibits a significant peak closely associated with van Hove singularities, as shown in Figs.~\ref{FigSM:Rutile}(d) and~\ref{FigSM:cuprates}(d). This suggests that these wavevectors—the M-point for Rutile and the X-point for cuprates—play a dominant role in light-matter coupling. Consequently, we focus our analysis on these dominant wavevectors. Neglecting the $\mathbf{k}$-index, the total effective Hamiltonian can be expressed as:
\begin{align}
    \frac{\hat{\mathcal{H}}}{\hbar} &= \omega_c \hat{c}^{\dagger} \hat{c} + \omega_{m} \left[ \alpha^{\dagger}\alpha + \beta^{\dagger}\beta+1\right] + \left[\mathcal{R}\left(\hat{\alpha}^{\dagger}\hat{\alpha}+\hat{\beta}^{\dagger}\hat{\beta}+1\right)\right.\nonumber \\
    &\quad\left.+\mathcal{T}\left(\hat{\alpha}\hat{\beta}+\hat{\alpha}^{\dagger}\hat{\beta}^{\dagger}\right)\right] \hat{c}^{\dagger}\hat{c}.
    \label{Total_H}
\end{align}

\subsection{Perelomov representation}
\label{subsec:Perelomov}
As discussed in the previous section, within our model magnon excitations exclusively manifest in pairs of $\hat{\alpha}_{\bf k}$ and $\hat{\beta}_{-\bf k}$ magnons. This observation leads us to rewrite the Hamiltonian based on two-magnon or equivalently Perelomov operators defined by~\cite{novaes2004some, gerry1991squeezed, gerry1985dynamics} 
\begin{equation} \begin{aligned}
\hat{K}^{+}&=\hat{\alpha}^{\dagger}\hat{\beta}^{\dagger},\qquad\hat{K}^{-}=\hat{\alpha}\hat{\beta}, \\
\hat{K}^{z}&=\frac{1}{2}\left(\hat{\alpha}^{\dagger}\hat{\alpha}+\hat{\beta}^{\dagger}\hat{\beta}+1\right), \label{Def:Coh}
\end{aligned} \end{equation}
which fulfill the commutation relations of a SU(1,1) Lie algebra: $\left[\hat{K}^{z},\hat{K}^{\pm}\right]=\pm\hat{K}^{\pm}$ and $\left[\hat{K}^{-},\hat{K}^{+}\right] = 2\hat{K}^z$ with $\hat{K}^{\pm} = \hat{K}^x \pm i \hat{K}^y$. The Casimir operator $\hat C = ({\hat K}^z)^2 - \frac{1}{2}({\hat K}^+{\hat K}^- + {\hat K}^-{\hat K}^+)$ which commutes with all $\hat{K}$ operators, is a constant of motion. In this representation, the Hamiltonian in Eq.~\eqref{Total_H} reads
\begin{equation}
    \frac{\hat{\mathcal{H}}}{\hbar} = \omega_c \hat{c}^{\dagger} \hat{c} + 2\omega_m\hat{K}^{z} +  \hat{\cal K} \hat{c}^{\dagger}\hat{c}, \label{Ham:eff}
\end{equation}
where we have defined
\begin{equation}
    \hat{\cal K} = 2{\cal R} \hat{K}^z + 2{\cal T} \hat{K}^x.
    \label{cal_K}
\end{equation}
For future convenience, we can also write Eq.~\eqref{cal_K} as $\hat{\cal K} = 2\mathbf{e} \linedot  \hat{\mathbf{K}}$ with $\mathbf{e} = {\cal T} \mathbf{x} - {\cal R} \mathbf{z}$, $\hat{\mathbf{K}}$ is the vector $(K^x,K^y,K^z)$ and we define a scalar product in SU(1,1) as $\mathbf{a} \linedot \mathbf{b} = a_xb_x + a_yb_y - a_zb_z$. 

The SU(1,1) basis states are $|m, k\rangle$ such that $\hat K^z |m, k\rangle = (m + k) |m, k\rangle$  where $m = 0, 1, 2, \ldots$ and $k$ is the Bargmann index. The eigenvalues of the Casimir operator $\hat{C} = (\hat{K}^z)^2 - (\hat{K}^x)^2 - (\hat{K}^y)^2$ are $k(k-1)$ with $k > 0$. Using Eq.~(\ref{Def:Coh}), we can show the relation $2k - 1 = |n_{\alpha} - n_{\beta}|$ where $n_{\alpha,\beta}$ are the number of $\{\alpha,\beta\}$-magnons.

The SU(1,1) Perelomov coherent states are defined as ~\cite{Aravind:88, robert2021coherent,perelomov1972coherent, gerry1991classical}
\begin{equation}
|\zeta, k \rangle = S(z) |0, k \rangle, \label{eq:2.4}
\end{equation}
where 
\begin{equation}
S(z) = \exp(z K^+ - z^* K^-), \label{eq:2.5}
\end{equation}
and \(z = -(\theta/2)e^{-i\phi}\), \(\zeta = -\tanh \left( \frac{\theta}{2} \right) e^{-i\phi}\). The angles \(\theta\) and \(\phi\) parametrize the SU(1,1) group manifold and have ranges \(-\infty < \theta < \infty\) and \(0 \leq \phi \leq 2\pi\).

The expectation values of $\hat{K}^{x,y,z}$ in Perelomov coherent states, denoted as $K^i \equiv \langle \hat{K}^i \rangle_{\text{PCS}}$, are:
\begin{equation}
K^z = K \cosh \vartheta, \quad K^x - iK^y = K \sinh \vartheta e^{-i\varphi}, 
\label{params}
\end{equation}
where the squares of the expectation values fulfill the relation:
\begin{equation}
 (K^x)^2 + (K^y)^2-(K^z)^2 = const. , \label{Ksq:const}
\end{equation}
or in the notation of the SU(1,1) dot product, $K^2 = -\mathbf{K} \linedot \mathbf{K}$, where we define the vector $\mathbf{K} = (K^x, K^y, K^z)$. Using the definition of the Casimir operator above, we can interpret Eq.~(\ref{Ksq:const}) as the constancy of the difference of number of $\alpha$ and $\beta$ magnons. In what follows we will obtain the equations of motion for the expectation values of the Perelomov operators.

\subsection{Coupled equations of motion for a driven system}
In this subsection, we derive the coupled Langevin equations of motion for photons and magnons governed by the Hamiltonian Eq.~(\ref{Ham:eff}), adding a driving source for the cavity and dissipation for both the cavity and magnons.

We consider a cavity subject to an input pump laser with an amplitude of $s_d$ and a frequency of $\omega_d$, giving an input power of $P_d = \hbar\omega_d s_d^2$. Using input-output theory~\cite{GardColl}, the equation of motion for the intracavity photon field in the rotating frame of the drive can be written as
\begin{equation}
    \frac{d\hat{c}}{dt} = -i\left(\Delta + \hat{\cal K} - \frac{i\kappa}{2}\right) \hat{c} + \sqrt{\rho \kappa} s_{d} + \sqrt{\kappa} \hat{c}_{\text{noise}}, \label{EOM:c}
\end{equation}
where $\Delta = \omega_c - \omega_d$ is the drive detuning, $\kappa$ is the cavity linewidth including intrinsic and extrinsic damping, $\rho$ is the ratio of extrinsic to intrinsic damping, $\hat{c}_{\rm noise}$ is the vacuum noise satisfying $\langle \hat{c}_{\rm noise} \rangle = 0$, and $\hat{\cal K}$ is defined in Eq.~\eqref{cal_K}. The dynamics of the magnon annihilation operators are given by
\begin{equation}
    \frac{d\hat{\xi}}{dt} = \frac{i}{\hbar} \left[\hat{\cal H},\hat{\xi}\right] - \frac{\kappa_G}{2} \hat{\xi} - \sqrt{\kappa_G}\hat{\xi}_{\rm noise},
\end{equation}
where $\xi\in\{\alpha,\beta\}$ represents one of the magnon modes, $\hat{\xi}_{\rm noise}$ is the bath noise for the $\xi$-magnons, and $\kappa_G$ is the intrinsic Gilbert damping, which is the same for both modes ~\cite{cottam1983raman, Rezende2020}. The noise satisfies the standard correlation functions~\cite{BP_OQS, RH_OQS}: $\langle \xi_{\rm noise}(t) \rangle = 0$ and $\langle \xi_{\rm noise}^{\dagger}(t) \xi_{\rm noise}(t') \rangle = n_{\rm th} \delta(t-t')$ where $n_{\rm th}$ is the Bose-Einstein distribution at ambient temperature.

From this, we can derive the equations of motion for the magnon-pair operators,
\begin{equation}
    \frac{d\left\langle \hat{K}^{i}\right\rangle }{dt}=\frac{i}{\hbar}\left\langle \left[\hat{{\cal H}},\hat{K}^{i}\right]\right\rangle -\kappa_{G}\left(\left\langle \hat{K}^{i}\right\rangle -\frac{\delta_{iz}}{2}\right),
\end{equation}
where $i\in{x,y,z}$ and expectations are taken over Perelomov coherent states.

In the following, we discuss the semi-classical equations of motion where we ignore quantum correlations between the spins and the photons, i.e. we approximate $\langle \hat{c} \hat{\cal K} \rangle \approx \langle \hat{c} \rangle \langle \hat{\cal K} \rangle $. $\langle \hat{\cal K} \rangle$ is the expectation value of the $\mathbf{K}$-vector in Perelomov coherent states, as explained in Sec.~\ref{subsec:Perelomov}. Defining the averages $c = \langle \hat{c} \rangle$ and $\mathcal{K} = \langle \hat{\cal K} \rangle$ (along with $K^{x,y,z}$ defined above), we can write the coupled semi-classical equations of motion as 
\begin{equation}
    \frac{d c}{dt} = -i\left(\Delta + {\cal K} - \frac{i\kappa}{2}\right) c + \sqrt{\rho\kappa} s_{d}, \label{dcdt:Class}
\end{equation}
and 
\begin{equation}
    {\dot{\bf K}}={\bf{\Omega}} \underline{\times} \mathbf{K} - \kappa_G \left( \mathbf{K} - \frac{\bf z}{2} \right),
    \label{dKdt:Class}
\end{equation}
where $\mathbf{\Omega} = 2\omega_m \mathbf{z} - 2\mathbf{e} |c|^2$ with $\mathbf{e}$ defined below Eq.~(\ref{Ham:eff}). The vector product $\underline{\times}$ in SU(1,1) is defined as 
\begin{equation}
    \mathbf{a} \underline{\times} \mathbf{b} = \begin{pmatrix}
        a_{y}b_{z}-a_{z}b_{y}\\
        a_{z}b_{x}-a_{x}b_{z}\\
        a_{y}b_{x}-a_{x}b_{y}
        \end{pmatrix} .
\end{equation}
It differs from the standard cross-product in the sign of the $z$-component. 

\section{Fast cavity regime}
\label{sec:FastCavity}

In this section, we discuss the fast cavity limit, corresponding to $\kappa^2 \gg \dot{\cal K}$, wherein we can solve the optical equations of motion adiabatically by treating the pseudospin operators parametrically. This allows us to eliminate photons to obtain an effective equation of motion only for the pseudospin in Sec.~\ref{Sec_FC:Eff}. Using this equation, we discuss the fixed points and the stability of magnons in Sec.~\ref{Sec_FC:Stab}. Finally, we discuss numerical simulations showing the time evolution of magnons in Sec.~\ref{Sec_FC:Num}.

\subsection{Effective equation of motion} \label{Sec_FC:Eff}
We can find $c(t)$ in terms of $\mathcal{K}(t)$ following a procedure analogous to~\cite{kusminskiy2016coupled}, where the coupling of light to a SU(2) spin was studied. We expand the photon field in terms of time derivatives of $\mathcal{K}$,
\begin{equation} 
    c(t) = c_{0}(t) + c_{1}(t) + \ldots 
\end{equation}
where the index $i$ of the $c_i$ terms indicates the derivative order. For example, the second order term $c_2$ consists of $\dot{\cal K}^2$ and $\ddot{\cal K}$. 

In steady state, $\dot{c} = 0$. Using Eq.~(\ref{dcdt:Class}), the zeroeth order contribution reads
\begin{equation}
    c_0(t) = \frac{\sqrt{\rho\kappa}s_{d}}{i\left(\Delta + {\cal K}(t) - \frac{i\kappa}{2}\right)}. \label{eq:c0}
\end{equation}
To find the first order correction $c_{1}$, we note that the time derivative $\dot{c}_1$ is a second order term and thus can be ignored, giving
\begin{equation}
    c_1(t) = \frac{-ic_0(t)}{\left(\Delta + {\cal K}(t) - \frac{i\kappa}{2}\right)^{2}}\dot{\mathcal{K}}(t).
\end{equation}
The number of photons circulating in the cavity is given by $n_c=\left|c\left(t\right)\right|^{2}$ which can be approximated as $\left|c(t)\right|^{2} \approx \left|c_{0}(t)\right|^{2} + c_{0}^{*}(t) c_{1}(t) + c_{1}^{*}(t) c_{0}(t)$. Explicitly, 
\begin{equation}
    n_c \approx |c_{0}|^2\left[1 + \frac{2\kappa \left(\Delta + {\cal K} \right)\dot{\mathcal{K}}}{D({\cal K})^{2}}\right], \label{nc:approx}
\end{equation}
where we have defined $D({\cal K}) = \left(\Delta+\mathcal{K}\right)^{2}+\kappa^{2}/4$.

Substituting Eq.~\eqref{nc:approx} into the equations of motion for the pseudospin vector $\mathbf{K}$, Eq.~(\ref{dKdt:Class}), gives
\begin{equation}
    \dot{\mathbf{K}} = \boldsymbol{\Omega}_{\text{eff}} \underline{\times} \mathbf{K} - \eta_{\text{opt}} \frac{\mathbf{e} \linedot \dot{\mathbf{K}}}{|\mathbf{e}|^2} \left(\mathbf{e}\underline{\times} \mathbf{K}\right) - \kappa_G \left(\mathbf{K}-\frac{\boldsymbol{z}}{2}\right),
    \label{EffDyn}
\end{equation}
where $\boldsymbol{\Omega}_{\rm eff} = 2\omega_m \mathbf{z} - 2\mathbf{e}|c_0|^2 $ includes the two-magnon frequency shift due to the instantaneous interaction with the cavity field, and $\eta_{\text{opt}}$ is an optically induced dissipation due to dynamical backaction and is given by
\begin{equation}
    \eta_{\text{opt}}(\mathbf{K}) = 8\kappa\left(\mathcal{T}^{2} + \mathcal{R}^2\right) \left|c_{0}\right|^{2}\frac{\left(\Delta + 2\mathbf{e} \linedot \mathbf{K} \right)}{D(\mathcal{K})^2}.
\end{equation}
Within this approximation, light induces a cooperative dissipation of magnon-pairs. The form of damping has similarities with the Gilbert damping term $\alpha_{G}\dot{\mathbf{S}}\times\mathbf{S}$, well known in SU(2) spin dynamics~\cite{Gilbert2004}. The sign of $\eta_{\rm opt}$ depends on the pseudospin components $K^i$ and the detuning of the optical drive $\Delta$. The pseudospin decays for $\eta_{\rm opt}>0$, while $\eta_{\rm opt}<0 $ leads to an amplification of the pseudospin ultimately limited by nonlinearities. We discuss these conditions in more detail in the following section.

\subsection{Stability of the steady state} \label{Sec_FC:Stab}

In this section, we discuss the fixed points of the pseudospin governed by the equilibrium points of the effective equation of motion, Eq.~\eqref{EffDyn}. The damping rate in typical insulating AFMs is very small ($\sim \SI{1}{\GHz}$~\cite{poppinger1977temperatureII,PhysRevB.35.1973, ohlmann1961antiferromagnetic, barak1980parallel, cottam1983raman, kotthaus1972,vaidya2020subterahertz}) and thus, to focus on the optically induced dynamics, we consider time evolutions at time scales much smaller than $\kappa_G^{-1}$. At such times, the quasi-equilibrium condition for the pseudospin vector is given by $\boldsymbol{\Omega}_{\rm eff} \underline{\times} \mathbf{K}_e = 0$ implying $\mathbf{K}_e \propto \boldsymbol{\Omega}_{\rm eff}$. $\mathbf{K}_e$ indicates the quasi-equilibirum value of the pseudospin. The normalization is set by the initial condition via the conservation law $\mathbf{K}(t)\linedot\mathbf{K}(t) = -K^2$ satisfied by Eq.~(\ref{EffDyn}) when the Gilbert term is ignored. The equilibrium pseudospin lies in the x-z plane as $\boldsymbol{\Omega}_{\rm eff}\cdot\mathbf{y} = 0$. Using the definitions for the coherent state given in Eq.~\eqref{params}, in equilibrium we obtain $\varphi_e = \pi$ and 
\begin{equation}
    \tanh\vartheta_e = \frac{\mathcal{T} |c_0|^2}{\omega_m + \mathcal{R}|c_0|^2}.
    \label{attractor}
\end{equation}

We see that if $\mathcal{T} > \mathcal{R}$, i.e. two-magnon scattering is stronger than the light-induced frequency shift of magnons, for a sufficiently large number of photons we get an$\tanh\vartheta_e > 1$. Physically, this implies that the system becomes unstable~\cite{odagaki1973_part2}. To assess the stability of the system, we linearize the equations of motion around the equilibrium point by setting $K^i = K^i_e + \delta K^i$. In the resulting dynamical matrix for $\delta K^i$, the eigenvalues should be non-positive for the system to be stable. One eigenvalue of the dynamical matrix is always zero because of the conservation law $K=\text{const.}$ The other two eigenvalues $\lambda_{\pm}$ satisfy
\begin{equation} \begin{aligned}
    \lambda_+ + \lambda_- &= -\eta_e \frac{2\omega_{m} {\cal T} \left({\cal T}K_{e}^{z}-{\cal R}K_{e}^{x}\right)}{{\cal T}^{2}+{\cal R}^{2}}, \\
    \lambda_+ \lambda_- &= \left(\omega_m + {\cal R}|c_0|^2 \right)^{2} - |c_0|^4 {\cal T}^{2} \\
    &\phantom{=} -\left({\cal T}K_{e}^{z}-{\cal R}K_{e}^{x}\right){\cal T}\omega_{M}\frac{\partial |c_{0}|^{2}}{\partial \mathcal{K}},
\end{aligned} \label{EigsDyn} \end{equation}
where $\eta_e \equiv \eta_{\rm opt}(\mathbf{K}_e)$ is the damping constant at equilibrium.

For stability, $\lambda_{+}\lambda_{-}> 0$ which is satisfied for small photon numbers. Moreover, $\lambda_{+}+\lambda_{-} < 0$ implying $\eta_e > 0$ that corresponds to $\Delta > -2\mathbf{e}\linedot \mathbf{K}$. At the equilibrium point $\mathbf{K}_e \propto \boldsymbol{\Omega}_{\rm eff}$ with $K_{e}^{x}\leq0$, the condition $\eta_e>0$ is never fulfilled for blue-detuning $\Delta<0$. We can therefore conclude that there are no fixed point attractors at blue-detuning. For red-detuning $\Delta>0$, there can be fixed point attractors when $\eta_e>0$ is fulfilled. 

\subsection{Numerical simulations}
\label{Sec_FC:Num}
In this section, we numerically analyze the nonlinear dynamics of the pseudospin $\mathbf{K}$ governed by Eq.~(\ref{EffDyn}) and guided by the stability analysis from the previous subsection. To facilitate the numerical analysis, parameters $\Delta$, $\cal T$, $\cal R$, $\kappa$, and $\kappa_G$ are expressed in dimensionless units of $2\omega_m$ with primes, such as $\Delta' = \Delta/(2\omega_m)$. Moreover, we define time $t'=2\omega_m t$ and light amplitude ${\Tilde{s}}_d=s_d\sqrt{\rho/2\omega_m}$. We neglect the magnon frequency shift induced by two-magnon Raman scattering, as it is negligible compared to the inherent magnon frequencies (\({\cal R}' \ll 1 \)). Future work can explore its influence. To have an estimation of \({\cal T}'\) in Rutiles, we consider \(B \approx 0.5 \times 10^{-20}\)~\cite{PENG2004306} and a refractive index \(n \approx 1.5\)~\cite{jahn1973linear}, a cavity frequency of 650 THz, and a volume \(V = 10^{-18} \, \text{m}^3\), resulting in \({\cal T}' \approx 0.2\) for magnons at the M-point. In cuprates, where \(\mathcal{G}\) is approximated as \(SJ_1\), numerical evaluations lead to \({\cal T}' = {\cal R}' = 0.2-0.3\) for magnons at the X-point. We set \({\cal T}' \approx 0.2\) and \({\cal R}' \approx 0\) for computational simplicity and compatibility across both materials.

\begin{figure}[t!]
  \centering
  \includegraphics[width=0.49\textwidth]{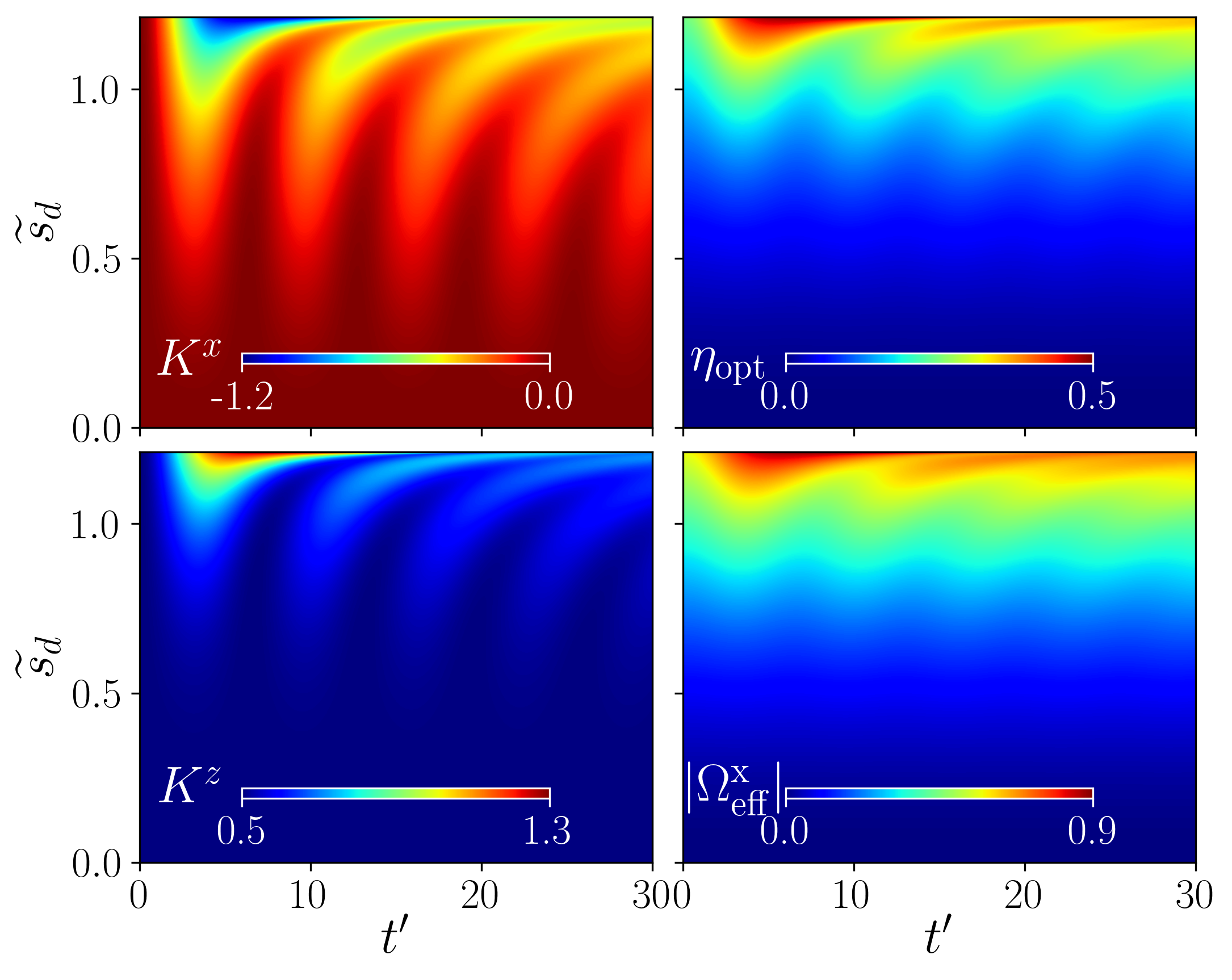}
  \caption{$K_x$, $K_z$, $\mathrm{\eta_{opt}}$, and $\Omega_{\rm opt}^x$ as functions of the drive field power and time in the fast cavity regime. The initial conditions for all the simulations are $K^z = 1/2$ and $K^x = K^y = 0$. The parameters are $\Delta'=1$ (red-detuned), $\kappa' = 2$, $\mathcal{T}'=0.2$, $\mathcal{R}'=0$, $\kappa_G'=0$ and $\rho=0.5$. Here we sweep $\tilde{s}_d$ from zero up to the threshold where the system becomes unstable, determined as  $\tilde{s}_d=1.2126$. For a light frequency $\omega_c=650$THz and a magnon frequency either at the M- or X-points this corresponds to input powers of $P_d\approx3.5\mu W$ and  $P_d\approx9.3\mu W$ respectively.}
\label{fig:FastCavity}
\end{figure}

\begin{figure*}
  \centering
  \includegraphics[width=0.49\textwidth, height=0.27\textwidth]{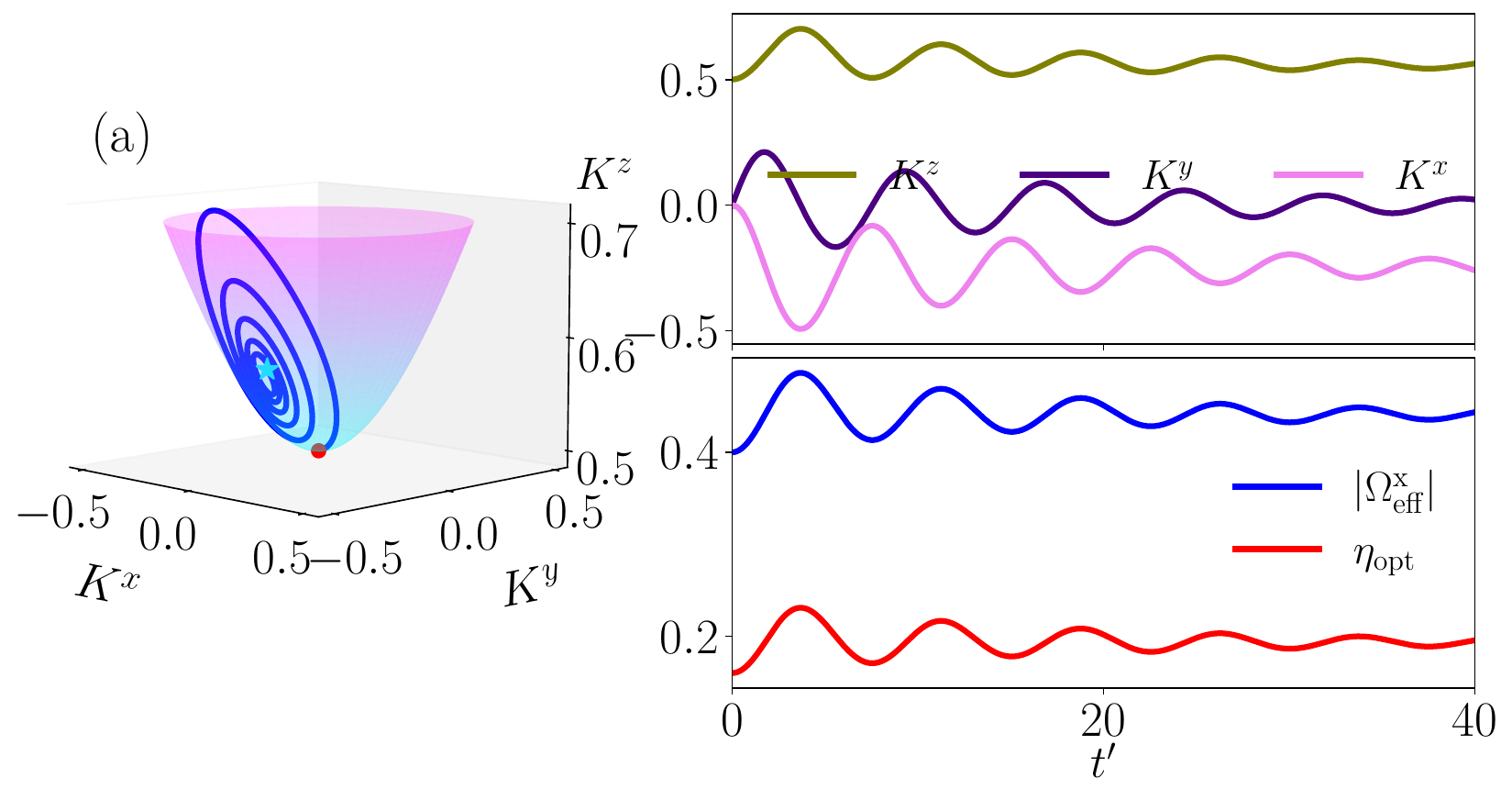}
  \includegraphics[width=0.49\textwidth, height=0.27\textwidth]{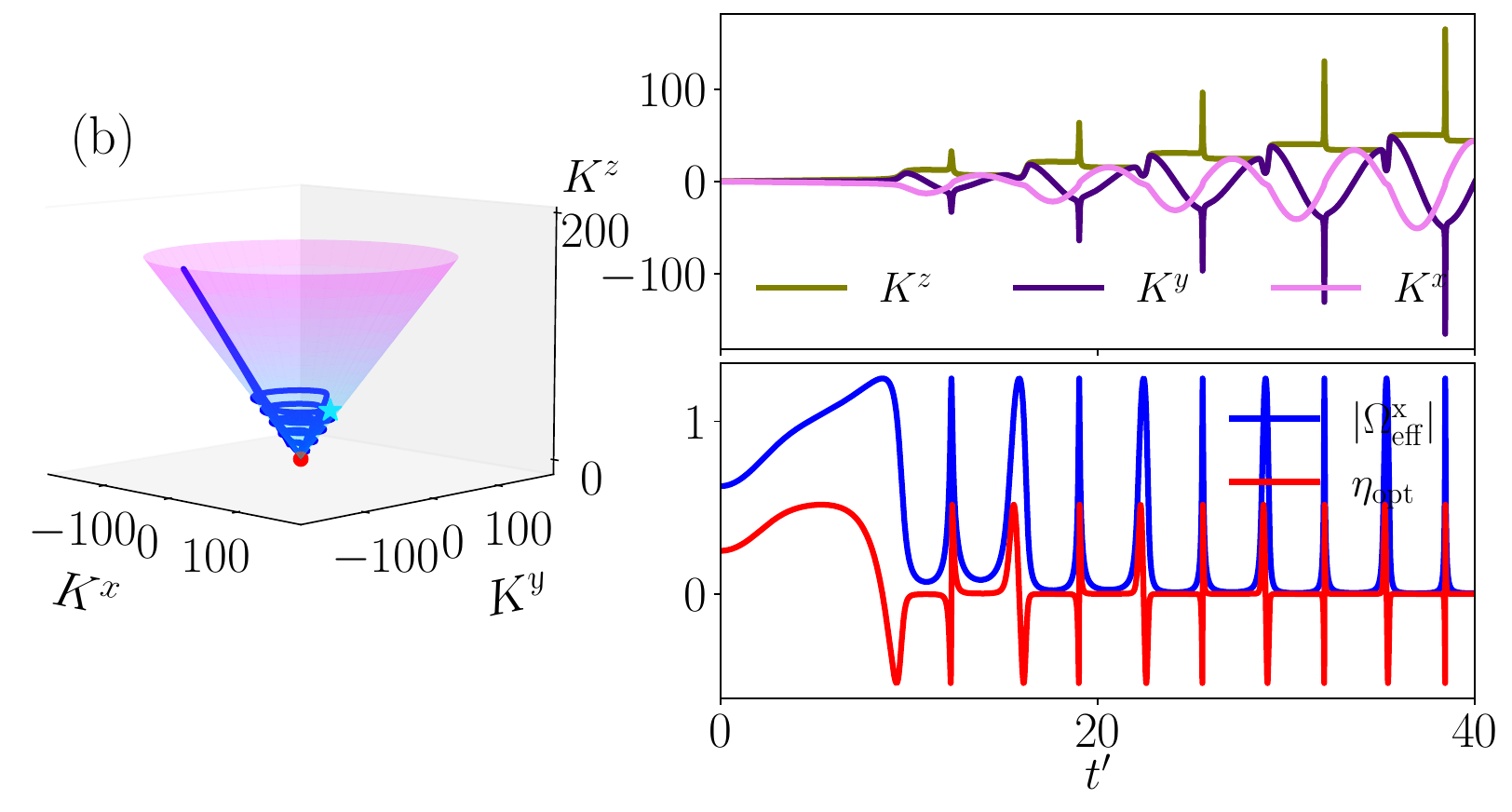}
  \caption{Trajectories of the pseudospin tip on the upper hyperboloid sheet, along with the temporal evolution of the pseudospin tip coordinates, as well as the time-dependent profiles of the optically induced frequency shift ($|\Omega_{\rm eff}^x|$) and dissipation coefficient ($\eta_{\rm opt}$). For (a) ${\Tilde{s}}_d=1.0$ (below power threshold) and (b) ${\Tilde{s}}_d=1.25$ (above power threshold). The red sphere and star mark the start and end of the trajectory, respectively. Parameters are similar to those in Fig.\ref{fig:FastCavity}.}
\label{fig:TrajFastCav}
\end{figure*}

The time evolution of \( K_x \), \( K_z \), \( \eta_{\text{opt}} \), and \( \Omega_{\text{opt}}^x \) under a red-detuned pump (\(\Delta' = 1\)) and varying drive power (${\Tilde{s}}_d$) in the fast cavity limit is illustrated in Fig.~\ref{fig:FastCavity}. Below a threshold power, chosen as the maximum value of $\Tilde{s}_d$ in the plots, the system dynamics resemble those of a damped harmonic oscillator, where after several oscillations, the K-components settle at specific equilibrium values, indicating a constant number of magnon pairs. For the input amplitudes chosen here, $\Delta - 2{\cal T}|c_0|^2 > 0$ is fulfilled implying $\eta_e > 0$. Looking at the stability conditions given in Eqs.~(\ref{EigsDyn}), $\eta_e>0$ implies $\lambda_+ + \lambda_- > 0$. However, beyond this power threshold, the product of eigenvalues can become negative $\lambda_+\lambda_- < 0$ making the fixed point unstable. This behavior is shown in Fig.~\ref{fig:TrajFastCav}, where we considered two power levels: one below the threshold in the plot (a) and one above the threshold in plot (b). As seen in Fig.~\ref{fig:TrajFastCav}(b), above the threshold, the pseudospin undergoes undamped oscillations with increasing amplitude over time, signifying a growing instability in the system. This behavior indicates that the magnon number continues to rise and does not settle into an equilibrium value. The photon number is maximized at $K^x = 0$ (see Eq.~(\ref{eq:c0})), at which point a sharp spike in $K^z$ and $K^y$ occurs. The optically induced damping ($\eta_{\rm opt}$) and rotation ($\Omega^x_{\rm eff}$) show oscillations. The transition from stable to unstable behavior as the drive power increases is analogous to a bifurcation in dynamical systems theory~\cite{jezek1990nonlinear, chen2000bifurcation}. This suggests the potential for rich dynamical phases as we explore different pump parameters, opening avenues for phase control and transitions, especially in quantum materials for quantum information applications and material studies~\cite{collado2018population, zhang2021nonequilibrium}.

In the blue-detuned regime, $\Delta'=-1$, the $K$-components do not settle into stable equilibrium values; instead, they continuously increase over time due to the persistent energy injection from the pump laser.

\subsection{Squeezing}
\label{subsec:Squeeze}
\begin{figure}[t!]
  \centering
  \includegraphics[width=0.49\textwidth]{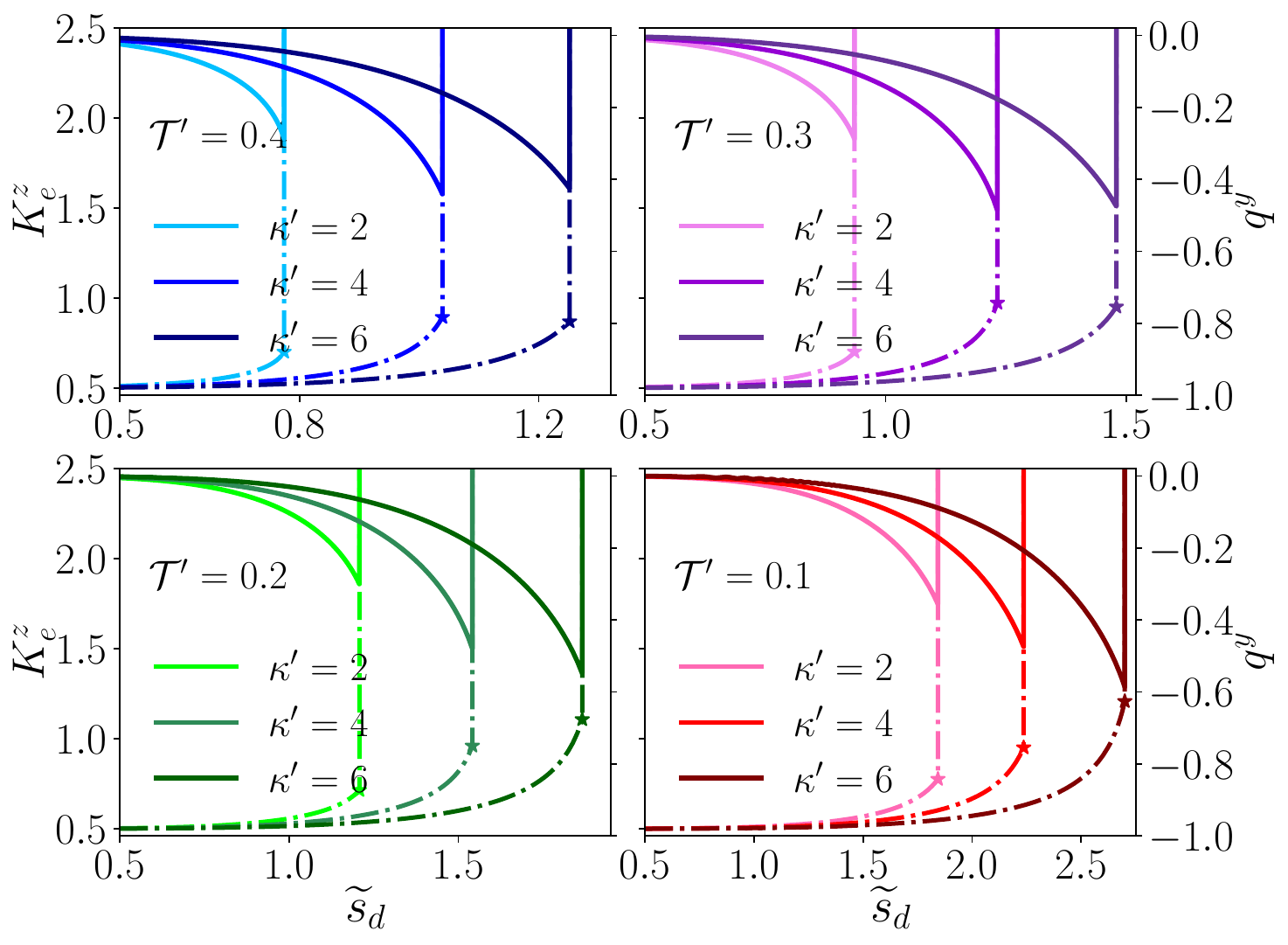}
  \caption{$K^z_e$ component of pseudospin (dash-dotted line) and squeezing factor $q^y$ (solid line) as a function of the drive laser power in the fast cavity regime for different $\mathcal{T}^\prime$ and $\kappa^\prime$.}
\label{fig:Squeeze}
\end{figure}

The Perelomov coherent states, defined in Eq.~(\ref{Def:Coh}), generated optically are squeezed~\cite{Kastrup2007,Aravind:88,bossini2019laser, perelomov1972coherent} in the following sense. The two noncommuting quadrature operators, $\hat{K}^{x}$ and $\hat{K}^{y}$ satisfy the Heisenberg uncertainty relation $2\sigma_{K^x} \sigma_{K^y} \ge |\braket{\hat{K}^z}|$, where the standard deviation is defined as 
\begin{equation} 
    \sigma_{K^i}^2 =\left\langle \left(\hat{K}^{i}\right)^{2}\right\rangle -\left\langle \hat{K}^{i}\right\rangle ^{2},
\end{equation}
for $i\in \{x, y\}$. We define the squeezing factors as
\begin{equation}
q^{i} =\frac{\sigma_{K^i}^2 -\frac{1}{2}\left|\left\langle \hat{K}^{z}\right\rangle \right|}{\frac{1}{2}\left|\left\langle \hat{K}^{z}\right\rangle \right|},\label{eq:squeezing-factor}
\end{equation}
A state is squeezed when either $q^x < 0$ or $q^y < 0$. These factors are calculated for Perelomov coherent states in App.~\ref{app:Perelomov} to be,
\begin{equation}
    \sigma_{K^i}^2 = \frac{1 + \left(K^i\right)^2}{2}. 
\end{equation}
In terms of $\vartheta$ and $\varphi$, see Eq.~(\ref{params}), we find
\begin{equation}
    q^x = \frac{4\sinh^2(\vartheta/2)}{\cosh\vartheta} \left(\cosh^2\frac{\vartheta}{2} \cos^2\varphi - \frac{1}{2} \right),
\end{equation}
while the same expression gives $q^y$ by replacing $\cos\varphi\rightarrow\sin\varphi$. The maximum squeezing corresponds to \(q^i = -1\) (\(i \in \{x, y\}\)). In this regime, \(q^x\) exhibits a positive value. In Fig.~\ref{fig:Squeeze}, we present the dependence of the squeezing factor (\(q^y\)) and equilibrium magnon-pair numbers (\(K_e^z\)) on drive field power for four sets of parameters (\(\mathcal{T}^\prime\), \(\kappa^\prime\)). As drive power increases, \(K_e^z\) rises and \(q^y\) decreases, reaching a maximum and minimum, respectively, at the power threshold. However, above the threshold, \(K_e^z\) goes to infinity and \(q^y\) jumps to positive values, indicating anti-squeezing or increased fluctuations, which means the system is out of equilibrium. Moreover, for a given value of \(\mathcal{T}^\prime\), a larger cavity damping results in a higher threshold power, a larger magnon pair population, and a smaller squeezing factor. 

Magnon squeezing in AFM insulators reduces spin fluctuations within the crystallographic unit cell below the level of the vacuum fluctuation~\cite{zhao2004magnon}. This noise reduction paves the way for coherent control of spin dynamics in AFMs, potentially enabling robust quantum memories and spin-based quantum computing platforms~\cite{yuan2021recent, romling2024quantum, wang2024ultrastrong, makihara2021ultrastrong, azimi2020hierarchy}. Furthermore, the diagonal components of the spin susceptibility tensor, $\chi_{ii}$ ($i=x,y,z$), are directly proportional to spin-component fluctuations, as described by the fluctuation-dissipation theorem~\cite{kubo1966fluctuation, abaimov2015statistical}. In magnon-squeezed states, the spin susceptibility reaches a local minima along the squeezed direction. This effect may extend to the electrical susceptibility $\chi_{e}$ due to spin-charge coupling in strongly correlated systems, though this requires further investigation. During squeezed state generation, suppressed spin fluctuations may influence magnon-phonon scattering, magnon-magnon interactions, and superconducting mechanisms, all of which merit further experimental and theoretical studies.

\section{Full Dynamic Response}
\label{IV}
In this section, we numerically solve the coupled system of nonlinear differential equations (Eqs.~(\ref{EOM:c},\ref{dKdt:Class})) to explore the full dynamics of the system. We consider three detunings $\Delta' \in \{1,0,-1\}$ respectively red, zero, and blue-detuning. Our numerical simulations reveal rich dynamics, including fixed points, limit cycles, and chaotic behavior, depending on the input power and detuning.

\begin{figure}
  \centering
  \includegraphics[width=0.49\textwidth]{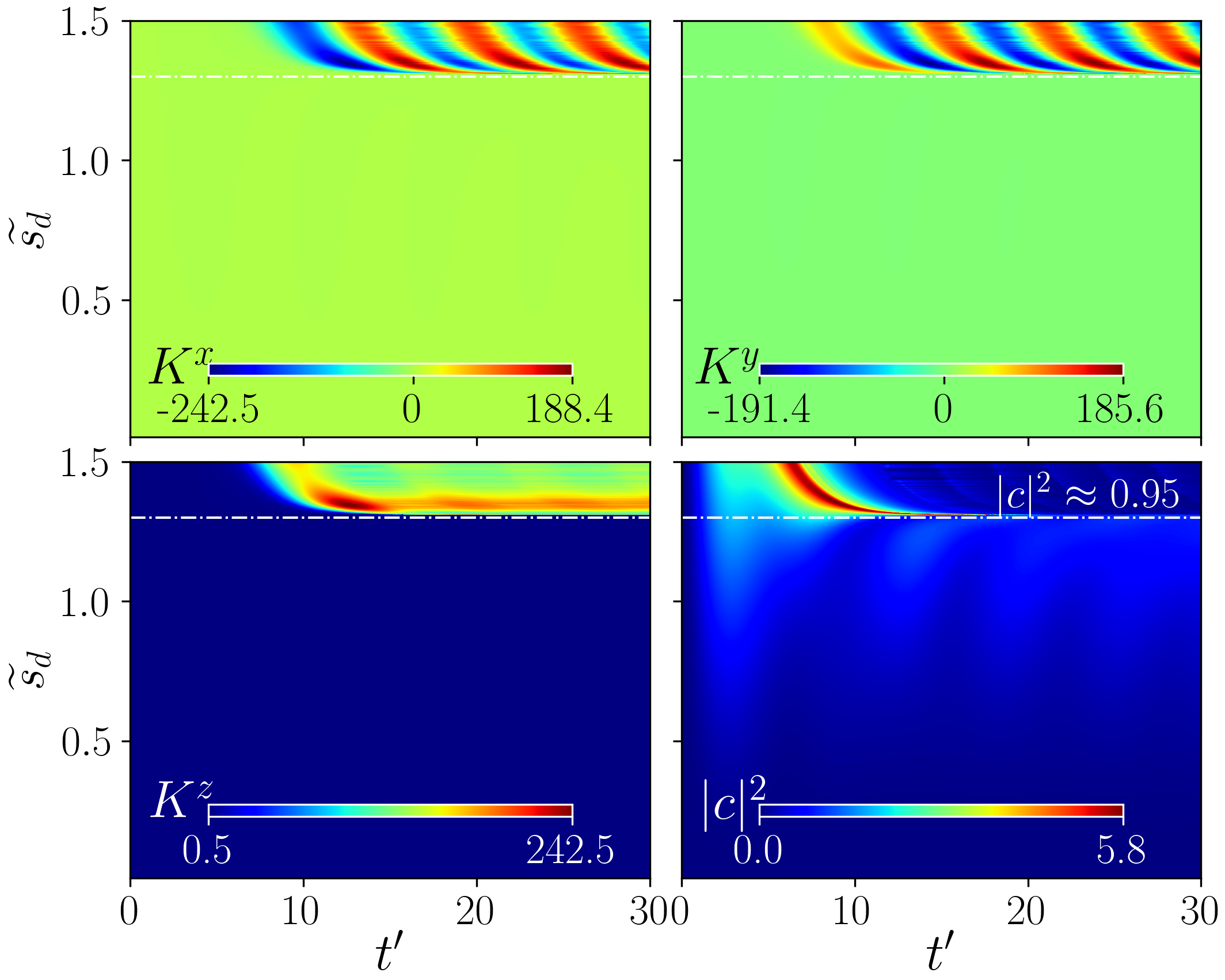}
  \caption{Temporal evaluation of pseudospin components and the photon field under varying drive power $\tilde{s}_d$, for red-detuning $\Delta'=1$.  We use similar parameters as in Fig.~\ref{fig:FastCavity} except for a smaller cavity linewidth: $\kappa_G' = 10^{-4}$, $\kappa' = 0.5$, $\mathcal{T}' = 0.2$, and $\mathcal{R}' = 0$. Dashed lines show a threshold power ${\Tilde{s}}_d\approx1.3$ (corresponding to power $P_d\approx4.0\mu W$ and  $P_d\approx10.7\mu W$ respectively for magnon frequency of M- and X-points, for $\omega_c=650$THz and $\rho=0.5$) below which the dynamics reach a fixed point.}
\label{fig:FullDyn:Red}
\end{figure}

\begin{figure}
  \centering
  \includegraphics[width=0.49\textwidth]{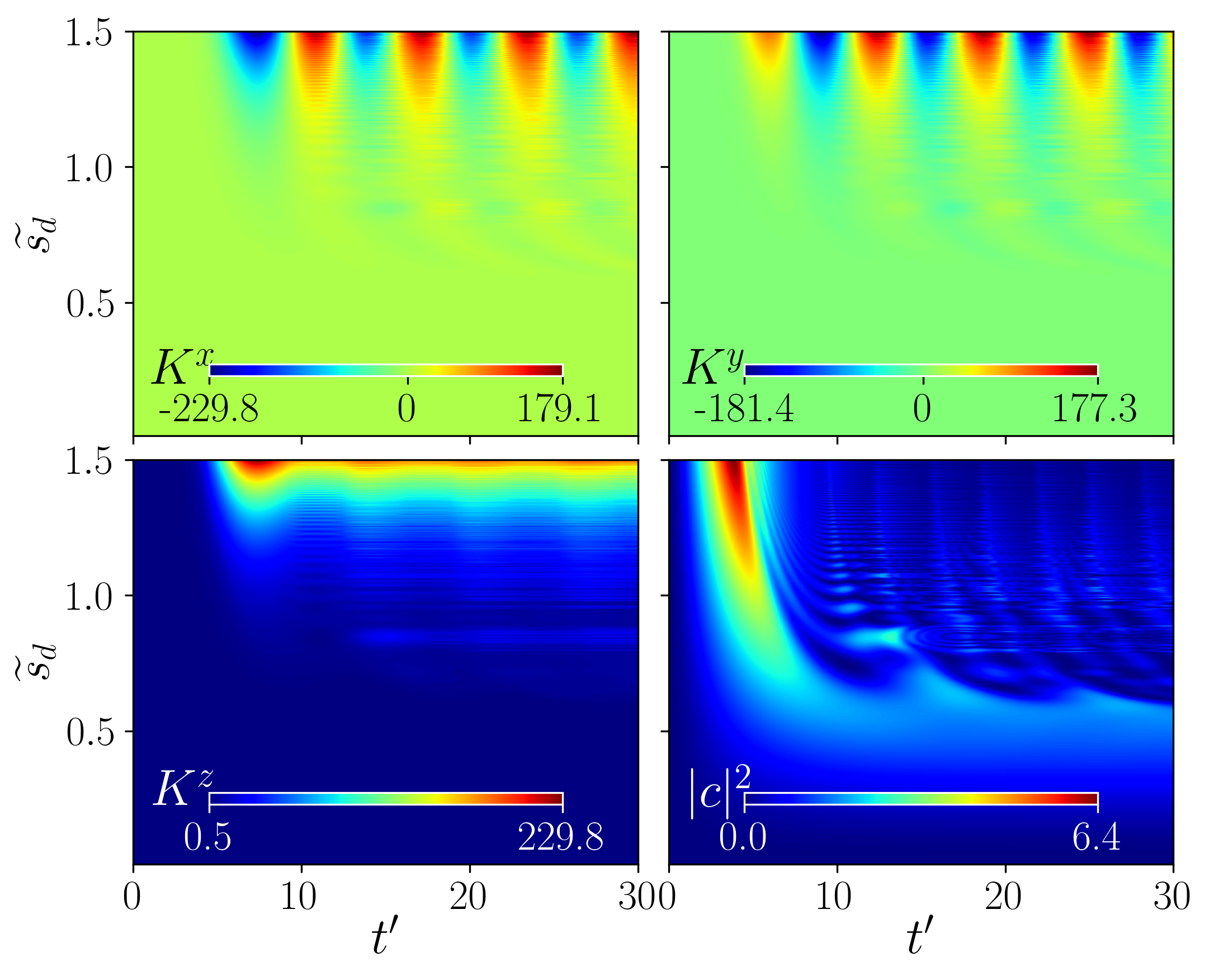}
  \caption{Temporal evaluation of pseudospin components and the photon field under varying drive power, $\tilde{s}_d$ for a resonant input $\Delta^\prime=0$ with parameters as in Fig.~\ref{fig:FullDyn:Red}}
\label{fig:FullDyn:Zero}
\end{figure}

\begin{figure}[t]
  \centering
  \includegraphics[width=0.49\textwidth]{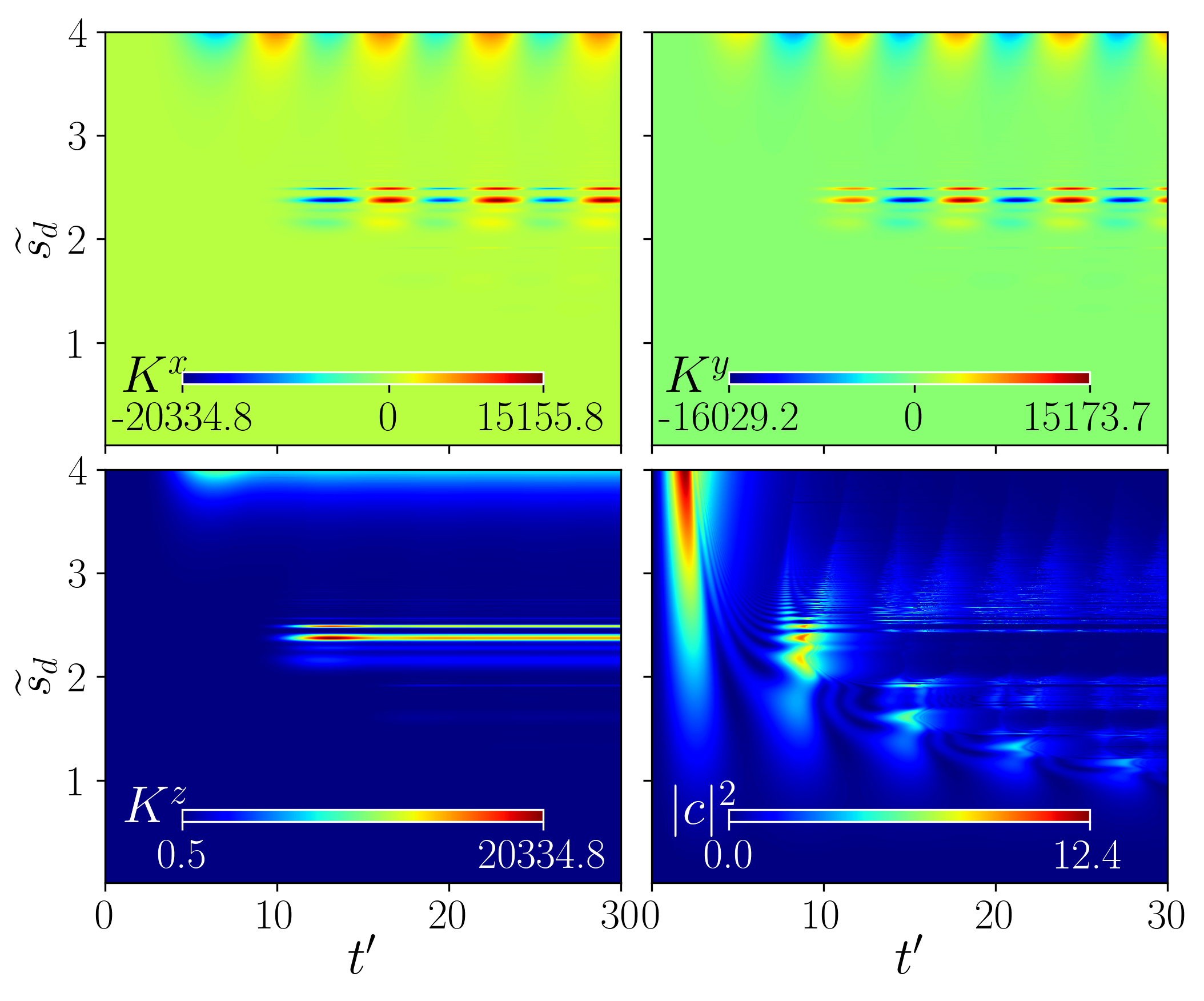}
  \caption{Temporal evaluation of pseudospin components and the photon field under varying drive power, $\tilde{s}_d$ for blue-detuning $\Delta^\prime=-1$ with parameters as in Fig.~\ref{fig:FullDyn:Red}. }
\label{fig:FullDyn:Blue}
\end{figure} 

Fig.~\ref{fig:FullDyn:Red} shows the temporal behavior of the system under varying drive power for the red-detuned regime. At low drive powers (below a threshold of approximately 1.3 in normalized units), the pseudospin vector $\mathbf{K} = (K_x, K_y, K_z)$ exhibits damped oscillations and eventually reaches a stable fixed point for each specific power level (see App.~\ref{app:DROC} for more details). This indicates that for each power value, there is a unique combination of $K_x$, $K_y$, $K_z$, and a corresponding photon number ($|c|^2$) that the system settles into. In this regime, the amplification of $|c|^2$ is power-dependent, reaching its maximum value at the threshold power. However, the behavior changes significantly when the drive power exceeds this threshold. Initially, $|c|^2$ increases rapidly, but the backaction by $\mathbf{K}$ on $|c|^2$ reduces it to nearly zero (a bit before $t' = 10$). With the optical cavity being essentially empty, the pseudospin $\mathbf{K}$ precesses at its natural frequency of approximately $2\omega_m$. These oscillations persist due to the small intrinsic damping of the magnons. Thus, the optical input excites self-sustained oscillations in the magnon pairs beyond the threshold power.

The system's dynamics for the resonant ($\Delta'=0$) and blue-detuned ($\Delta'=-1$) regimes are illustrated in Figs.~\ref{fig:FullDyn:Zero} and~\ref{fig:FullDyn:Blue}, respectively. In both cases, stable oscillations are observed when the input power remains below a critical threshold value of 1. Further details on the analysis can be found in App.~\ref{app:DROC}. For input powers exceeding this critical value, the system undergoes a transition to a chaotic regime. This transition is attributed to a significant energy transfer to the AFM, manifested by a continuous increase in the $K^z$ component over time. This increasing $K^z$ signifies the onset of a dynamical instability in the system.

\section{Conclusion}

In summary, we theoretically explored the potential of using high-finesse optical cavities to precisely manipulate and control the dynamics of magnons in AFM insulators, including Rutiles and cuprates. We focused on materials where two-magnon Raman scattering is crucial in coupling the photon cavity with AFM spin sublattices. This mechanism facilitates the generation of pure magnon pairs and induces shifts in the magnon frequency spectrum. We demonstrated that when the drive laser and cavity photons are red-detuned, there exists a certain power threshold for the pump. Below this threshold, magnon-pair squeezing is evident, while above it, magnon pairs transition into a limit cycle. To characterize the features of attractor points and determine stability conditions in the squeezing states, we used the fast cavity approximation. 

Within these AFMs, magnons have an intrinsic capability to generate squeezed states due to interference between sublattice spin waves. Using an optical cavity, we can modulate these pre-existing squeezing states for applications in quantum information. Higher squeezing factors suggest the potential for cavity control of AFM magnons, e.g. for controlling the phase diagram of materials where spin fluctuations are believed to play an important role, such as high-T$_c$ cuprate superconductors. Furthermore,  squeezing leads to an increased precision in sensing a conjugate variable (here, proportional to the photon number). 

The squeezing factor, a key metric of our study, can be experimentally measured using Brillouin light scattering techniques. Employing Eq.~(\ref{Ham:eff}) and the input-output formalism, the scattered light $\hat{c}_{\rm out}$ can be expressed in terms of the incoming light $\hat{c}_{\rm in}$ and the Perelomov operators as
\begin{equation}
    \hat{c}_{\rm out} - \hat{c}_{\rm in} \propto 2\mathcal{R} \hat{K}^z + \mathcal{T} \hat{K}^- + \mathcal{T}^* \hat{K}^+\,.
\end{equation}
Since the phase of $\mathcal{T}$ depends on the polarization of the incident light, derived via Eq.~(\ref{Hint:Micro2}), by measuring both the mean and the fluctuations of the output field, we obtain analogous information for an arbitrary pseudospin component, say $\sum_i \xi_i\hat{K}^i$ for some constants $\xi_i$. By systematically varying the values of $\xi_i$, one can derive the averages up to the second order in terms of $\hat{K}^i$, thereby revealing the squeezing parameter $q^i$. 

Furthermore, we showed that when the pump and cavity photons are at resonance or blue-detuned, depending on the pump power, magnon-pairs exhibit nonlinear dynamics such as auto-oscillations and chaos. This can open the door to novel spintronic applications, including spin-torque nano-oscillators (STNOs) and magnonic logic devices.

\section{Acknowledgement}
S.S. and S.V.K. acknowledge funding from the Bundesministerium für Bildung und Forschung (BMBF) under the project QECHQS (Grant No. 16KIS1590K) and the Deutsche Forschungsgemeinschaft (DFG, German Research Foundation)—Project-ID 429529648—TRR 306 QuCoLiMa (“Quantum Cooperativity of Light and Matter”). A.-L.E.R. received the support of a fellowship from the ”la Caixa” Foundation (ID 100010434). The fellowship code is LCF/BQ/DI22/11940029.

\appendix

\section{Antiferromagnetic Hamiltonian} \label{app:AFM_Ham}

\begin{figure}[t!]
    \includegraphics[width=0.4\textwidth]{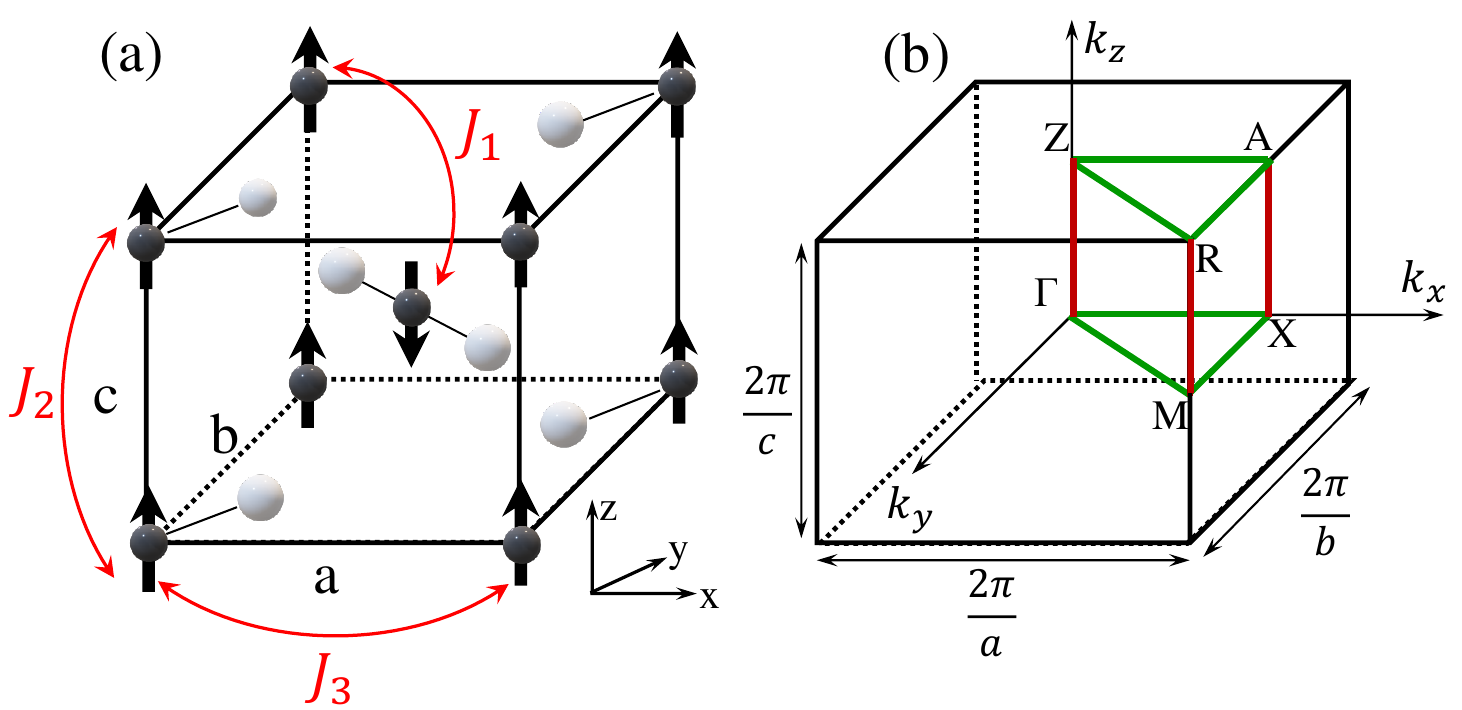}\\
    \includegraphics[width=0.48\textwidth]{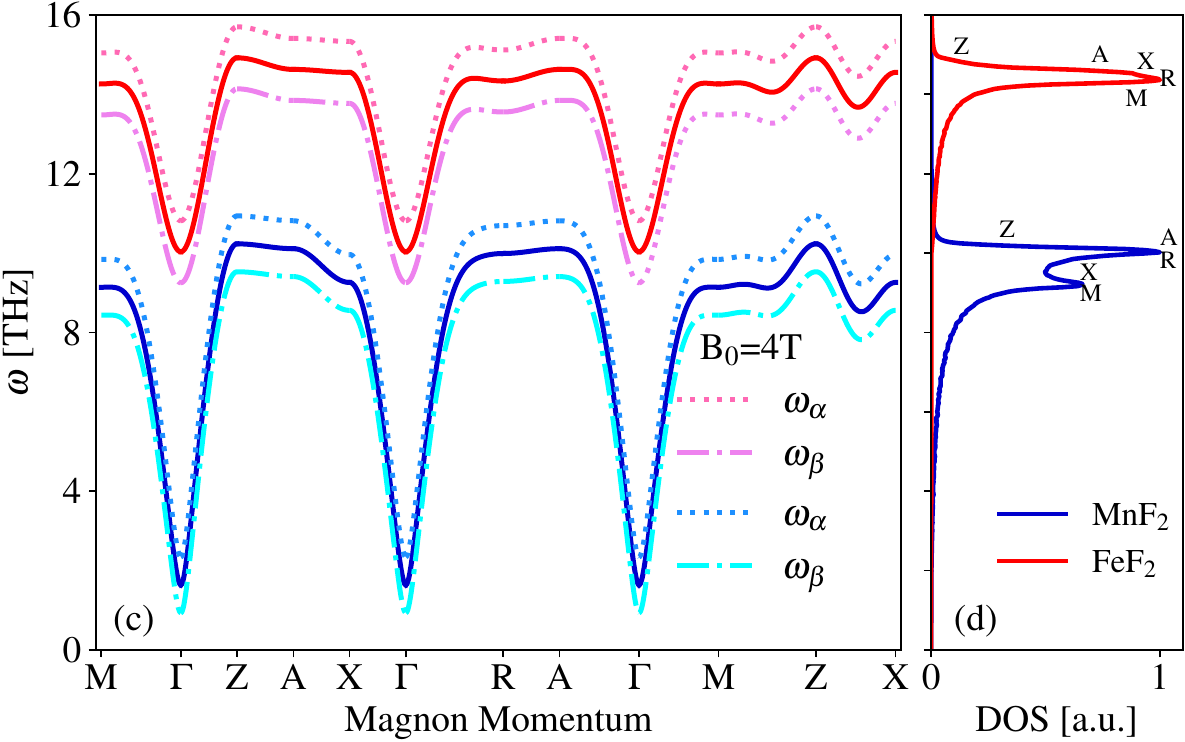}
    \caption{(a) Crystal and magnetic structure of the rutile AFMs MnF$_2$ and FeF$_2$. The large circles show the position of Mn$^{2+}$ (or Fe$^{2+}$) ions with their magnetic moment directions shown as arrows. The small circles represent the fluorine atoms. (b) The primitive Brillouin zone, critical points, and symmetric lines of the rutile AFMs. (c)  Magnon dispersion for $B_0=0$ (solid lines) and $B_0=4 T$ (dotted and dash-dotted lines) and (d) one-magnon density of states (DOS) in MnF$_2$ (blue) and FeF$_2$ (red). $\omega_{\mathrm M}^{\mathrm {MnF_2}}=9.25$THz and $\omega_{\mathrm M}^{\mathrm {FeF_2}}=14.55$THz.}
    \label{FigSM:Rutile}
\end{figure}

In this appendix, we consider the diagonalization of the AFM Hamiltonian in two distinct categories of AFM insulators. The first class consists of materials with rutile crystal structures, such as MnF$_2$ and FeF$_2$. These materials feature a body-centered tetragonal lattice, where magnetic ions reside at the corners and body-centered positions, as depicted in Fig.~\ref{FigSM:Rutile}(a). Additionally, they exhibit uniaxial (easy-axis) anisotropy. Below their respective Néel temperatures (66.5 K for MnF$_2$ and 78.4 K for FeF$_2$), and in the absence of an external magnetic field, the spins order into two sublattices with opposite orientations along the easy anisotropy direction (c-axis). 

\begin{figure}[t!]
   \includegraphics[width=0.48\textwidth]{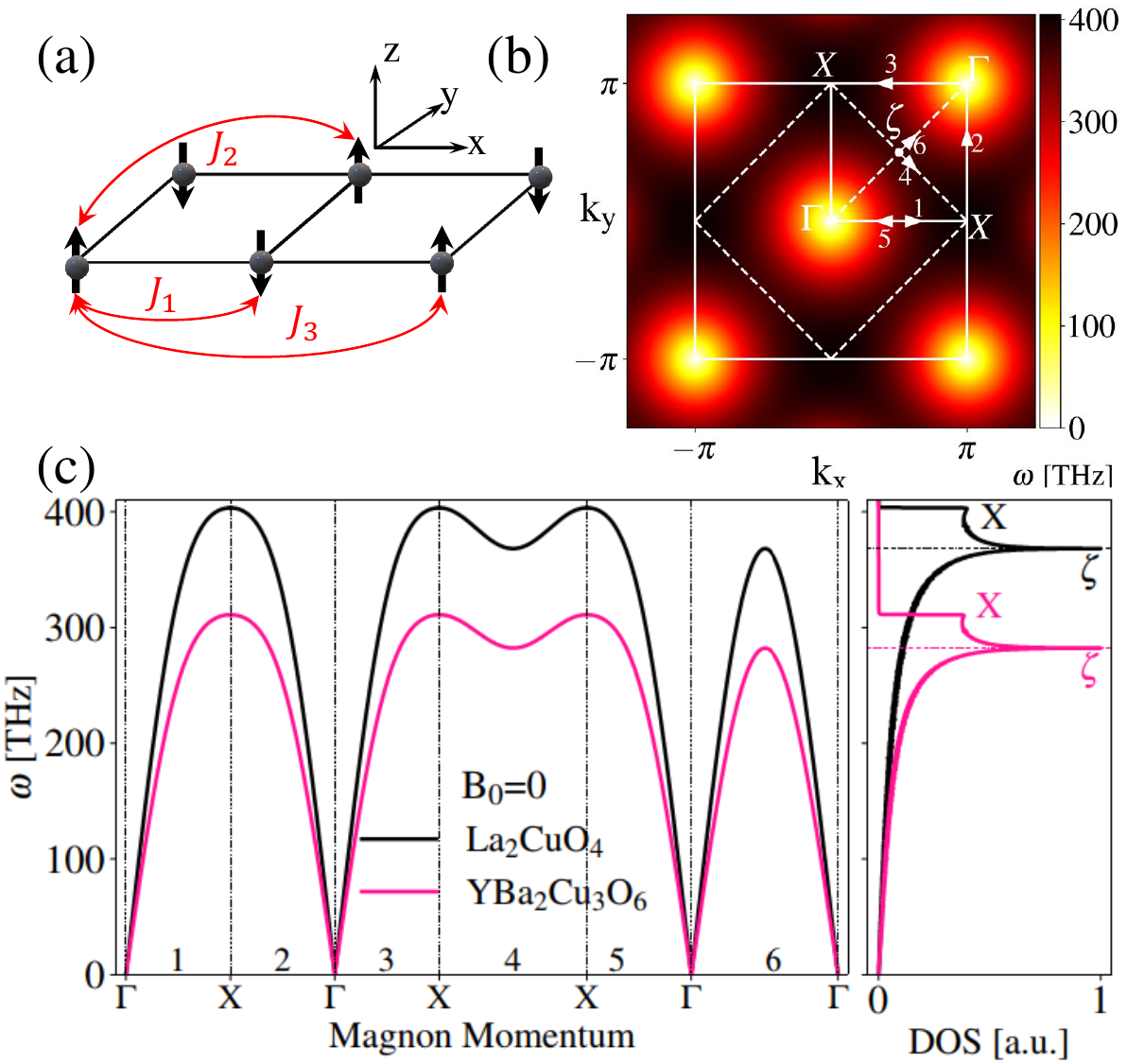}
   \caption{(a) Magnetic structure and definition of short-range exchange constants $J_{l}$ ($l=1,2,3$) up to the third nearest-neighbor (NNN) for two-dimensional cuprates, (b) Magnon dispersion and  first Brillouin zone in La$_2$CuO$_4$, (c) Magnon dispersion and density of states for monolayers of La$_2$CuO$_4$ and YBa$_2$Cu$_3$O$_6$.}
\label{FigSM:cuprates}
\end{figure}

The second class of AFM insulators considered in this study comprises cuprate parent compounds, including La$_2$CuO$_4$, YBa$_2$Cu$_3$O$_6$, and Tl$_2$Ba$_2$CuO$_6$. These high-$T_c$ compounds possess a layered structure composed of intercalated copper-oxygen planes. For simplicity, we disregard interlayer couplings and instead consider quasi-two-dimensional (2D) constituents, as depicted in Fig.~\ref{FigSM:cuprates}(a). 

In the Heisenberg spin model, the Hamiltonian is given by
\begin{equation}
   \hat{\mathcal{H}}_m = \sum_{\mathbf{R}} \left[ \sum_{\mathbf{R}'} J_{\mathbf{R'}} \hat{\boldsymbol{S}}_{\mathbf{R}} \cdot \hat{\boldsymbol{S}}_{\mathbf{R} + \mathbf{R}'} - D \left(\hat{S}_{\mathbf{R}}^{z}\right)^{2} - B_0 \hat{S}_{\mathbf{R}}^{z} \right],
   \label{App_Hm0}
\end{equation}
where $\hat{\boldsymbol{S}}_{\mathbf{R}}$ denotes a spin-$S$ particle located at site $\mathbf{R}$, $J_{\mathbf{R}'}$ is the exchange interaction between spins displaced by $\mathbf{R}'$, and $B_0$ is an external magnetic field.

We consider up to the third-nearest neighbor with exchange constants $J_1$, $J_2$, and $J_3$. The number of neighbors corresponding to each exchange is $z_1=8$, $z_{2}=2$, and $z_{3}=4$ for Rutile-AFMs and $z_1=z_2=z_3=4$ for cuprates. The second term in the Hamiltonian represents the easy-axis anisotropy term, which is related to the anisotropy field $H_A$ through the equation $D = g\mu_BH_A/(2S)$, where $g$ represents the spectroscopic splitting factor, and $\mu_B$ is the Bohr magneton. This uniaxial anisotropy aligns the Néel vector in the $\mathbf{z}$ direction. 

The low-energy excitations of $\hat{\mathcal{H}}_{0}$ to the lowest order in $1/S$ are found by Holstein-Primakoff linear spin-wave theory. Transforming to Fourier space the Hamiltonian can be recast as $\hat{\mathcal{H}}_{0}=\sum_{{\bf k}}\psi_{{\bf k}}^{\dagger}\mathcal{H}_{{\bf k}}\psi_{{\bf k}}$ in the Nambu basis $\psi_{\bf{k}}^{\dagger}=(a_{\mathbf{k}}^{\dagger},b_{-{\bf k}})$. $\mathcal{H}_{{\bf k}}= h_{0}I + h_{x}\sigma_{x}+h_{z}\sigma_{z}$ where $\sigma$ is the vector of Pauli matrices and $h_{0}=J_{{\bf k}}+2SD$, $h_{x}=z_1SJ_{1}\gamma_{1{\bf k}}$, $h_{z}=g\mu_{B}B_0$, with $J_{{\bf k}}=z_1SJ_{1}+z_2SJ_{2}\left(\gamma_{2\mathbf{k}}-1\right)+z_3SJ_{3}\left(\gamma_{3\mathbf{k}}-1\right)$, and geometric structure factors 
\begin{align}
 \gamma_{\bf{1k}} &= \cos(ak_x/2) \cos(bk_y/2) \cos(ck_z/2), \nonumber\\
 \gamma_{\bf{2k}} &= \cos\left(k_{z}c\right), \nonumber\\
 \gamma_{3\bf{k}} &= [\cos(ak_x) + \cos(bk_y)]/2,
 \label{Coeff}
\end{align} 
The lattice constants are $a=b=\SI{4.874}{\angstrom} $ and $c=\SI{3.31}{\angstrom}$ for MnF$_2$ and $a=b = \SI{4.697}{\angstrom}$ and $c = \SI{3.309}{\angstrom}$ for FeF$_2$~\cite{hutchings1970spin, rezende1978antiferromagnetic, ohlmann1961antiferromagnetic}. The unperturbed Hamiltonian is diagonalized to $\hat{\mathcal{H}}_{0}=E_0+\sum_{{\bf k}}\left[\varepsilon_{\alpha{\bf k}}\alpha_{{\bf k}}^{\dagger}\alpha_{{\bf k}}+\varepsilon_{\beta{\bf k}}\beta_{-{\bf k}}^{\dagger}\beta_{-{\bf k}}\right]$ via Bogoliubov transformation~\cite{rezende2019, parvini2020, bostrom2021all}
\begin{equation}
    \begin{pmatrix}
        a_{\mathbf{k}} \\ b_{\mathbf{-k}}^{\dagger}
    \end{pmatrix} = \begin{pmatrix}
        \cosh(\theta_{\bf k}/2) & -\sinh(\theta_{\bf k}/2) \\
        -\sinh(\theta_{\bf k}/2) & \cosh(\theta_{\bf k}/2)
    \end{pmatrix} \begin{pmatrix}
        \alpha_{\mathbf{k}} \\ \beta_{\mathbf{-k}}^{\dagger}
    \end{pmatrix},
\label{Bogoliubov}
\end{equation}
the Bogoliubov angle is $\theta_{{\bf k}} = \tanh^{-1} \left(h_x/h_0\right)$, and the dispersion of the upper and lower magnon branches is $\varepsilon_{\alpha/\beta}=\pm h_{z}+\sqrt{h_{0}^{2}-h_{x}^{2}}$. Only at ${\bf k}=0$, the dispersion terms associated with $J_2$ and $J_3$ vanish. Consequently, retaining these two terms in the Hamiltonian is crucial to study two-magnon Raman scattering, where any type of wavevector can be excited. Table~\ref{table I} lists the exchange constants, spin values, and uniaxial anisotropy constants of FeF$_2$ and MnF$_2$. The dipolar interaction mainly governs the magnetic anisotropy in MnF$_2$, whereas in FeF$_2$ it arises mainly from spin-orbit coupling and exhibits a substantial magnitude. Similarly, the experimental critical temperature $T_c$ (K) and the calculated exchange interactions (meV) for these quasi 2D cuprates are presented in Table~\ref{table:MO coeff}. 

The magnon dispersion curves and density of states (DOS) for MnF$_2$ and FeF$_2$ are shown in Fig.~\ref{FigSM:Rutile}. The magnon modes exhibit degeneracy (solid lines) in the absence of an external magnetic field. The higher-frequency peak corresponds to magnons near the Brillouin zone's end faces along the z-direction. At $B_0=0$, the frequency of the $\Gamma$-magnon in MnF$_2$ is $\SI{1.64}{\tera\hertz}$. As the wave number increases, the frequency increases owing to higher exchange energies, eventually reaching nearly $\SI{10}{\tera\hertz}$ at the zone boundary. In contrast, FeF$_2$ exhibits a substantially higher predicted frequency of $\SI{10}{\tera\hertz}$ for the $\Gamma$-magnon, mainly because of its significant anisotropy. At the zone boundary, the frequency value is projected to further amplify to approximately $\SI{15}{\tera\hertz}$. The effect of the magnetic field on dispersion in FeF$_2$ is negligible, as shown in Fig.~\ref{FigSM:Rutile}, and is completely absent in $\omega_\alpha+\omega_\beta$ due to the neglected orthorhombic anisotropy. In addition, the density of magnon states shows prominent peaks near the van Hove singularities, which are located close to the Brillouin zone boundary. 

\begin{table}[ht]
\caption{spin value, g-factor, exchange constants, and uniaxial anisotropy fields (meV) associated with Rutile AFMs~\cite{nikotin1969magnon, hutchings1970spin}.}
\centering 
\begin{tabular}{c c c c c c c c c c c c c c c c c c c c}
\hline\hline
AFM     &&&& S   &&& g     &&&  $J_1$  &&&  $J_2$  &&&  $J_3$  &&&  $g\mu_BB_A$      \\ [0.9ex] \hline
MnF$_2$ &&&& 5/2 &&& 2     &&&  0.304  &&&  -0.056 &&&   0.008  &&&  0.095   \\ 
FeF$_2$ &&&& 2   &&& 2.22  &&&  0.45  &&&  -0.0062  &&&  0.024  &&&  2.57    \\
\hline
\label{table I} 
\end{tabular} 
\end{table}

\begin{table}[ht]
\caption{Experimental critical temperature T$_c$(K), exchange constants (meV), and lattice constant \(a\) (\(\text{\AA}\)) for parent cuprates-AFMs~\cite{wan2009calculated, tam2022charge, keimer1992neel, armstrong2010variable, canali1992theory, sandvik1998numerical}.}
\centering 
\begin{tabular}{c c c c c c c c c c c c c c c c c}
\hline\hline
cuprates-AFM            &&&&  $J_1$  &&&  $J_2$  &&& $J_3$  &&&  $T_c$ &&& a\\ [0.9ex] \hline 
La$_2$CuO$_4$       &&&&  108.8  &&&  -12.0  &&&   -0.2 &&&  42&&&  3.81 \\
YBa$_2$Cu$_3$O$_6$  &&&&  93.0   &&&  -4.7   &&&   2.4  &&&  90 &&& 3.85 \\
\hline
\label{table:MO coeff} 
\end{tabular} 
\end{table}

For spin-up and spin-down modes, the X-point of the energy spectrum is a local maximum, and the midpoint between X points is a saddle point, as shown in Fig.~\ref{FigSM:cuprates}(c). This feature gives rise to Van Hove singularity in the magnon DOS, $\zeta$ point. Compared with rutile AFMs, Fig.~\ref{FigSM:Rutile}, the magnon modes in these materials have a higher frequency.

\section{Full-Dynamic Response of Optomagnonic Cavity}
\label{app:DROC}

\begin{figure*}[ht!]
  \centering
  \begin{minipage}{0.5\textwidth}
    \includegraphics[width=\textwidth]{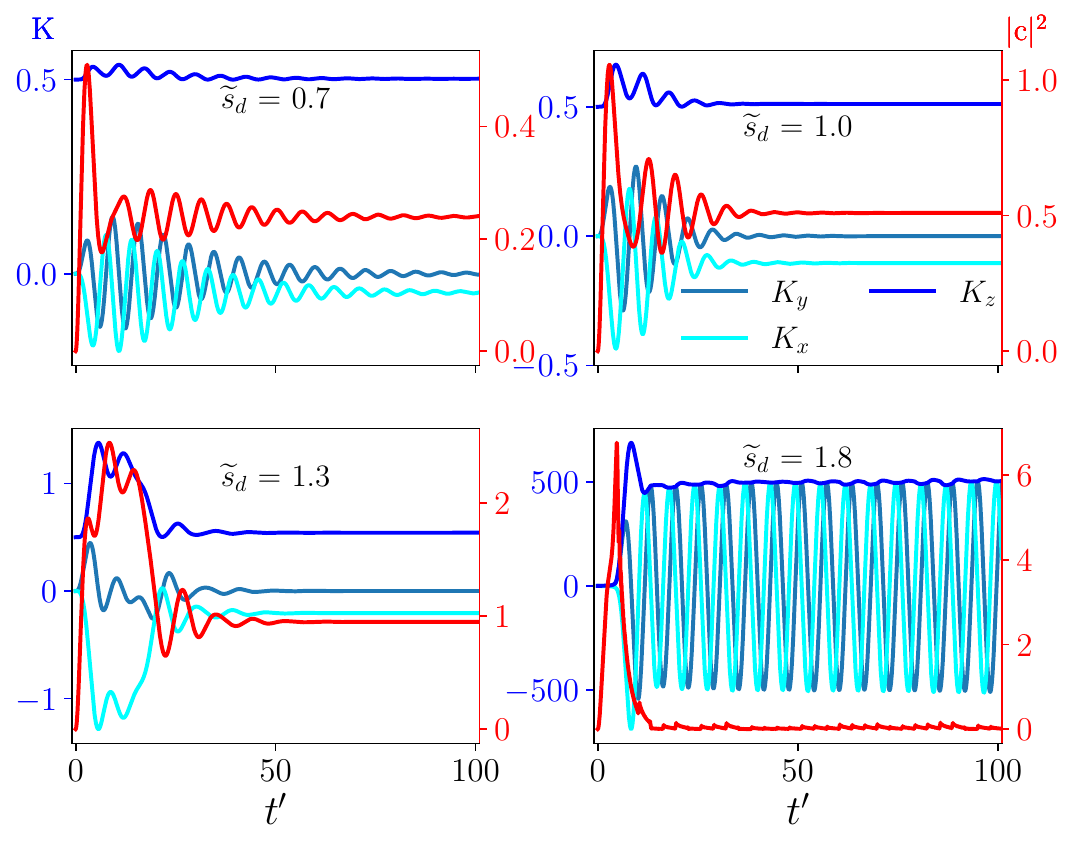}
  \end{minipage}%
  \begin{minipage}{0.5\textwidth}
    \includegraphics[width=\linewidth]{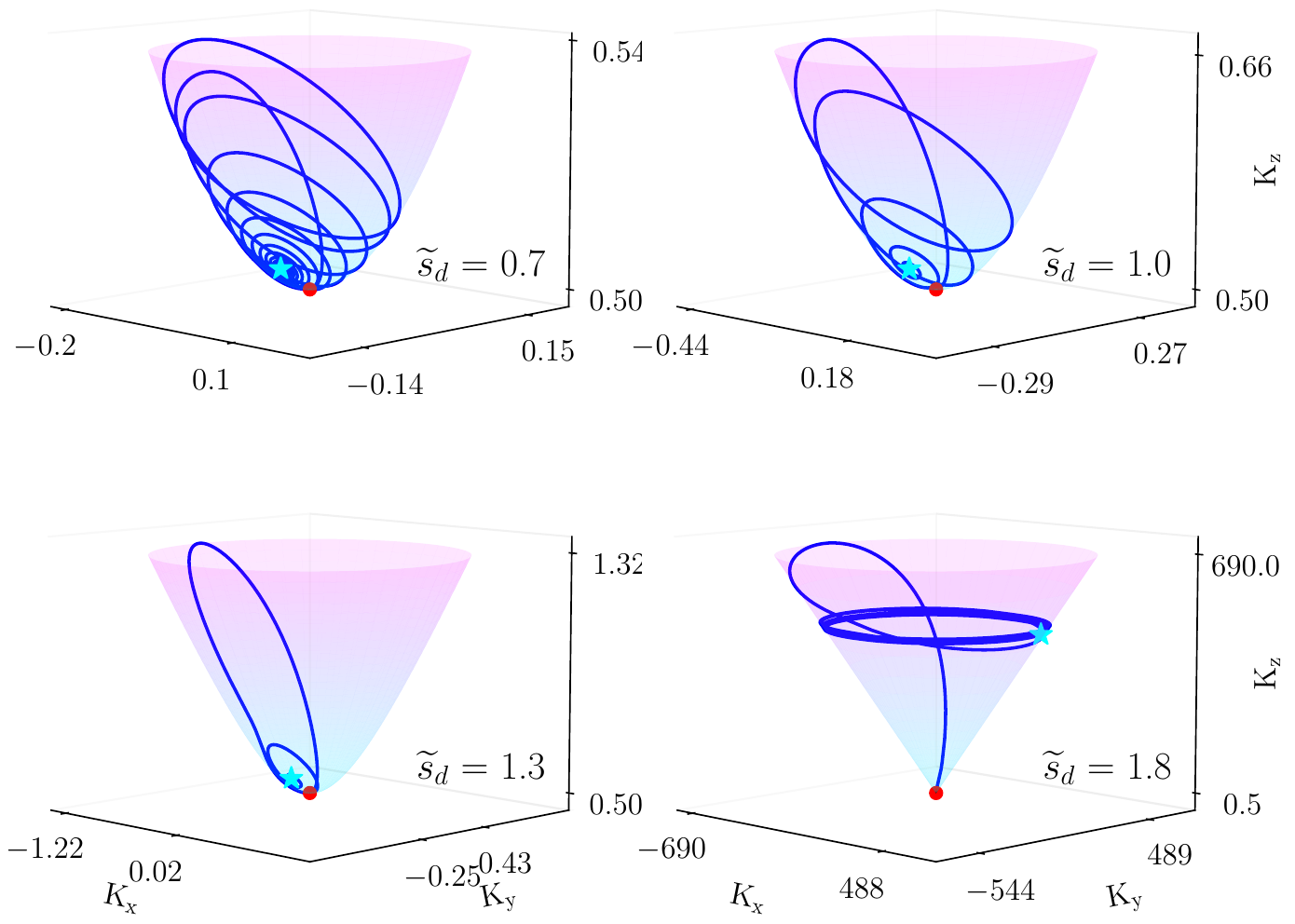}
  \end{minipage}
  \caption{Temporal evolution of pseudospin components $K_x$, $K_y$ and $K_z$ and the photon field under varying normalized derive laser power. $\Delta^\prime=1$, $\kappa_G^\prime=10^{-4}$, and initial pseudospin state ${\bf K}_0=\frac{1}{2}[0,0,1]$ and $c(0)=0$.}
  \label{appfig:Kvec:TimeEvol}
\end{figure*}

\begin{figure}[t!]
  \centering
    \centering
    \includegraphics[width=0.48\textwidth]{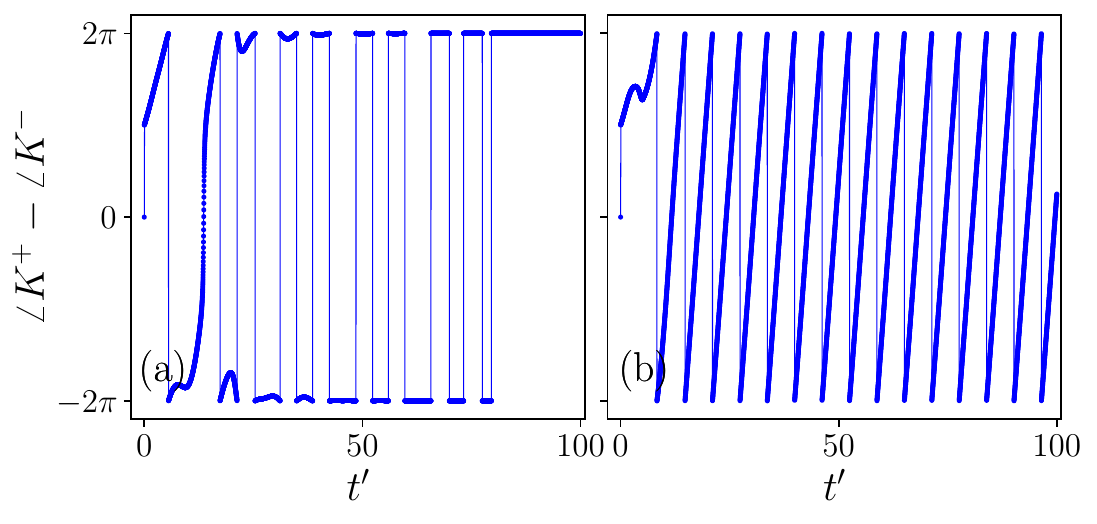}
    \caption{Phase difference between magnon creation and annihilation operators for (a) $\Tilde{s}_d=1.3$ and (b) $\Tilde{s}_d=1.8$.}
    \label{PhaseDifference}
\end{figure}

Fig.~\ref{appfig:Kvec:TimeEvol} shows the temporal evolution of the pseudospin components ($K^x$, $K^y$, and $K^z$) and the photon field under varying driving laser power for a specific case: a photon cavity and drive laser detuning of $\Delta^\prime=1$. The figure also presents the corresponding trajectories of the pseudospin tip on the upper hyperboloid sheet. The short simulation time ($t^\prime=100$) allows for a detailed observation of the dynamic behavior. The numerical simulations reveal a distinct power threshold for the driving laser, identified at approximately 1.3. Below this threshold, the pseudospin components converge towards a stable equilibrium point (attractor). However, beyond this critical value, the pseudospin dynamics transition into a sustained oscillatory regime. In this regime, $K^z$ reaches a constant value, while the photon field decays to zero amplitude. Notably, within the damped oscillator regime (below the threshold), the phase difference between the magnon creation and annihilation operators stabilizes at $2\pi$ after reaching the attractor point. Conversely, in the oscillatory state (above the threshold), this phase difference exhibits dynamic fluctuations within the range of $-2\pi$ to $2\pi$, as shown in Fig.~\ref{PhaseDifference}.

\begin{figure}[t!]
  \centering
    \centering
    \includegraphics[width=0.48\textwidth]{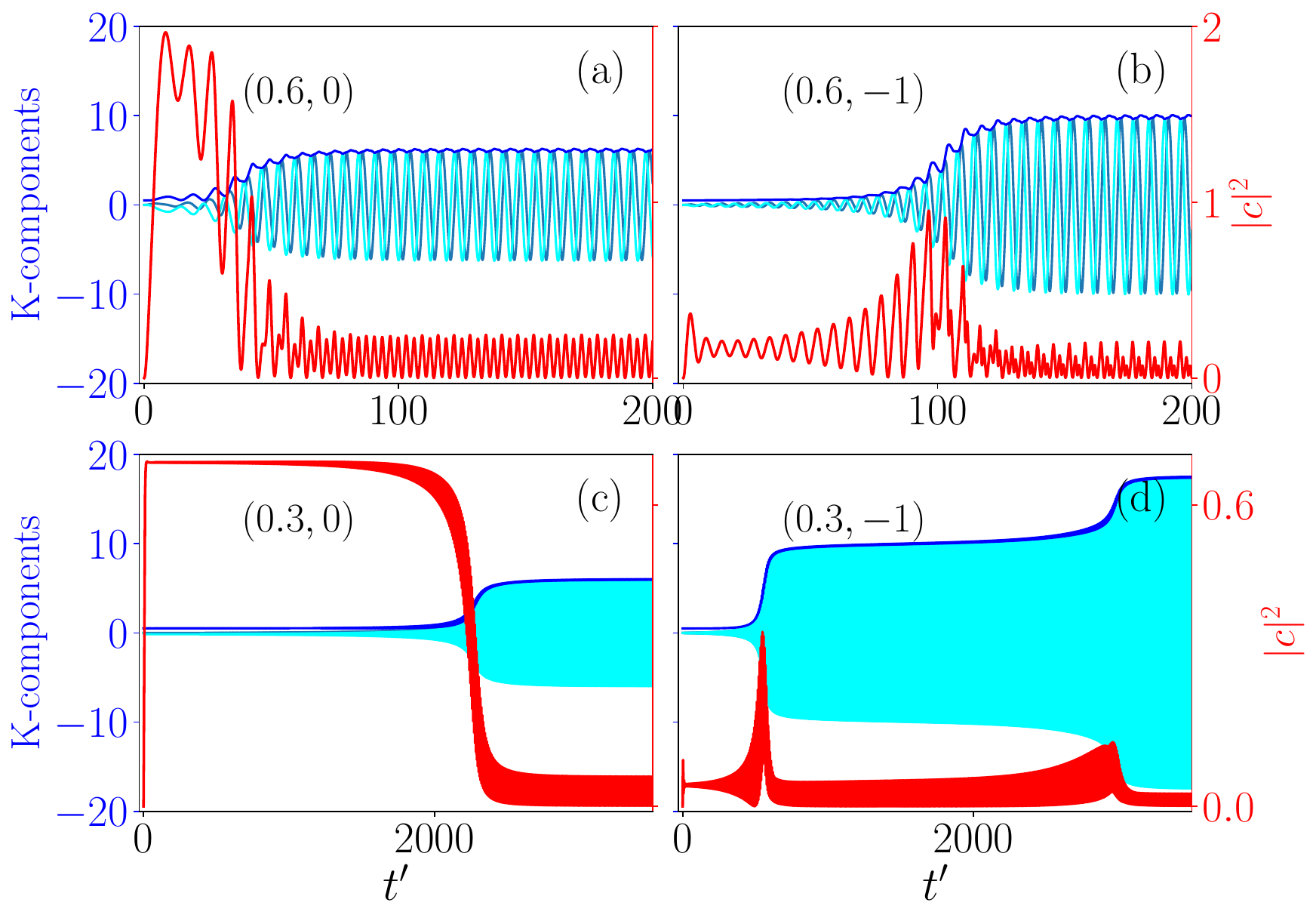}
    \caption{Temporal evolution of pseudospin components $K_x$, $K_y$, and $K_z$, along with the photon field, for varying pairs of parameters $(\Tilde{s}_d, \Delta')$.}
    \label{Fig5SM}
\end{figure}

The temporal evolution of the pseudospin and photon field for $\Delta^\prime=0$ ($\omega_d=\omega_c$) and $\Delta^\prime=-1$ ($\omega_d>\omega_c$) are shown in Fig.~\ref{Fig5SM}. The results show the emergence of auto-oscillation states with varying amplitudes and frequencies depending on the pump laser power. These auto-oscillatory behaviors hold promise for applications in modern spin-torque oscillators (STTNOs), signal processing, and other related fields ~\cite{2023enhancing, 2010fractional}.
 
\section{Squeezing of Perelomov Coherent States} \label{app:Perelomov}

This section discusses the basic mathematical results of the Perelomov coherent states that help derive various averages in the main text. For a fixed pseudo-spin $K$, they are defined by $\Ket{K,\vartheta,\varphi} = \hat{\cal D}(\vartheta,\varphi) \Ket{0,K}$ where $\hat{\cal D} = \exp(-\vartheta \hat{A})$ with
\begin{equation}
	\hat{A} = \frac{ -e^{-i\varphi} \hat{K}^+ + e^{i\varphi} \hat{K}^- }{2}.
\end{equation}
For later convenience, we also define
\begin{equation}
	\hat{B} = \frac{ e^{-i\varphi} \hat{K}^+ + e^{i\varphi} \hat{K}^- }{2}.
\end{equation}
We have the commutation relations $\left[\hat{A},\hat{K}^{z}\right]=\hat{B}$, $\left[\hat{B},\hat{K}^{z}\right]=\hat{A}$, and $\left[\hat{A},\hat{B}\right]=\hat{K}^{z}$. Below, we will use the Baker-Campbell-Hausdorff (BCH) formula 
\begin{equation}
    e^{\hat{X}}\hat{Y}e^{-\hat{X}} = \hat{Y} + \frac{[\hat{X},\hat{Y}]}{1!}+\frac{[\hat{X},[\hat{X},\hat{Y}]]}{2!}+\dots
\end{equation}

We now find how the operators $\{\hat{A},\hat{B},\hat{K}^z\}$ transform under $\hat{\cal D}$. We already know that $\hat{\cal D}^{\dagger} \hat{A} \hat{\cal D} = \hat{A}$ because $\hat{A}$ commutes with $\hat{\cal D}$. We can find $\hat{\cal D}^{\dagger} \hat{B} \hat{\cal D}$ using the BCH formula and the commutation relations written above,
\begin{equation}
    \hat{\cal D}^{\dagger} \hat{B} \hat{\cal D} = \cosh\vartheta\hat{B}+\sinh\vartheta\hat{K}^{z}.
\end{equation}
Similarly, we can derive
\begin{equation}
    \hat{\cal D}^{\dagger} \hat{K}^z \hat{\cal D} = \sinh\vartheta \hat{B} + \cosh\vartheta \hat{K}^z .
\end{equation}
Now, we know the averages in the ground state as $\langle \hat{K}^z \rangle_0 = K$ and $\langle \hat{K}^{\pm} \rangle_0 = 0$ where $\langle \hat{X} \rangle_0 = \Braket{K,0 | \hat{X} |K,0}$. The averages in the Perelomov coherent states, defined as $\langle \hat{X} \rangle = \Braket{K,\vartheta,\varphi | \hat{X} |K,\vartheta,\varphi}$, can be found using the above transformations: $\langle\hat{K}^z\rangle = K\cosh\vartheta$, $\langle\hat{A} \rangle = 0$ and $\langle\hat{B}\rangle = K\sinh\vartheta$. The last two expressions also imply $\langle\hat{K}^{\pm} \rangle = Ke^{\pm i\varphi} \sinh\vartheta$. A useful relation derived from the above is
\begin{equation}
	\hat{\cal D}^{\dagger} \left(\hat{K}^- - e^{-i\varphi} \tanh\frac{\vartheta}{2} \hat{K}^z \right) \hat{\cal D} = \hat{K}^{-} + e^{-i\varphi}\hat{K}^{z}\tanh\frac{\vartheta}{2}.
\end{equation}
Using $\hat{K}^z\Ket{K,0} = K\Ket{K,0}$ and $\hat{K}^-\Ket{K,0} = 0$, this gives the effect of $\hat{K}^-$ on coherent states as $\hat{K}^{-}\Ket{K,\vartheta,\varphi}=\lambda\left(K+\hat{K}^{z}\right)\Ket{K,\vartheta,\varphi}$ where $\lambda = e^{-i\varphi} \tanh\frac{\vartheta}{2}$.

To calculate the squeezing parameters, we want to find variances of $\hat{K}^x$ and $\hat{K}^y$. To this end, we again use the transformations
\begin{equation}
    \left\langle \hat{A}^2 \right\rangle = \left\langle \hat{A}^2 \right\rangle_0 = \frac{-K}{2}, 
\end{equation}
where we used $\hat{K}^-\hat{K}^+ \Ket{K,0} = 2K\Ket{K,0}$. Similarly,
\begin{equation}
    \left\langle \hat{B}^2 \right\rangle = \frac{K}{2} \cosh^2\vartheta + K^2 \sinh^2\vartheta,
\end{equation}
and the cross-correlation
\begin{equation}
    \left\langle \hat{A} \hat{B} \right\rangle = \frac{K}{2} \cosh\vartheta .
\end{equation}
Using $\hat{K}^x = \hat{B}\cos\varphi - i\hat{A}\sin\varphi$, we find the variance
\begin{equation}
    \left\langle \left(\Delta \hat{K}^x \right)^2 \right\rangle = \frac{K}{2}\left(1+\sinh^{2}\vartheta\cos^{2}\varphi\right)
\end{equation}
and similar result holds with $\hat{K}^x \rightarrow \hat{K}^y$ and $\cos\varphi \rightarrow \sin\varphi$. These calculations and formulations lay the groundwork for exploring squeezing phenomena in this paper.

\bibliography{literature.bib} 

\begin{thebibliography}{148}%
\makeatletter
\providecommand \@ifxundefined [1]{%
 \@ifx{#1\undefined}
}%
\providecommand \@ifnum [1]{%
 \ifnum #1\expandafter \@firstoftwo
 \else \expandafter \@secondoftwo
 \fi
}%
\providecommand \@ifx [1]{%
 \ifx #1\expandafter \@firstoftwo
 \else \expandafter \@secondoftwo
 \fi
}%
\providecommand \natexlab [1]{#1}%
\providecommand \enquote  [1]{``#1''}%
\providecommand \bibnamefont  [1]{#1}%
\providecommand \bibfnamefont [1]{#1}%
\providecommand \citenamefont [1]{#1}%
\providecommand \href@noop [0]{\@secondoftwo}%
\providecommand \href [0]{\begingroup \@sanitize@url \@href}%
\providecommand \@href[1]{\@@startlink{#1}\@@href}%
\providecommand \@@href[1]{\endgroup#1\@@endlink}%
\providecommand \@sanitize@url [0]{\catcode `\\12\catcode `\$12\catcode
  `\&12\catcode `\#12\catcode `\^12\catcode `\_12\catcode `\%12\relax}%
\providecommand \@@startlink[1]{}%
\providecommand \@@endlink[0]{}%
\providecommand \url  [0]{\begingroup\@sanitize@url \@url }%
\providecommand \@url [1]{\endgroup\@href {#1}{\urlprefix }}%
\providecommand \urlprefix  [0]{URL }%
\providecommand \Eprint [0]{\href }%
\providecommand \doibase [0]{http://dx.doi.org/}%
\providecommand \selectlanguage [0]{\@gobble}%
\providecommand \bibinfo  [0]{\@secondoftwo}%
\providecommand \bibfield  [0]{\@secondoftwo}%
\providecommand \translation [1]{[#1]}%
\providecommand \BibitemOpen [0]{}%
\providecommand \bibitemStop [0]{}%
\providecommand \bibitemNoStop [0]{.\EOS\space}%
\providecommand \EOS [0]{\spacefactor3000\relax}%
\providecommand \BibitemShut  [1]{\csname bibitem#1\endcsname}%
\let\auto@bib@innerbib\@empty
\bibitem [{\citenamefont {KEYES}(1981)}]{keyes1981limitations}%
  \BibitemOpen
  \bibfield  {author} {\bibinfo {author} {\bibfnamefont {R.}~\bibnamefont
  {KEYES}},\ }in\ \href {\doibase
  https://doi.org/10.1016/B978-0-12-234101-4.50012-6} {\emph {\bibinfo
  {booktitle} {VLSI Electronics: Microstructure Science}}},\ Vol.~\bibinfo
  {volume} {1},\ \bibinfo {editor} {edited by\ \bibinfo {editor} {\bibfnamefont
  {N.~G.}\ \bibnamefont {Einspruch}}}\ (\bibinfo  {publisher} {Elsevier},\
  \bibinfo {year} {1981})\ pp.\ \bibinfo {pages} {185--230}\BibitemShut
  {NoStop}%
\bibitem [{\citenamefont {Vashchenko}\ and\ \citenamefont
  {Sinkevitch}(2008)}]{vashchenko2008physical}%
  \BibitemOpen
  \bibfield  {author} {\bibinfo {author} {\bibfnamefont {V.~A.}\ \bibnamefont
  {Vashchenko}}\ and\ \bibinfo {author} {\bibfnamefont {V.~F.}\ \bibnamefont
  {Sinkevitch}},\ }\href@noop {} {\emph {\bibinfo {title} {Physical limitations
  of semiconductor devices}}},\ Vol.\ \bibinfo {volume} {340}\ (\bibinfo
  {publisher} {Springer},\ \bibinfo {year} {2008})\BibitemShut {NoStop}%
\bibitem [{\citenamefont {Keyes}(2001)}]{keyes2001fundamental}%
  \BibitemOpen
  \bibfield  {author} {\bibinfo {author} {\bibfnamefont {R.~W.}\ \bibnamefont
  {Keyes}},\ }\href {\doibase 10.1109/5.915372} {\bibfield  {journal} {\bibinfo
   {journal} {Proc. IEEE}\ }\textbf {\bibinfo {volume} {89}},\ \bibinfo {pages}
  {227} (\bibinfo {year} {2001})}\BibitemShut {NoStop}%
\bibitem [{\citenamefont {Flebus}\ \emph {et~al.}(2024)\citenamefont {Flebus},
  \citenamefont {Grundler}, \citenamefont {Rana}, \citenamefont {Otani},
  \citenamefont {Barsukov}, \citenamefont {Barman}, \citenamefont {Gubbiotti},
  \citenamefont {Landeros}, \citenamefont {Akerman}, \citenamefont {Ebels},
  \citenamefont {Pirro}, \citenamefont {Demidov}, \citenamefont {Schultheiss},
  \citenamefont {Csaba}, \citenamefont {Wang}, \citenamefont {Ciubotaru},
  \citenamefont {Nikonov}, \citenamefont {Che}, \citenamefont {Hertel},
  \citenamefont {Ono}, \citenamefont {Afanasiev}, \citenamefont {Mentink},
  \citenamefont {Rasing}, \citenamefont {Hillebrands}, \citenamefont
  {Kusminskiy}, \citenamefont {Zhang}, \citenamefont {Du}, \citenamefont
  {Finco}, \citenamefont {van~der Sar}, \citenamefont {Luo}, \citenamefont
  {Shiota}, \citenamefont {Sklenar}, \citenamefont {Yu},\ and\ \citenamefont
  {Rao}}]{Flebus_2024}%
  \BibitemOpen
  \bibfield  {author} {\bibinfo {author} {\bibfnamefont {B.}~\bibnamefont
  {Flebus}}, \bibinfo {author} {\bibfnamefont {D.}~\bibnamefont {Grundler}},
  \bibinfo {author} {\bibfnamefont {B.}~\bibnamefont {Rana}}, \bibinfo {author}
  {\bibfnamefont {Y.}~\bibnamefont {Otani}}, \bibinfo {author} {\bibfnamefont
  {I.}~\bibnamefont {Barsukov}}, \bibinfo {author} {\bibfnamefont
  {A.}~\bibnamefont {Barman}}, \bibinfo {author} {\bibfnamefont
  {G.}~\bibnamefont {Gubbiotti}}, \bibinfo {author} {\bibfnamefont
  {P.}~\bibnamefont {Landeros}}, \bibinfo {author} {\bibfnamefont
  {J.}~\bibnamefont {Akerman}}, \bibinfo {author} {\bibfnamefont
  {U.}~\bibnamefont {Ebels}}, \bibinfo {author} {\bibfnamefont
  {P.}~\bibnamefont {Pirro}}, \bibinfo {author} {\bibfnamefont {V.~E.}\
  \bibnamefont {Demidov}}, \bibinfo {author} {\bibfnamefont {K.}~\bibnamefont
  {Schultheiss}}, \bibinfo {author} {\bibfnamefont {G.}~\bibnamefont {Csaba}},
  \bibinfo {author} {\bibfnamefont {Q.}~\bibnamefont {Wang}}, \bibinfo {author}
  {\bibfnamefont {F.}~\bibnamefont {Ciubotaru}}, \bibinfo {author}
  {\bibfnamefont {D.~E.}\ \bibnamefont {Nikonov}}, \bibinfo {author}
  {\bibfnamefont {P.}~\bibnamefont {Che}}, \bibinfo {author} {\bibfnamefont
  {R.}~\bibnamefont {Hertel}}, \bibinfo {author} {\bibfnamefont
  {T.}~\bibnamefont {Ono}}, \bibinfo {author} {\bibfnamefont {D.}~\bibnamefont
  {Afanasiev}}, \bibinfo {author} {\bibfnamefont {J.}~\bibnamefont {Mentink}},
  \bibinfo {author} {\bibfnamefont {T.}~\bibnamefont {Rasing}}, \bibinfo
  {author} {\bibfnamefont {B.}~\bibnamefont {Hillebrands}}, \bibinfo {author}
  {\bibfnamefont {S.~V.}\ \bibnamefont {Kusminskiy}}, \bibinfo {author}
  {\bibfnamefont {W.}~\bibnamefont {Zhang}}, \bibinfo {author} {\bibfnamefont
  {C.~R.}\ \bibnamefont {Du}}, \bibinfo {author} {\bibfnamefont
  {A.}~\bibnamefont {Finco}}, \bibinfo {author} {\bibfnamefont
  {T.}~\bibnamefont {van~der Sar}}, \bibinfo {author} {\bibfnamefont {Y.~K.}\
  \bibnamefont {Luo}}, \bibinfo {author} {\bibfnamefont {Y.}~\bibnamefont
  {Shiota}}, \bibinfo {author} {\bibfnamefont {J.}~\bibnamefont {Sklenar}},
  \bibinfo {author} {\bibfnamefont {T.}~\bibnamefont {Yu}}, \ and\ \bibinfo
  {author} {\bibfnamefont {J.}~\bibnamefont {Rao}},\ }\href {\doibase
  10.1088/1361-648X/ad399c} {\bibfield  {journal} {\bibinfo  {journal} {J.
  Phys. Condens. Matter}\ }\textbf {\bibinfo {volume} {36}},\ \bibinfo {pages}
  {363501} (\bibinfo {year} {2024})}\BibitemShut {NoStop}%
\bibitem [{\citenamefont {Chumak}\ \emph {et~al.}(2022)\citenamefont {Chumak},
  \citenamefont {Kabos}, \citenamefont {Wu}, \citenamefont {Abert},
  \citenamefont {Adelmann}, \citenamefont {Adeyeye}, \citenamefont {Åkerman},
  \citenamefont {Aliev}, \citenamefont {Anane}, \citenamefont {Awad},
  \citenamefont {Back}, \citenamefont {Barman}, \citenamefont {Bauer},
  \citenamefont {Becherer}, \citenamefont {Beginin}, \citenamefont
  {Bittencourt}, \citenamefont {Blanter}, \citenamefont {Bortolotti},
  \citenamefont {Boventer}, \citenamefont {Bozhko}, \citenamefont {Bunyaev},
  \citenamefont {Carmiggelt}, \citenamefont {Cheenikundil}, \citenamefont
  {Ciubotaru}, \citenamefont {Cotofana}, \citenamefont {Csaba}, \citenamefont
  {Dobrovolskiy}, \citenamefont {Dubs}, \citenamefont {Elyasi}, \citenamefont
  {Fripp}, \citenamefont {Fulara}, \citenamefont {Golovchanskiy}, \citenamefont
  {Gonzalez-Ballestero}, \citenamefont {Graczyk}, \citenamefont {Grundler},
  \citenamefont {Gruszecki}, \citenamefont {Gubbiotti}, \citenamefont
  {Guslienko}, \citenamefont {Haldar}, \citenamefont {Hamdioui}, \citenamefont
  {Hertel}, \citenamefont {Hillebrands}, \citenamefont {Hioki}, \citenamefont
  {Houshang}, \citenamefont {Hu}, \citenamefont {Huebl}, \citenamefont {Huth},
  \citenamefont {Iacocca}, \citenamefont {Jungfleisch}, \citenamefont
  {Kakazei}, \citenamefont {Khitun}, \citenamefont {Khymyn}, \citenamefont
  {Kikkawa}, \citenamefont {Kläui}, \citenamefont {Klein}, \citenamefont
  {Kłos}, \citenamefont {Knauer}, \citenamefont {Koraltan}, \citenamefont
  {Kostylev}, \citenamefont {Krawczyk}, \citenamefont {Krivorotov},
  \citenamefont {Kruglyak}, \citenamefont {Lachance-Quirion}, \citenamefont
  {Ladak}, \citenamefont {Lebrun}, \citenamefont {Li}, \citenamefont {Lindner},
  \citenamefont {Macêdo}, \citenamefont {Mayr}, \citenamefont {Melkov},
  \citenamefont {Mieszczak}, \citenamefont {Nakamura}, \citenamefont {Nembach},
  \citenamefont {Nikitin}, \citenamefont {Nikitov}, \citenamefont {Novosad},
  \citenamefont {Otálora}, \citenamefont {Otani}, \citenamefont {Papp},
  \citenamefont {Pigeau}, \citenamefont {Pirro}, \citenamefont {Porod},
  \citenamefont {Porrati}, \citenamefont {Qin}, \citenamefont {Rana},
  \citenamefont {Reimann}, \citenamefont {Riente}, \citenamefont
  {Romero-Isart}, \citenamefont {Ross}, \citenamefont {Sadovnikov},
  \citenamefont {Safin}, \citenamefont {Saitoh}, \citenamefont {Schmidt},
  \citenamefont {Schultheiss}, \citenamefont {Schultheiss}, \citenamefont
  {Serga}, \citenamefont {Sharma}, \citenamefont {Shaw}, \citenamefont {Suess},
  \citenamefont {Surzhenko}, \citenamefont {Szulc}, \citenamefont {Taniguchi},
  \citenamefont {Urbánek}, \citenamefont {Usami}, \citenamefont {Ustinov},
  \citenamefont {van~der Sar}, \citenamefont {van Dijken}, \citenamefont
  {Vasyuchka}, \citenamefont {Verba}, \citenamefont {Viola~Kusminskiy},
  \citenamefont {Wang}, \citenamefont {Weides}, \citenamefont {Weiler},
  \citenamefont {Wintz}, \citenamefont {Wolski},\ and\ \citenamefont
  {Zhang}}]{ChumakIEEE2022}%
  \BibitemOpen
  \bibfield  {author} {\bibinfo {author} {\bibfnamefont {A.~V.}\ \bibnamefont
  {Chumak}}, \bibinfo {author} {\bibfnamefont {P.}~\bibnamefont {Kabos}},
  \bibinfo {author} {\bibfnamefont {M.}~\bibnamefont {Wu}}, \bibinfo {author}
  {\bibfnamefont {C.}~\bibnamefont {Abert}}, \bibinfo {author} {\bibfnamefont
  {C.}~\bibnamefont {Adelmann}}, \bibinfo {author} {\bibfnamefont {A.~O.}\
  \bibnamefont {Adeyeye}}, \bibinfo {author} {\bibfnamefont {J.}~\bibnamefont
  {Åkerman}}, \bibinfo {author} {\bibfnamefont {F.~G.}\ \bibnamefont {Aliev}},
  \bibinfo {author} {\bibfnamefont {A.}~\bibnamefont {Anane}}, \bibinfo
  {author} {\bibfnamefont {A.}~\bibnamefont {Awad}}, \bibinfo {author}
  {\bibfnamefont {C.~H.}\ \bibnamefont {Back}}, \bibinfo {author}
  {\bibfnamefont {A.}~\bibnamefont {Barman}}, \bibinfo {author} {\bibfnamefont
  {G.~E.~W.}\ \bibnamefont {Bauer}}, \bibinfo {author} {\bibfnamefont
  {M.}~\bibnamefont {Becherer}}, \bibinfo {author} {\bibfnamefont {E.~N.}\
  \bibnamefont {Beginin}}, \bibinfo {author} {\bibfnamefont {V.~A. S.~V.}\
  \bibnamefont {Bittencourt}}, \bibinfo {author} {\bibfnamefont {Y.~M.}\
  \bibnamefont {Blanter}}, \bibinfo {author} {\bibfnamefont {P.}~\bibnamefont
  {Bortolotti}}, \bibinfo {author} {\bibfnamefont {I.}~\bibnamefont
  {Boventer}}, \bibinfo {author} {\bibfnamefont {D.~A.}\ \bibnamefont
  {Bozhko}}, \bibinfo {author} {\bibfnamefont {S.~A.}\ \bibnamefont {Bunyaev}},
  \bibinfo {author} {\bibfnamefont {J.~J.}\ \bibnamefont {Carmiggelt}},
  \bibinfo {author} {\bibfnamefont {R.~R.}\ \bibnamefont {Cheenikundil}},
  \bibinfo {author} {\bibfnamefont {F.}~\bibnamefont {Ciubotaru}}, \bibinfo
  {author} {\bibfnamefont {S.}~\bibnamefont {Cotofana}}, \bibinfo {author}
  {\bibfnamefont {G.}~\bibnamefont {Csaba}}, \bibinfo {author} {\bibfnamefont
  {O.~V.}\ \bibnamefont {Dobrovolskiy}}, \bibinfo {author} {\bibfnamefont
  {C.}~\bibnamefont {Dubs}}, \bibinfo {author} {\bibfnamefont {M.}~\bibnamefont
  {Elyasi}}, \bibinfo {author} {\bibfnamefont {K.~G.}\ \bibnamefont {Fripp}},
  \bibinfo {author} {\bibfnamefont {H.}~\bibnamefont {Fulara}}, \bibinfo
  {author} {\bibfnamefont {I.~A.}\ \bibnamefont {Golovchanskiy}}, \bibinfo
  {author} {\bibfnamefont {C.}~\bibnamefont {Gonzalez-Ballestero}}, \bibinfo
  {author} {\bibfnamefont {P.}~\bibnamefont {Graczyk}}, \bibinfo {author}
  {\bibfnamefont {D.}~\bibnamefont {Grundler}}, \bibinfo {author}
  {\bibfnamefont {P.}~\bibnamefont {Gruszecki}}, \bibinfo {author}
  {\bibfnamefont {G.}~\bibnamefont {Gubbiotti}}, \bibinfo {author}
  {\bibfnamefont {K.}~\bibnamefont {Guslienko}}, \bibinfo {author}
  {\bibfnamefont {A.}~\bibnamefont {Haldar}}, \bibinfo {author} {\bibfnamefont
  {S.}~\bibnamefont {Hamdioui}}, \bibinfo {author} {\bibfnamefont
  {R.}~\bibnamefont {Hertel}}, \bibinfo {author} {\bibfnamefont
  {B.}~\bibnamefont {Hillebrands}}, \bibinfo {author} {\bibfnamefont
  {T.}~\bibnamefont {Hioki}}, \bibinfo {author} {\bibfnamefont
  {A.}~\bibnamefont {Houshang}}, \bibinfo {author} {\bibfnamefont {C.-M.}\
  \bibnamefont {Hu}}, \bibinfo {author} {\bibfnamefont {H.}~\bibnamefont
  {Huebl}}, \bibinfo {author} {\bibfnamefont {M.}~\bibnamefont {Huth}},
  \bibinfo {author} {\bibfnamefont {E.}~\bibnamefont {Iacocca}}, \bibinfo
  {author} {\bibfnamefont {M.~B.}\ \bibnamefont {Jungfleisch}}, \bibinfo
  {author} {\bibfnamefont {G.~N.}\ \bibnamefont {Kakazei}}, \bibinfo {author}
  {\bibfnamefont {A.}~\bibnamefont {Khitun}}, \bibinfo {author} {\bibfnamefont
  {R.}~\bibnamefont {Khymyn}}, \bibinfo {author} {\bibfnamefont
  {T.}~\bibnamefont {Kikkawa}}, \bibinfo {author} {\bibfnamefont
  {M.}~\bibnamefont {Kläui}}, \bibinfo {author} {\bibfnamefont
  {O.}~\bibnamefont {Klein}}, \bibinfo {author} {\bibfnamefont {J.~W.}\
  \bibnamefont {Kłos}}, \bibinfo {author} {\bibfnamefont {S.}~\bibnamefont
  {Knauer}}, \bibinfo {author} {\bibfnamefont {S.}~\bibnamefont {Koraltan}},
  \bibinfo {author} {\bibfnamefont {M.}~\bibnamefont {Kostylev}}, \bibinfo
  {author} {\bibfnamefont {M.}~\bibnamefont {Krawczyk}}, \bibinfo {author}
  {\bibfnamefont {I.~N.}\ \bibnamefont {Krivorotov}}, \bibinfo {author}
  {\bibfnamefont {V.~V.}\ \bibnamefont {Kruglyak}}, \bibinfo {author}
  {\bibfnamefont {D.}~\bibnamefont {Lachance-Quirion}}, \bibinfo {author}
  {\bibfnamefont {S.}~\bibnamefont {Ladak}}, \bibinfo {author} {\bibfnamefont
  {R.}~\bibnamefont {Lebrun}}, \bibinfo {author} {\bibfnamefont
  {Y.}~\bibnamefont {Li}}, \bibinfo {author} {\bibfnamefont {M.}~\bibnamefont
  {Lindner}}, \bibinfo {author} {\bibfnamefont {R.}~\bibnamefont {Macêdo}},
  \bibinfo {author} {\bibfnamefont {S.}~\bibnamefont {Mayr}}, \bibinfo {author}
  {\bibfnamefont {G.~A.}\ \bibnamefont {Melkov}}, \bibinfo {author}
  {\bibfnamefont {S.}~\bibnamefont {Mieszczak}}, \bibinfo {author}
  {\bibfnamefont {Y.}~\bibnamefont {Nakamura}}, \bibinfo {author}
  {\bibfnamefont {H.~T.}\ \bibnamefont {Nembach}}, \bibinfo {author}
  {\bibfnamefont {A.~A.}\ \bibnamefont {Nikitin}}, \bibinfo {author}
  {\bibfnamefont {S.~A.}\ \bibnamefont {Nikitov}}, \bibinfo {author}
  {\bibfnamefont {V.}~\bibnamefont {Novosad}}, \bibinfo {author} {\bibfnamefont
  {J.~A.}\ \bibnamefont {Otálora}}, \bibinfo {author} {\bibfnamefont
  {Y.}~\bibnamefont {Otani}}, \bibinfo {author} {\bibfnamefont
  {A.}~\bibnamefont {Papp}}, \bibinfo {author} {\bibfnamefont {B.}~\bibnamefont
  {Pigeau}}, \bibinfo {author} {\bibfnamefont {P.}~\bibnamefont {Pirro}},
  \bibinfo {author} {\bibfnamefont {W.}~\bibnamefont {Porod}}, \bibinfo
  {author} {\bibfnamefont {F.}~\bibnamefont {Porrati}}, \bibinfo {author}
  {\bibfnamefont {H.}~\bibnamefont {Qin}}, \bibinfo {author} {\bibfnamefont
  {B.}~\bibnamefont {Rana}}, \bibinfo {author} {\bibfnamefont {T.}~\bibnamefont
  {Reimann}}, \bibinfo {author} {\bibfnamefont {F.}~\bibnamefont {Riente}},
  \bibinfo {author} {\bibfnamefont {O.}~\bibnamefont {Romero-Isart}}, \bibinfo
  {author} {\bibfnamefont {A.}~\bibnamefont {Ross}}, \bibinfo {author}
  {\bibfnamefont {A.~V.}\ \bibnamefont {Sadovnikov}}, \bibinfo {author}
  {\bibfnamefont {A.~R.}\ \bibnamefont {Safin}}, \bibinfo {author}
  {\bibfnamefont {E.}~\bibnamefont {Saitoh}}, \bibinfo {author} {\bibfnamefont
  {G.}~\bibnamefont {Schmidt}}, \bibinfo {author} {\bibfnamefont
  {H.}~\bibnamefont {Schultheiss}}, \bibinfo {author} {\bibfnamefont
  {K.}~\bibnamefont {Schultheiss}}, \bibinfo {author} {\bibfnamefont {A.~A.}\
  \bibnamefont {Serga}}, \bibinfo {author} {\bibfnamefont {S.}~\bibnamefont
  {Sharma}}, \bibinfo {author} {\bibfnamefont {J.~M.}\ \bibnamefont {Shaw}},
  \bibinfo {author} {\bibfnamefont {D.}~\bibnamefont {Suess}}, \bibinfo
  {author} {\bibfnamefont {O.}~\bibnamefont {Surzhenko}}, \bibinfo {author}
  {\bibfnamefont {K.}~\bibnamefont {Szulc}}, \bibinfo {author} {\bibfnamefont
  {T.}~\bibnamefont {Taniguchi}}, \bibinfo {author} {\bibfnamefont
  {M.}~\bibnamefont {Urbánek}}, \bibinfo {author} {\bibfnamefont
  {K.}~\bibnamefont {Usami}}, \bibinfo {author} {\bibfnamefont {A.~B.}\
  \bibnamefont {Ustinov}}, \bibinfo {author} {\bibfnamefont {T.}~\bibnamefont
  {van~der Sar}}, \bibinfo {author} {\bibfnamefont {S.}~\bibnamefont {van
  Dijken}}, \bibinfo {author} {\bibfnamefont {V.~I.}\ \bibnamefont
  {Vasyuchka}}, \bibinfo {author} {\bibfnamefont {R.}~\bibnamefont {Verba}},
  \bibinfo {author} {\bibfnamefont {S.}~\bibnamefont {Viola~Kusminskiy}},
  \bibinfo {author} {\bibfnamefont {Q.}~\bibnamefont {Wang}}, \bibinfo {author}
  {\bibfnamefont {M.}~\bibnamefont {Weides}}, \bibinfo {author} {\bibfnamefont
  {M.}~\bibnamefont {Weiler}}, \bibinfo {author} {\bibfnamefont
  {S.}~\bibnamefont {Wintz}}, \bibinfo {author} {\bibfnamefont {S.~P.}\
  \bibnamefont {Wolski}}, \ and\ \bibinfo {author} {\bibfnamefont
  {X.}~\bibnamefont {Zhang}},\ }\href {\doibase 10.1109/TMAG.2022.3149664}
  {\bibfield  {journal} {\bibinfo  {journal} {IEEE Trans. Magn.}\ }\textbf
  {\bibinfo {volume} {58}},\ \bibinfo {pages} {1} (\bibinfo {year}
  {2022})}\BibitemShut {NoStop}%
\bibitem [{\citenamefont {Baltz}\ \emph {et~al.}(2018)\citenamefont {Baltz},
  \citenamefont {Manchon}, \citenamefont {Tsoi}, \citenamefont {Moriyama},
  \citenamefont {Ono},\ and\ \citenamefont {Tserkovnyak}}]{AFM_Rev1}%
  \BibitemOpen
  \bibfield  {author} {\bibinfo {author} {\bibfnamefont {V.}~\bibnamefont
  {Baltz}}, \bibinfo {author} {\bibfnamefont {A.}~\bibnamefont {Manchon}},
  \bibinfo {author} {\bibfnamefont {M.}~\bibnamefont {Tsoi}}, \bibinfo {author}
  {\bibfnamefont {T.}~\bibnamefont {Moriyama}}, \bibinfo {author}
  {\bibfnamefont {T.}~\bibnamefont {Ono}}, \ and\ \bibinfo {author}
  {\bibfnamefont {Y.}~\bibnamefont {Tserkovnyak}},\ }\href {\doibase
  10.1103/RevModPhys.90.015005} {\bibfield  {journal} {\bibinfo  {journal}
  {Rev. Mod. Phys.}\ }\textbf {\bibinfo {volume} {90}},\ \bibinfo {pages}
  {015005} (\bibinfo {year} {2018})}\BibitemShut {NoStop}%
\bibitem [{\citenamefont {Jungwirth}\ \emph {et~al.}(2018)\citenamefont
  {Jungwirth}, \citenamefont {Sinova}, \citenamefont {Manchon}, \citenamefont
  {Marti}, \citenamefont {Wunderlich},\ and\ \citenamefont
  {Felser}}]{AFM_Rev2}%
  \BibitemOpen
  \bibfield  {author} {\bibinfo {author} {\bibfnamefont {T.}~\bibnamefont
  {Jungwirth}}, \bibinfo {author} {\bibfnamefont {J.}~\bibnamefont {Sinova}},
  \bibinfo {author} {\bibfnamefont {A.}~\bibnamefont {Manchon}}, \bibinfo
  {author} {\bibfnamefont {X.}~\bibnamefont {Marti}}, \bibinfo {author}
  {\bibfnamefont {J.}~\bibnamefont {Wunderlich}}, \ and\ \bibinfo {author}
  {\bibfnamefont {C.}~\bibnamefont {Felser}},\ }\href {\doibase
  10.1038/s41567-018-0063-6} {\bibfield  {journal} {\bibinfo  {journal} {Nat.
  Phys.}\ }\textbf {\bibinfo {volume} {14}},\ \bibinfo {pages} {200} (\bibinfo
  {year} {2018})}\BibitemShut {NoStop}%
\bibitem [{\citenamefont {Spaldin}(2010)}]{spaldin2010magnetic}%
  \BibitemOpen
  \bibfield  {author} {\bibinfo {author} {\bibfnamefont {N.~A.}\ \bibnamefont
  {Spaldin}},\ }\href@noop {} {\emph {\bibinfo {title} {Magnetic materials:
  fundamentals and applications}}}\ (\bibinfo  {publisher} {Cambridge
  university press},\ \bibinfo {year} {2010})\BibitemShut {NoStop}%
\bibitem [{\citenamefont {Duine}\ \emph {et~al.}(2018)\citenamefont {Duine},
  \citenamefont {Lee}, \citenamefont {Parkin},\ and\ \citenamefont
  {Stiles}}]{duine2018synthetic}%
  \BibitemOpen
  \bibfield  {author} {\bibinfo {author} {\bibfnamefont {R.}~\bibnamefont
  {Duine}}, \bibinfo {author} {\bibfnamefont {K.-J.}\ \bibnamefont {Lee}},
  \bibinfo {author} {\bibfnamefont {S.~S.}\ \bibnamefont {Parkin}}, \ and\
  \bibinfo {author} {\bibfnamefont {M.~D.}\ \bibnamefont {Stiles}},\ }\href
  {\doibase https://doi.org/10.1038/s41567-018-0050-y} {\bibfield  {journal}
  {\bibinfo  {journal} {Nat. Phys.}\ }\textbf {\bibinfo {volume} {14}},\
  \bibinfo {pages} {217} (\bibinfo {year} {2018})}\BibitemShut {NoStop}%
\bibitem [{\citenamefont {Han}\ \emph {et~al.}(2023)\citenamefont {Han},
  \citenamefont {Cheng}, \citenamefont {Liu}, \citenamefont {Ohno},\ and\
  \citenamefont {Fukami}}]{han2023coherent}%
  \BibitemOpen
  \bibfield  {author} {\bibinfo {author} {\bibfnamefont {J.}~\bibnamefont
  {Han}}, \bibinfo {author} {\bibfnamefont {R.}~\bibnamefont {Cheng}}, \bibinfo
  {author} {\bibfnamefont {L.}~\bibnamefont {Liu}}, \bibinfo {author}
  {\bibfnamefont {H.}~\bibnamefont {Ohno}}, \ and\ \bibinfo {author}
  {\bibfnamefont {S.}~\bibnamefont {Fukami}},\ }\href {\doibase
  https://doi.org/10.1038/s41563-023-01492-6} {\bibfield  {journal} {\bibinfo
  {journal} {Nat. Mater.}\ }\textbf {\bibinfo {volume} {22}},\ \bibinfo {pages}
  {684} (\bibinfo {year} {2023})}\BibitemShut {NoStop}%
\bibitem [{\citenamefont {Jungwirth}\ \emph {et~al.}(2016)\citenamefont
  {Jungwirth}, \citenamefont {Marti}, \citenamefont {Wadley},\ and\
  \citenamefont {Wunderlich}}]{jungwirth2016antiferromagnetic}%
  \BibitemOpen
  \bibfield  {author} {\bibinfo {author} {\bibfnamefont {T.}~\bibnamefont
  {Jungwirth}}, \bibinfo {author} {\bibfnamefont {X.}~\bibnamefont {Marti}},
  \bibinfo {author} {\bibfnamefont {P.}~\bibnamefont {Wadley}}, \ and\ \bibinfo
  {author} {\bibfnamefont {J.}~\bibnamefont {Wunderlich}},\ }\href {\doibase
  https://doi.org/10.1038/nnano.2016.18} {\bibfield  {journal} {\bibinfo
  {journal} {Nat. Nanotechnol.}\ }\textbf {\bibinfo {volume} {11}},\ \bibinfo
  {pages} {231} (\bibinfo {year} {2016})}\BibitemShut {NoStop}%
\bibitem [{\citenamefont {Qaiumzadeh}\ \emph {et~al.}(2018)\citenamefont
  {Qaiumzadeh}, \citenamefont {Ado}, \citenamefont {Duine}, \citenamefont
  {Titov},\ and\ \citenamefont {Brataas}}]{qaiumzadeh2018}%
  \BibitemOpen
  \bibfield  {author} {\bibinfo {author} {\bibfnamefont {A.}~\bibnamefont
  {Qaiumzadeh}}, \bibinfo {author} {\bibfnamefont {I.~A.}\ \bibnamefont {Ado}},
  \bibinfo {author} {\bibfnamefont {R.~A.}\ \bibnamefont {Duine}}, \bibinfo
  {author} {\bibfnamefont {M.}~\bibnamefont {Titov}}, \ and\ \bibinfo {author}
  {\bibfnamefont {A.}~\bibnamefont {Brataas}},\ }\href {\doibase
  https://doi.org/10.1103/PhysRevLett.120.197202} {\bibfield  {journal}
  {\bibinfo  {journal} {Phys. Rev. Lett.}\ }\textbf {\bibinfo {volume} {120}},\
  \bibinfo {pages} {197202} (\bibinfo {year} {2018})}\BibitemShut {NoStop}%
\bibitem [{\citenamefont {Chumak}\ \emph {et~al.}(2015)\citenamefont {Chumak},
  \citenamefont {Vasyuchka}, \citenamefont {Serga},\ and\ \citenamefont
  {Hillebrands}}]{chumak2015magnon}%
  \BibitemOpen
  \bibfield  {author} {\bibinfo {author} {\bibfnamefont {A.~V.}\ \bibnamefont
  {Chumak}}, \bibinfo {author} {\bibfnamefont {V.~I.}\ \bibnamefont
  {Vasyuchka}}, \bibinfo {author} {\bibfnamefont {A.~A.}\ \bibnamefont
  {Serga}}, \ and\ \bibinfo {author} {\bibfnamefont {B.}~\bibnamefont
  {Hillebrands}},\ }\href {\doibase https://www.nature.com/articles/nphys3347}
  {\bibfield  {journal} {\bibinfo  {journal} {Nat. Phys.}\ }\textbf {\bibinfo
  {volume} {11}},\ \bibinfo {pages} {453} (\bibinfo {year} {2015})}\BibitemShut
  {NoStop}%
\bibitem [{\citenamefont {Chumak}(2019)}]{chumak2019magnon}%
  \BibitemOpen
  \bibfield  {author} {\bibinfo {author} {\bibfnamefont {A.~V.}\ \bibnamefont
  {Chumak}},\ }in\ \href@noop {} {\emph {\bibinfo {booktitle} {Spintronics
  Handbook, Second Edition: Spin Transport and Magnetism}}}\ (\bibinfo
  {publisher} {CRC Press},\ \bibinfo {year} {2019})\ pp.\ \bibinfo {pages}
  {247--302}\BibitemShut {NoStop}%
\bibitem [{\citenamefont {Wang}\ \emph {et~al.}(2020)\citenamefont {Wang},
  \citenamefont {Kewenig}, \citenamefont {Schneider}, \citenamefont {Verba},
  \citenamefont {Kohl}, \citenamefont {Heinz}, \citenamefont {Geilen},
  \citenamefont {Mohseni}, \citenamefont {L{\"a}gel}, \citenamefont {Ciubotaru}
  \emph {et~al.}}]{wang2020magnonic}%
  \BibitemOpen
  \bibfield  {author} {\bibinfo {author} {\bibfnamefont {Q.}~\bibnamefont
  {Wang}}, \bibinfo {author} {\bibfnamefont {M.}~\bibnamefont {Kewenig}},
  \bibinfo {author} {\bibfnamefont {M.}~\bibnamefont {Schneider}}, \bibinfo
  {author} {\bibfnamefont {R.}~\bibnamefont {Verba}}, \bibinfo {author}
  {\bibfnamefont {F.}~\bibnamefont {Kohl}}, \bibinfo {author} {\bibfnamefont
  {B.}~\bibnamefont {Heinz}}, \bibinfo {author} {\bibfnamefont
  {M.}~\bibnamefont {Geilen}}, \bibinfo {author} {\bibfnamefont
  {M.}~\bibnamefont {Mohseni}}, \bibinfo {author} {\bibfnamefont
  {B.}~\bibnamefont {L{\"a}gel}}, \bibinfo {author} {\bibfnamefont
  {F.}~\bibnamefont {Ciubotaru}},  \emph {et~al.},\ }\href {\doibase
  https://doi.org/10.1038/s41928-020-00485-6} {\bibfield  {journal} {\bibinfo
  {journal} {Nat. Electron.}\ }\textbf {\bibinfo {volume} {3}},\ \bibinfo
  {pages} {765} (\bibinfo {year} {2020})}\BibitemShut {NoStop}%
\bibitem [{\citenamefont {Hortensius}\ \emph {et~al.}(2021)\citenamefont
  {Hortensius}, \citenamefont {Afanasiev}, \citenamefont {Matthiesen},
  \citenamefont {Leenders}, \citenamefont {Citro}, \citenamefont {Kimel},
  \citenamefont {Mikhaylovskiy}, \citenamefont {Ivanov},\ and\ \citenamefont
  {Caviglia}}]{hortensius2021coherent}%
  \BibitemOpen
  \bibfield  {author} {\bibinfo {author} {\bibfnamefont {J.}~\bibnamefont
  {Hortensius}}, \bibinfo {author} {\bibfnamefont {D.}~\bibnamefont
  {Afanasiev}}, \bibinfo {author} {\bibfnamefont {M.}~\bibnamefont
  {Matthiesen}}, \bibinfo {author} {\bibfnamefont {R.}~\bibnamefont
  {Leenders}}, \bibinfo {author} {\bibfnamefont {R.}~\bibnamefont {Citro}},
  \bibinfo {author} {\bibfnamefont {A.}~\bibnamefont {Kimel}}, \bibinfo
  {author} {\bibfnamefont {R.}~\bibnamefont {Mikhaylovskiy}}, \bibinfo {author}
  {\bibfnamefont {B.}~\bibnamefont {Ivanov}}, \ and\ \bibinfo {author}
  {\bibfnamefont {A.}~\bibnamefont {Caviglia}},\ }\href {\doibase
  https://doi.org/10.1038/s41567-021-01290-4} {\bibfield  {journal} {\bibinfo
  {journal} {Nat. Phys.}\ }\textbf {\bibinfo {volume} {17}},\ \bibinfo {pages}
  {1001} (\bibinfo {year} {2021})}\BibitemShut {NoStop}%
\bibitem [{\citenamefont {Borovik-Romanov}\ and\ \citenamefont
  {Kreines}(1982)}]{BLS_Rev}%
  \BibitemOpen
  \bibfield  {author} {\bibinfo {author} {\bibfnamefont {A.}~\bibnamefont
  {Borovik-Romanov}}\ and\ \bibinfo {author} {\bibfnamefont {N.}~\bibnamefont
  {Kreines}},\ }\href {\doibase https://doi.org/10.1016/0370-1573(82)90118-1}
  {\bibfield  {journal} {\bibinfo  {journal} {Phys. Rep.}\ }\textbf {\bibinfo
  {volume} {81}},\ \bibinfo {pages} {351} (\bibinfo {year} {1982})}\BibitemShut
  {NoStop}%
\bibitem [{\citenamefont {{\v{Z}}elezn{\`y}}\ \emph {et~al.}(2014)\citenamefont
  {{\v{Z}}elezn{\`y}}, \citenamefont {Gao}, \citenamefont {V{\`y}born{\`y}},
  \citenamefont {Zemen}, \citenamefont {Ma{\v{s}}ek}, \citenamefont {Manchon},
  \citenamefont {Wunderlich}, \citenamefont {Sinova},\ and\ \citenamefont
  {Jungwirth}}]{vzelezny2014}%
  \BibitemOpen
  \bibfield  {author} {\bibinfo {author} {\bibfnamefont {J.}~\bibnamefont
  {{\v{Z}}elezn{\`y}}}, \bibinfo {author} {\bibfnamefont {H.}~\bibnamefont
  {Gao}}, \bibinfo {author} {\bibfnamefont {K.}~\bibnamefont
  {V{\`y}born{\`y}}}, \bibinfo {author} {\bibfnamefont {J.}~\bibnamefont
  {Zemen}}, \bibinfo {author} {\bibfnamefont {J.}~\bibnamefont {Ma{\v{s}}ek}},
  \bibinfo {author} {\bibfnamefont {A.}~\bibnamefont {Manchon}}, \bibinfo
  {author} {\bibfnamefont {J.}~\bibnamefont {Wunderlich}}, \bibinfo {author}
  {\bibfnamefont {J.}~\bibnamefont {Sinova}}, \ and\ \bibinfo {author}
  {\bibfnamefont {T.}~\bibnamefont {Jungwirth}},\ }\href {\doibase
  https://doi.org/10.1103/PhysRevLett.113.157201} {\bibfield  {journal}
  {\bibinfo  {journal} {Phys. Rev. Lett.}\ }\textbf {\bibinfo {volume} {113}},\
  \bibinfo {pages} {157201} (\bibinfo {year} {2014})}\BibitemShut {NoStop}%
\bibitem [{\citenamefont {Wadley}\ \emph {et~al.}(2016)\citenamefont {Wadley},
  \citenamefont {Howells}, \citenamefont {{\v{Z}}elezn{\`y}}, \citenamefont
  {Andrews}, \citenamefont {Hills}, \citenamefont {Campion}, \citenamefont
  {Nov{\'a}k}, \citenamefont {Olejn{\'\i}k}, \citenamefont {Maccherozzi},
  \citenamefont {Dhesi} \emph {et~al.}}]{wadley2016electrical}%
  \BibitemOpen
  \bibfield  {author} {\bibinfo {author} {\bibfnamefont {P.}~\bibnamefont
  {Wadley}}, \bibinfo {author} {\bibfnamefont {B.}~\bibnamefont {Howells}},
  \bibinfo {author} {\bibfnamefont {J.}~\bibnamefont {{\v{Z}}elezn{\`y}}},
  \bibinfo {author} {\bibfnamefont {C.}~\bibnamefont {Andrews}}, \bibinfo
  {author} {\bibfnamefont {V.}~\bibnamefont {Hills}}, \bibinfo {author}
  {\bibfnamefont {R.~P.}\ \bibnamefont {Campion}}, \bibinfo {author}
  {\bibfnamefont {V.}~\bibnamefont {Nov{\'a}k}}, \bibinfo {author}
  {\bibfnamefont {K.}~\bibnamefont {Olejn{\'\i}k}}, \bibinfo {author}
  {\bibfnamefont {F.}~\bibnamefont {Maccherozzi}}, \bibinfo {author}
  {\bibfnamefont {S.}~\bibnamefont {Dhesi}},  \emph {et~al.},\ }\href
  {https://www.science.org/doi/full/10.1126/science.aab1031} {\bibfield
  {journal} {\bibinfo  {journal} {Science}\ }\textbf {\bibinfo {volume}
  {351}},\ \bibinfo {pages} {587} (\bibinfo {year} {2016})}\BibitemShut
  {NoStop}%
\bibitem [{\citenamefont {Moriyama}\ \emph {et~al.}(2018)\citenamefont
  {Moriyama}, \citenamefont {Oda}, \citenamefont {Ohkochi}, \citenamefont
  {Kimata},\ and\ \citenamefont {Ono}}]{moriyama2018spin}%
  \BibitemOpen
  \bibfield  {author} {\bibinfo {author} {\bibfnamefont {T.}~\bibnamefont
  {Moriyama}}, \bibinfo {author} {\bibfnamefont {K.}~\bibnamefont {Oda}},
  \bibinfo {author} {\bibfnamefont {T.}~\bibnamefont {Ohkochi}}, \bibinfo
  {author} {\bibfnamefont {M.}~\bibnamefont {Kimata}}, \ and\ \bibinfo {author}
  {\bibfnamefont {T.}~\bibnamefont {Ono}},\ }\href {\doibase
  https://doi.org/10.1038/s41598-018-32508-w} {\bibfield  {journal} {\bibinfo
  {journal} {Scientific reports}\ }\textbf {\bibinfo {volume} {8}},\ \bibinfo
  {pages} {14167} (\bibinfo {year} {2018})}\BibitemShut {NoStop}%
\bibitem [{\citenamefont {Wienholdt}\ \emph {et~al.}(2012)\citenamefont
  {Wienholdt}, \citenamefont {Hinzke},\ and\ \citenamefont
  {Nowak}}]{wienholdt2012thz}%
  \BibitemOpen
  \bibfield  {author} {\bibinfo {author} {\bibfnamefont {S.}~\bibnamefont
  {Wienholdt}}, \bibinfo {author} {\bibfnamefont {D.}~\bibnamefont {Hinzke}}, \
  and\ \bibinfo {author} {\bibfnamefont {U.}~\bibnamefont {Nowak}},\ }\href
  {https://journals.aps.org/prl/pdf/10.1103/PhysRevLett.108.247207} {\bibfield
  {journal} {\bibinfo  {journal} {Phys. Rev. Lett.}\ }\textbf {\bibinfo
  {volume} {108}},\ \bibinfo {pages} {247207} (\bibinfo {year}
  {2012})}\BibitemShut {NoStop}%
\bibitem [{\citenamefont {Bossini}\ \emph {et~al.}(2019)\citenamefont
  {Bossini}, \citenamefont {Dal~Conte}, \citenamefont {Cerullo}, \citenamefont
  {Gomonay}, \citenamefont {Pisarev}, \citenamefont {Borovsak}, \citenamefont
  {Mihailovic}, \citenamefont {Sinova}, \citenamefont {Mentink}, \citenamefont
  {Rasing} \emph {et~al.}}]{bossini2019laser}%
  \BibitemOpen
  \bibfield  {author} {\bibinfo {author} {\bibfnamefont {D.}~\bibnamefont
  {Bossini}}, \bibinfo {author} {\bibfnamefont {S.}~\bibnamefont {Dal~Conte}},
  \bibinfo {author} {\bibfnamefont {G.}~\bibnamefont {Cerullo}}, \bibinfo
  {author} {\bibfnamefont {O.}~\bibnamefont {Gomonay}}, \bibinfo {author}
  {\bibfnamefont {R.}~\bibnamefont {Pisarev}}, \bibinfo {author} {\bibfnamefont
  {M.}~\bibnamefont {Borovsak}}, \bibinfo {author} {\bibfnamefont
  {D.}~\bibnamefont {Mihailovic}}, \bibinfo {author} {\bibfnamefont
  {J.}~\bibnamefont {Sinova}}, \bibinfo {author} {\bibfnamefont
  {J.}~\bibnamefont {Mentink}}, \bibinfo {author} {\bibfnamefont
  {T.}~\bibnamefont {Rasing}},  \emph {et~al.},\ }\href {\doibase
  10.1103/PhysRevB.100.024428} {\bibfield  {journal} {\bibinfo  {journal}
  {Phys. Rev. B}\ }\textbf {\bibinfo {volume} {100}},\ \bibinfo {pages}
  {024428} (\bibinfo {year} {2019})}\BibitemShut {NoStop}%
\bibitem [{\citenamefont {Olejn{\'\i}k}\ \emph {et~al.}(2018)\citenamefont
  {Olejn{\'\i}k}, \citenamefont {Seifert}, \citenamefont {Ka{\v{s}}par},
  \citenamefont {Nov{\'a}k}, \citenamefont {Wadley}, \citenamefont {Campion},
  \citenamefont {Baumgartner}, \citenamefont {Gambardella}, \citenamefont
  {N{\v{e}}mec}, \citenamefont {Wunderlich} \emph
  {et~al.}}]{olejnik2018terahertz}%
  \BibitemOpen
  \bibfield  {author} {\bibinfo {author} {\bibfnamefont {K.}~\bibnamefont
  {Olejn{\'\i}k}}, \bibinfo {author} {\bibfnamefont {T.}~\bibnamefont
  {Seifert}}, \bibinfo {author} {\bibfnamefont {Z.}~\bibnamefont
  {Ka{\v{s}}par}}, \bibinfo {author} {\bibfnamefont {V.}~\bibnamefont
  {Nov{\'a}k}}, \bibinfo {author} {\bibfnamefont {P.}~\bibnamefont {Wadley}},
  \bibinfo {author} {\bibfnamefont {R.~P.}\ \bibnamefont {Campion}}, \bibinfo
  {author} {\bibfnamefont {M.}~\bibnamefont {Baumgartner}}, \bibinfo {author}
  {\bibfnamefont {P.}~\bibnamefont {Gambardella}}, \bibinfo {author}
  {\bibfnamefont {P.}~\bibnamefont {N{\v{e}}mec}}, \bibinfo {author}
  {\bibfnamefont {J.}~\bibnamefont {Wunderlich}},  \emph {et~al.},\ }\href
  {\doibase DOI: 10.1126/sciadv.aar3566} {\bibfield  {journal} {\bibinfo
  {journal} {Sci. Adv.}\ }\textbf {\bibinfo {volume} {4}},\ \bibinfo {pages}
  {eaar3566} (\bibinfo {year} {2018})}\BibitemShut {NoStop}%
\bibitem [{\citenamefont {Satoh}\ \emph {et~al.}(2015)\citenamefont {Satoh},
  \citenamefont {Iida}, \citenamefont {Higuchi}, \citenamefont {Fiebig},\ and\
  \citenamefont {Shimura}}]{satoh2015writing}%
  \BibitemOpen
  \bibfield  {author} {\bibinfo {author} {\bibfnamefont {T.}~\bibnamefont
  {Satoh}}, \bibinfo {author} {\bibfnamefont {R.}~\bibnamefont {Iida}},
  \bibinfo {author} {\bibfnamefont {T.}~\bibnamefont {Higuchi}}, \bibinfo
  {author} {\bibfnamefont {M.}~\bibnamefont {Fiebig}}, \ and\ \bibinfo {author}
  {\bibfnamefont {T.}~\bibnamefont {Shimura}},\ }\href {\doibase
  https://doi.org/10.1038/nphoton.2014.273} {\bibfield  {journal} {\bibinfo
  {journal} {Nat. Photonics}\ }\textbf {\bibinfo {volume} {9}},\ \bibinfo
  {pages} {25} (\bibinfo {year} {2015})}\BibitemShut {NoStop}%
\bibitem [{\citenamefont {Kimel}\ \emph {et~al.}(2024)\citenamefont {Kimel},
  \citenamefont {Rasing},\ and\ \citenamefont {Ivanov}}]{KIMEL2024172039}%
  \BibitemOpen
  \bibfield  {author} {\bibinfo {author} {\bibfnamefont {A.}~\bibnamefont
  {Kimel}}, \bibinfo {author} {\bibfnamefont {T.}~\bibnamefont {Rasing}}, \
  and\ \bibinfo {author} {\bibfnamefont {B.}~\bibnamefont {Ivanov}},\ }\href
  {\doibase https://doi.org/10.1016/j.jmmm.2024.172039} {\bibfield  {journal}
  {\bibinfo  {journal} {J. Magn. Magn. Mater.}\ }\textbf {\bibinfo {volume}
  {598}},\ \bibinfo {pages} {172039} (\bibinfo {year} {2024})}\BibitemShut
  {NoStop}%
\bibitem [{\citenamefont {N{\v{e}}mec}\ \emph {et~al.}(2018)\citenamefont
  {N{\v{e}}mec}, \citenamefont {Fiebig}, \citenamefont {Kampfrath},\ and\
  \citenamefont {Kimel}}]{nvemec2018antiferromagnetic}%
  \BibitemOpen
  \bibfield  {author} {\bibinfo {author} {\bibfnamefont {P.}~\bibnamefont
  {N{\v{e}}mec}}, \bibinfo {author} {\bibfnamefont {M.}~\bibnamefont {Fiebig}},
  \bibinfo {author} {\bibfnamefont {T.}~\bibnamefont {Kampfrath}}, \ and\
  \bibinfo {author} {\bibfnamefont {A.~V.}\ \bibnamefont {Kimel}},\ }\href
  {\doibase https://doi.org/10.1038/s41567-018-0051-x} {\bibfield  {journal}
  {\bibinfo  {journal} {Nat. Phys.}\ }\textbf {\bibinfo {volume} {14}},\
  \bibinfo {pages} {229} (\bibinfo {year} {2018})}\BibitemShut {NoStop}%
\bibitem [{\citenamefont {Surynek}\ \emph {et~al.}(2024)\citenamefont
  {Surynek}, \citenamefont {Zubac}, \citenamefont {Olejnik}, \citenamefont
  {Farkas}, \citenamefont {Krizek}, \citenamefont {Nadvornik}, \citenamefont
  {Kubascik}, \citenamefont {Trojanek}, \citenamefont {Campion}, \citenamefont
  {Novak} \emph {et~al.}}]{surynek2024picosecond}%
  \BibitemOpen
  \bibfield  {author} {\bibinfo {author} {\bibfnamefont {M.}~\bibnamefont
  {Surynek}}, \bibinfo {author} {\bibfnamefont {J.}~\bibnamefont {Zubac}},
  \bibinfo {author} {\bibfnamefont {K.}~\bibnamefont {Olejnik}}, \bibinfo
  {author} {\bibfnamefont {A.}~\bibnamefont {Farkas}}, \bibinfo {author}
  {\bibfnamefont {F.}~\bibnamefont {Krizek}}, \bibinfo {author} {\bibfnamefont
  {L.}~\bibnamefont {Nadvornik}}, \bibinfo {author} {\bibfnamefont
  {P.}~\bibnamefont {Kubascik}}, \bibinfo {author} {\bibfnamefont
  {F.}~\bibnamefont {Trojanek}}, \bibinfo {author} {\bibfnamefont
  {R.}~\bibnamefont {Campion}}, \bibinfo {author} {\bibfnamefont
  {V.}~\bibnamefont {Novak}},  \emph {et~al.},\ }\href@noop {} {\bibfield
  {journal} {\bibinfo  {journal} {arXiv preprint arXiv:2401.17370}\ } (\bibinfo
  {year} {2024})}\BibitemShut {NoStop}%
\bibitem [{\citenamefont {Viola~Kusminskiy}(2019)}]{Silvia_Book}%
  \BibitemOpen
  \bibfield  {author} {\bibinfo {author} {\bibfnamefont {S.}~\bibnamefont
  {Viola~Kusminskiy}},\ }\href {\doibase 10.1007/978-3-030-13345-0} {\emph
  {\bibinfo {title} {Quantum Magnetism, Spin Waves, and Optical Cavities}}}\
  (\bibinfo  {publisher} {Springer Cham},\ \bibinfo {year} {2019})\BibitemShut
  {NoStop}%
\bibitem [{\citenamefont {Almpanis}(2021)}]{OMag_Book}%
  \BibitemOpen
  \bibfield  {author} {\bibinfo {author} {\bibfnamefont {E.}~\bibnamefont
  {Almpanis}},\ }\href@noop {} {\emph {\bibinfo {title} {Optomagnonic
  Structures: Novel Architectures for Simultaneous Control of Light and Spin
  Waves}}}\ (\bibinfo  {publisher} {World Scientific},\ \bibinfo {year}
  {2021})\BibitemShut {NoStop}%
\bibitem [{\citenamefont {Rameshti}\ \emph {et~al.}(2022)\citenamefont
  {Rameshti}, \citenamefont {Viola~Kusminskiy}, \citenamefont {Haigh},
  \citenamefont {Usami}, \citenamefont {Lachance-Quirion}, \citenamefont
  {Nakamura}, \citenamefont {Hu}, \citenamefont {Tang}, \citenamefont {Bauer},\
  and\ \citenamefont {Blanter}}]{rameshti2022cavity}%
  \BibitemOpen
  \bibfield  {author} {\bibinfo {author} {\bibfnamefont {B.~Z.}\ \bibnamefont
  {Rameshti}}, \bibinfo {author} {\bibfnamefont {S.}~\bibnamefont
  {Viola~Kusminskiy}}, \bibinfo {author} {\bibfnamefont {J.~A.}\ \bibnamefont
  {Haigh}}, \bibinfo {author} {\bibfnamefont {K.}~\bibnamefont {Usami}},
  \bibinfo {author} {\bibfnamefont {D.}~\bibnamefont {Lachance-Quirion}},
  \bibinfo {author} {\bibfnamefont {Y.}~\bibnamefont {Nakamura}}, \bibinfo
  {author} {\bibfnamefont {C.-M.}\ \bibnamefont {Hu}}, \bibinfo {author}
  {\bibfnamefont {H.~X.}\ \bibnamefont {Tang}}, \bibinfo {author}
  {\bibfnamefont {G.~E.}\ \bibnamefont {Bauer}}, \ and\ \bibinfo {author}
  {\bibfnamefont {Y.~M.}\ \bibnamefont {Blanter}},\ }\href {\doibase
  10.1016/j.physrep.2022.06.001} {\bibfield  {journal} {\bibinfo  {journal}
  {Phys. Rep.}\ }\textbf {\bibinfo {volume} {979}},\ \bibinfo {pages} {1}
  (\bibinfo {year} {2022})}\BibitemShut {NoStop}%
\bibitem [{\citenamefont {Lee}\ \emph {et~al.}(2023)\citenamefont {Lee},
  \citenamefont {Lee},\ and\ \citenamefont {Hwang}}]{lee2023cavity}%
  \BibitemOpen
  \bibfield  {author} {\bibinfo {author} {\bibfnamefont {J.~M.}\ \bibnamefont
  {Lee}}, \bibinfo {author} {\bibfnamefont {H.-W.}\ \bibnamefont {Lee}}, \ and\
  \bibinfo {author} {\bibfnamefont {M.-J.}\ \bibnamefont {Hwang}},\ }\href
  {\doibase https://doi.org/10.1103/PhysRevB.108.L241404} {\bibfield  {journal}
  {\bibinfo  {journal} {Phys. Rev. B}\ }\textbf {\bibinfo {volume} {108}},\
  \bibinfo {pages} {L241404} (\bibinfo {year} {2023})}\BibitemShut {NoStop}%
\bibitem [{\citenamefont {Sharma}\ \emph {et~al.}(2022)\citenamefont {Sharma},
  \citenamefont {Bittencourt},\ and\ \citenamefont
  {Viola~Kusminskiy}}]{sharma2022protocol}%
  \BibitemOpen
  \bibfield  {author} {\bibinfo {author} {\bibfnamefont {S.}~\bibnamefont
  {Sharma}}, \bibinfo {author} {\bibfnamefont {V.~A.}\ \bibnamefont
  {Bittencourt}}, \ and\ \bibinfo {author} {\bibfnamefont {S.}~\bibnamefont
  {Viola~Kusminskiy}},\ }\href {\doibase 10.1088/2515-7639/ac81f0} {\bibfield
  {journal} {\bibinfo  {journal} {Journal of Physics: Materials}\ }\textbf
  {\bibinfo {volume} {5}},\ \bibinfo {pages} {034006} (\bibinfo {year}
  {2022})}\BibitemShut {NoStop}%
\bibitem [{\citenamefont {Zhang}\ \emph {et~al.}(2014)\citenamefont {Zhang},
  \citenamefont {Zou}, \citenamefont {Jiang},\ and\ \citenamefont
  {Tang}}]{zhang2014strongly}%
  \BibitemOpen
  \bibfield  {author} {\bibinfo {author} {\bibfnamefont {X.}~\bibnamefont
  {Zhang}}, \bibinfo {author} {\bibfnamefont {C.-L.}\ \bibnamefont {Zou}},
  \bibinfo {author} {\bibfnamefont {L.}~\bibnamefont {Jiang}}, \ and\ \bibinfo
  {author} {\bibfnamefont {H.~X.}\ \bibnamefont {Tang}},\ }\href {\doibase
  https://doi.org/10.1103/PhysRevLett.113.156401} {\bibfield  {journal}
  {\bibinfo  {journal} {Phys. Rev. Lett.}\ }\textbf {\bibinfo {volume} {113}},\
  \bibinfo {pages} {156401} (\bibinfo {year} {2014})}\BibitemShut {NoStop}%
\bibitem [{\citenamefont {Hisatomi}\ \emph {et~al.}(2016)\citenamefont
  {Hisatomi}, \citenamefont {Osada}, \citenamefont {Tabuchi}, \citenamefont
  {Ishikawa}, \citenamefont {Noguchi}, \citenamefont {Yamazaki}, \citenamefont
  {Usami},\ and\ \citenamefont {Nakamura}}]{hisatomi2016bidirectional}%
  \BibitemOpen
  \bibfield  {author} {\bibinfo {author} {\bibfnamefont {R.}~\bibnamefont
  {Hisatomi}}, \bibinfo {author} {\bibfnamefont {A.}~\bibnamefont {Osada}},
  \bibinfo {author} {\bibfnamefont {Y.}~\bibnamefont {Tabuchi}}, \bibinfo
  {author} {\bibfnamefont {T.}~\bibnamefont {Ishikawa}}, \bibinfo {author}
  {\bibfnamefont {A.}~\bibnamefont {Noguchi}}, \bibinfo {author} {\bibfnamefont
  {R.}~\bibnamefont {Yamazaki}}, \bibinfo {author} {\bibfnamefont
  {K.}~\bibnamefont {Usami}}, \ and\ \bibinfo {author} {\bibfnamefont
  {Y.}~\bibnamefont {Nakamura}},\ }\href {\doibase
  https://doi.org/10.1103/PhysRevB.93.174427} {\bibfield  {journal} {\bibinfo
  {journal} {Phys. Rev. B}\ }\textbf {\bibinfo {volume} {93}},\ \bibinfo
  {pages} {174427} (\bibinfo {year} {2016})}\BibitemShut {NoStop}%
\bibitem [{\citenamefont {Viola~Kusminskiy}\ \emph {et~al.}(2016)\citenamefont
  {Viola~Kusminskiy}, \citenamefont {Tang},\ and\ \citenamefont
  {Marquardt}}]{kusminskiy2016coupled}%
  \BibitemOpen
  \bibfield  {author} {\bibinfo {author} {\bibfnamefont {S.}~\bibnamefont
  {Viola~Kusminskiy}}, \bibinfo {author} {\bibfnamefont {H.~X.}\ \bibnamefont
  {Tang}}, \ and\ \bibinfo {author} {\bibfnamefont {F.}~\bibnamefont
  {Marquardt}},\ }\href
  {https://journals.aps.org/pra/abstract/10.1103/PhysRevA.94.033821} {\bibfield
   {journal} {\bibinfo  {journal} {Phys. Rev. A}\ }\textbf {\bibinfo {volume}
  {94}},\ \bibinfo {pages} {033821} (\bibinfo {year} {2016})}\BibitemShut
  {NoStop}%
\bibitem [{\citenamefont {Liu}\ \emph {et~al.}(2016)\citenamefont {Liu},
  \citenamefont {Zhang}, \citenamefont {Tang},\ and\ \citenamefont
  {Flatt{\'e}}}]{liu2016optomagnonics}%
  \BibitemOpen
  \bibfield  {author} {\bibinfo {author} {\bibfnamefont {T.}~\bibnamefont
  {Liu}}, \bibinfo {author} {\bibfnamefont {X.}~\bibnamefont {Zhang}}, \bibinfo
  {author} {\bibfnamefont {H.~X.}\ \bibnamefont {Tang}}, \ and\ \bibinfo
  {author} {\bibfnamefont {M.~E.}\ \bibnamefont {Flatt{\'e}}},\ }\href
  {https://journals.aps.org/prb/abstract/10.1103/PhysRevB.94.060405} {\bibfield
   {journal} {\bibinfo  {journal} {Phys. Rev. B}\ }\textbf {\bibinfo {volume}
  {94}},\ \bibinfo {pages} {060405} (\bibinfo {year} {2016})}\BibitemShut
  {NoStop}%
\bibitem [{\citenamefont {Osada}\ \emph {et~al.}(2018)\citenamefont {Osada},
  \citenamefont {Gloppe}, \citenamefont {Hisatomi}, \citenamefont {Noguchi},
  \citenamefont {Yamazaki}, \citenamefont {Nomura}, \citenamefont {Nakamura},\
  and\ \citenamefont {Usami}}]{osada2018brillouin}%
  \BibitemOpen
  \bibfield  {author} {\bibinfo {author} {\bibfnamefont {A.}~\bibnamefont
  {Osada}}, \bibinfo {author} {\bibfnamefont {A.}~\bibnamefont {Gloppe}},
  \bibinfo {author} {\bibfnamefont {R.}~\bibnamefont {Hisatomi}}, \bibinfo
  {author} {\bibfnamefont {A.}~\bibnamefont {Noguchi}}, \bibinfo {author}
  {\bibfnamefont {R.}~\bibnamefont {Yamazaki}}, \bibinfo {author}
  {\bibfnamefont {M.}~\bibnamefont {Nomura}}, \bibinfo {author} {\bibfnamefont
  {Y.}~\bibnamefont {Nakamura}}, \ and\ \bibinfo {author} {\bibfnamefont
  {K.}~\bibnamefont {Usami}},\ }\href {\doibase
  https://doi.org/10.1103/PhysRevLett.120.133602} {\bibfield  {journal}
  {\bibinfo  {journal} {Phys. Rev. Lett.}\ }\textbf {\bibinfo {volume} {120}},\
  \bibinfo {pages} {133602} (\bibinfo {year} {2018})}\BibitemShut {NoStop}%
\bibitem [{\citenamefont {Liang}\ \emph {et~al.}(2023)\citenamefont {Liang},
  \citenamefont {Li},\ and\ \citenamefont {Wu}}]{liang2023all}%
  \BibitemOpen
  \bibfield  {author} {\bibinfo {author} {\bibfnamefont {Z.}~\bibnamefont
  {Liang}}, \bibinfo {author} {\bibfnamefont {J.}~\bibnamefont {Li}}, \ and\
  \bibinfo {author} {\bibfnamefont {Y.}~\bibnamefont {Wu}},\ }\href {\doibase
  https://doi.org/10.1103/PhysRevA.107.033701} {\bibfield  {journal} {\bibinfo
  {journal} {Phys. Rev. A}\ }\textbf {\bibinfo {volume} {107}},\ \bibinfo
  {pages} {033701} (\bibinfo {year} {2023})}\BibitemShut {NoStop}%
\bibitem [{\citenamefont {Sharma}\ \emph {et~al.}(2019)\citenamefont {Sharma},
  \citenamefont {Rameshti}, \citenamefont {Blanter},\ and\ \citenamefont
  {Bauer}}]{sharma2019optimal}%
  \BibitemOpen
  \bibfield  {author} {\bibinfo {author} {\bibfnamefont {S.}~\bibnamefont
  {Sharma}}, \bibinfo {author} {\bibfnamefont {B.~Z.}\ \bibnamefont
  {Rameshti}}, \bibinfo {author} {\bibfnamefont {Y.~M.}\ \bibnamefont
  {Blanter}}, \ and\ \bibinfo {author} {\bibfnamefont {G.~E.}\ \bibnamefont
  {Bauer}},\ }\href {\doibase https://doi.org/10.1103/PhysRevB.99.214423}
  {\bibfield  {journal} {\bibinfo  {journal} {Phys. Rev. B}\ }\textbf {\bibinfo
  {volume} {99}},\ \bibinfo {pages} {214423} (\bibinfo {year}
  {2019})}\BibitemShut {NoStop}%
\bibitem [{\citenamefont {Haigh}\ \emph {et~al.}(2016)\citenamefont {Haigh},
  \citenamefont {Nunnenkamp}, \citenamefont {Ramsay},\ and\ \citenamefont
  {Ferguson}}]{haigh2016triple}%
  \BibitemOpen
  \bibfield  {author} {\bibinfo {author} {\bibfnamefont {J.}~\bibnamefont
  {Haigh}}, \bibinfo {author} {\bibfnamefont {A.}~\bibnamefont {Nunnenkamp}},
  \bibinfo {author} {\bibfnamefont {A.}~\bibnamefont {Ramsay}}, \ and\ \bibinfo
  {author} {\bibfnamefont {A.}~\bibnamefont {Ferguson}},\ }\href {\doibase
  https://doi.org/10.1103/PhysRevLett.117.133602} {\bibfield  {journal}
  {\bibinfo  {journal} {Phys. Rev. Lett.}\ }\textbf {\bibinfo {volume} {117}},\
  \bibinfo {pages} {133602} (\bibinfo {year} {2016})}\BibitemShut {NoStop}%
\bibitem [{\citenamefont {\v{S}imi\'{c}}\ \emph {et~al.}(2020)\citenamefont
  {\v{S}imi\'{c}}, \citenamefont {Sharma}, \citenamefont {Blanter},\ and\
  \citenamefont {Bauer}}]{simic2020coherent}%
  \BibitemOpen
  \bibfield  {author} {\bibinfo {author} {\bibfnamefont {F.}~\bibnamefont
  {\v{S}imi\'{c}}}, \bibinfo {author} {\bibfnamefont {S.}~\bibnamefont
  {Sharma}}, \bibinfo {author} {\bibfnamefont {Y.~M.}\ \bibnamefont {Blanter}},
  \ and\ \bibinfo {author} {\bibfnamefont {G.~E.~W.}\ \bibnamefont {Bauer}},\
  }\href {\doibase 10.1103/PhysRevB.101.100401} {\bibfield  {journal} {\bibinfo
   {journal} {Phys. Rev. B}\ }\textbf {\bibinfo {volume} {101}},\ \bibinfo
  {pages} {100401} (\bibinfo {year} {2020})}\BibitemShut {NoStop}%
\bibitem [{\citenamefont {Sharma}\ \emph {et~al.}(2024)\citenamefont {Sharma},
  \citenamefont {Viola~Kusminskiy},\ and\ \citenamefont
  {Bittencourt}}]{sharma2024tomo}%
  \BibitemOpen
  \bibfield  {author} {\bibinfo {author} {\bibfnamefont {S.}~\bibnamefont
  {Sharma}}, \bibinfo {author} {\bibfnamefont {S.}~\bibnamefont
  {Viola~Kusminskiy}}, \ and\ \bibinfo {author} {\bibfnamefont {V.~A. S.~V.}\
  \bibnamefont {Bittencourt}},\ }\href {\doibase 10.1103/PhysRevB.110.014416}
  {\bibfield  {journal} {\bibinfo  {journal} {Phys. Rev. B}\ }\textbf {\bibinfo
  {volume} {110}},\ \bibinfo {pages} {014416} (\bibinfo {year}
  {2024})}\BibitemShut {NoStop}%
\bibitem [{\citenamefont {Parvini}\ \emph {et~al.}(2020)\citenamefont
  {Parvini}, \citenamefont {Bittencourt},\ and\ \citenamefont
  {Viola~Kusminskiy}}]{parvini2020}%
  \BibitemOpen
  \bibfield  {author} {\bibinfo {author} {\bibfnamefont {T.~S.}\ \bibnamefont
  {Parvini}}, \bibinfo {author} {\bibfnamefont {V.~A.}\ \bibnamefont
  {Bittencourt}}, \ and\ \bibinfo {author} {\bibfnamefont {S.}~\bibnamefont
  {Viola~Kusminskiy}},\ }\href {\doibase 10.1103/PhysRevResearch.2.022027}
  {\bibfield  {journal} {\bibinfo  {journal} {Phys. Rev. Res.}\ }\textbf
  {\bibinfo {volume} {2}},\ \bibinfo {pages} {022027} (\bibinfo {year}
  {2020})}\BibitemShut {NoStop}%
\bibitem [{\citenamefont {Xiao}\ \emph {et~al.}(2019)\citenamefont {Xiao},
  \citenamefont {Yan}, \citenamefont {Zhang}, \citenamefont {Grigoryan},
  \citenamefont {Hu}, \citenamefont {Guo},\ and\ \citenamefont
  {Xia}}]{xiao2019magnon}%
  \BibitemOpen
  \bibfield  {author} {\bibinfo {author} {\bibfnamefont {Y.}~\bibnamefont
  {Xiao}}, \bibinfo {author} {\bibfnamefont {X.}~\bibnamefont {Yan}}, \bibinfo
  {author} {\bibfnamefont {Y.}~\bibnamefont {Zhang}}, \bibinfo {author}
  {\bibfnamefont {V.}~\bibnamefont {Grigoryan}}, \bibinfo {author}
  {\bibfnamefont {C.}~\bibnamefont {Hu}}, \bibinfo {author} {\bibfnamefont
  {H.}~\bibnamefont {Guo}}, \ and\ \bibinfo {author} {\bibfnamefont
  {K.}~\bibnamefont {Xia}},\ }\href {\doibase
  https://doi.org/10.1103/PhysRevB.99.094407} {\bibfield  {journal} {\bibinfo
  {journal} {Phys. Rev. B}\ }\textbf {\bibinfo {volume} {99}},\ \bibinfo
  {pages} {094407} (\bibinfo {year} {2019})}\BibitemShut {NoStop}%
\bibitem [{\citenamefont {Curtis}\ \emph {et~al.}(2022)\citenamefont {Curtis},
  \citenamefont {Grankin}, \citenamefont {Poniatowski}, \citenamefont
  {Galitski}, \citenamefont {Narang},\ and\ \citenamefont
  {Demler}}]{curtis2022cavity}%
  \BibitemOpen
  \bibfield  {author} {\bibinfo {author} {\bibfnamefont {J.~B.}\ \bibnamefont
  {Curtis}}, \bibinfo {author} {\bibfnamefont {A.}~\bibnamefont {Grankin}},
  \bibinfo {author} {\bibfnamefont {N.~R.}\ \bibnamefont {Poniatowski}},
  \bibinfo {author} {\bibfnamefont {V.~M.}\ \bibnamefont {Galitski}}, \bibinfo
  {author} {\bibfnamefont {P.}~\bibnamefont {Narang}}, \ and\ \bibinfo {author}
  {\bibfnamefont {E.}~\bibnamefont {Demler}},\ }\href {\doibase
  https://doi.org/10.1103/PhysRevResearch.4.013101} {\bibfield  {journal}
  {\bibinfo  {journal} {Phys. Rev. Res.}\ }\textbf {\bibinfo {volume} {4}},\
  \bibinfo {pages} {013101} (\bibinfo {year} {2022})}\BibitemShut {NoStop}%
\bibitem [{\citenamefont {Boventer}\ \emph {et~al.}(2023)\citenamefont
  {Boventer}, \citenamefont {Simensen}, \citenamefont {Brekke}, \citenamefont
  {Weides}, \citenamefont {Anane}, \citenamefont {Kl{\"a}ui}, \citenamefont
  {Brataas},\ and\ \citenamefont {Lebrun}}]{boventer2023antiferromagnetic}%
  \BibitemOpen
  \bibfield  {author} {\bibinfo {author} {\bibfnamefont {I.}~\bibnamefont
  {Boventer}}, \bibinfo {author} {\bibfnamefont {H.}~\bibnamefont {Simensen}},
  \bibinfo {author} {\bibfnamefont {B.}~\bibnamefont {Brekke}}, \bibinfo
  {author} {\bibfnamefont {M.}~\bibnamefont {Weides}}, \bibinfo {author}
  {\bibfnamefont {A.}~\bibnamefont {Anane}}, \bibinfo {author} {\bibfnamefont
  {M.}~\bibnamefont {Kl{\"a}ui}}, \bibinfo {author} {\bibfnamefont
  {A.}~\bibnamefont {Brataas}}, \ and\ \bibinfo {author} {\bibfnamefont
  {R.}~\bibnamefont {Lebrun}},\ }\href {\doibase
  https://doi.org/10.1103/PhysRevApplied.19.014071} {\bibfield  {journal}
  {\bibinfo  {journal} {Phys. Rev. Appl.}\ }\textbf {\bibinfo {volume} {19}},\
  \bibinfo {pages} {014071} (\bibinfo {year} {2023})}\BibitemShut {NoStop}%
\bibitem [{\citenamefont {Bia{\l}ek}\ \emph {et~al.}(2021)\citenamefont
  {Bia{\l}ek}, \citenamefont {Zhang}, \citenamefont {Yu},\ and\ \citenamefont
  {Ansermet}}]{bialek2021strong}%
  \BibitemOpen
  \bibfield  {author} {\bibinfo {author} {\bibfnamefont {M.}~\bibnamefont
  {Bia{\l}ek}}, \bibinfo {author} {\bibfnamefont {J.}~\bibnamefont {Zhang}},
  \bibinfo {author} {\bibfnamefont {H.}~\bibnamefont {Yu}}, \ and\ \bibinfo
  {author} {\bibfnamefont {J.-P.}\ \bibnamefont {Ansermet}},\ }\href {\doibase
  https://doi.org/10.1103/PhysRevApplied.15.044018} {\bibfield  {journal}
  {\bibinfo  {journal} {Phys. Rev. Appl.}\ }\textbf {\bibinfo {volume} {15}},\
  \bibinfo {pages} {044018} (\bibinfo {year} {2021})}\BibitemShut {NoStop}%
\bibitem [{\citenamefont {Zhang}\ \emph
  {et~al.}(2021{\natexlab{a}})\citenamefont {Zhang}, \citenamefont {Sun},
  \citenamefont {Lu}, \citenamefont {Guo}, \citenamefont {Xue}, \citenamefont
  {Chen}, \citenamefont {Tian}, \citenamefont {Yan},\ and\ \citenamefont
  {Bai}}]{zhang2021zero}%
  \BibitemOpen
  \bibfield  {author} {\bibinfo {author} {\bibfnamefont {Q.}~\bibnamefont
  {Zhang}}, \bibinfo {author} {\bibfnamefont {Y.}~\bibnamefont {Sun}}, \bibinfo
  {author} {\bibfnamefont {Z.}~\bibnamefont {Lu}}, \bibinfo {author}
  {\bibfnamefont {J.}~\bibnamefont {Guo}}, \bibinfo {author} {\bibfnamefont
  {J.}~\bibnamefont {Xue}}, \bibinfo {author} {\bibfnamefont {Y.}~\bibnamefont
  {Chen}}, \bibinfo {author} {\bibfnamefont {Y.}~\bibnamefont {Tian}}, \bibinfo
  {author} {\bibfnamefont {S.}~\bibnamefont {Yan}}, \ and\ \bibinfo {author}
  {\bibfnamefont {L.}~\bibnamefont {Bai}},\ }\href
  {https://pubs.aip.org/aip/apl/article/119/10/102402/39618/Zero-field-magnon-photon-coupling-in}
  {\bibfield  {journal} {\bibinfo  {journal} {Appl. Phys. Lett.}\ }\textbf
  {\bibinfo {volume} {119}} (\bibinfo {year} {2021}{\natexlab{a}})}\BibitemShut
  {NoStop}%
\bibitem [{\citenamefont {Fleury}\ and\ \citenamefont
  {Loudon}(1968)}]{PhysRev.Fleury}%
  \BibitemOpen
  \bibfield  {author} {\bibinfo {author} {\bibfnamefont {P.~A.}\ \bibnamefont
  {Fleury}}\ and\ \bibinfo {author} {\bibfnamefont {R.}~\bibnamefont
  {Loudon}},\ }\href {\doibase 10.1103/PhysRev.166.514} {\bibfield  {journal}
  {\bibinfo  {journal} {Phys. Rev.}\ }\textbf {\bibinfo {volume} {166}},\
  \bibinfo {pages} {514} (\bibinfo {year} {1968})}\BibitemShut {NoStop}%
\bibitem [{\citenamefont {Fleury}\ \emph {et~al.}(1966)\citenamefont {Fleury},
  \citenamefont {Porto}, \citenamefont {Cheesman},\ and\ \citenamefont
  {Guggenheim}}]{fleury1966light}%
  \BibitemOpen
  \bibfield  {author} {\bibinfo {author} {\bibfnamefont {P.}~\bibnamefont
  {Fleury}}, \bibinfo {author} {\bibfnamefont {S.}~\bibnamefont {Porto}},
  \bibinfo {author} {\bibfnamefont {L.}~\bibnamefont {Cheesman}}, \ and\
  \bibinfo {author} {\bibfnamefont {H.}~\bibnamefont {Guggenheim}},\ }\href
  {\doibase 10.1103/PhysRevLett.17.84} {\bibfield  {journal} {\bibinfo
  {journal} {Phys. Rev. Lett.}\ }\textbf {\bibinfo {volume} {17}},\ \bibinfo
  {pages} {84} (\bibinfo {year} {1966})}\BibitemShut {NoStop}%
\bibitem [{\citenamefont {Loudon}(1968)}]{loudon1968theory}%
  \BibitemOpen
  \bibfield  {author} {\bibinfo {author} {\bibfnamefont {R.}~\bibnamefont
  {Loudon}},\ }\href {\doibase https://doi.org/10.1080/00018736800101296}
  {\bibfield  {journal} {\bibinfo  {journal} {Adv. Phys.}\ }\textbf {\bibinfo
  {volume} {17}},\ \bibinfo {pages} {243} (\bibinfo {year} {1968})}\BibitemShut
  {NoStop}%
\bibitem [{\citenamefont {Fleury}\ \emph {et~al.}(1967)\citenamefont {Fleury},
  \citenamefont {Porto},\ and\ \citenamefont {Loudon}}]{fleury1967two}%
  \BibitemOpen
  \bibfield  {author} {\bibinfo {author} {\bibfnamefont {P.~A.}\ \bibnamefont
  {Fleury}}, \bibinfo {author} {\bibfnamefont {S.~P.~S.}\ \bibnamefont
  {Porto}}, \ and\ \bibinfo {author} {\bibfnamefont {R.}~\bibnamefont
  {Loudon}},\ }\href {\doibase 10.1103/PhysRevLett.18.658} {\bibfield
  {journal} {\bibinfo  {journal} {Phys. Rev. Lett.}\ }\textbf {\bibinfo
  {volume} {18}},\ \bibinfo {pages} {658} (\bibinfo {year} {1967})}\BibitemShut
  {NoStop}%
\bibitem [{\citenamefont {Lockwood}\ and\ \citenamefont
  {Cottam}(1987)}]{PhysRevB.35.1973}%
  \BibitemOpen
  \bibfield  {author} {\bibinfo {author} {\bibfnamefont {D.~J.}\ \bibnamefont
  {Lockwood}}\ and\ \bibinfo {author} {\bibfnamefont {M.~G.}\ \bibnamefont
  {Cottam}},\ }\href {\doibase 10.1103/PhysRevB.35.1973} {\bibfield  {journal}
  {\bibinfo  {journal} {Phys. Rev. B}\ }\textbf {\bibinfo {volume} {35}},\
  \bibinfo {pages} {1973} (\bibinfo {year} {1987})}\BibitemShut {NoStop}%
\bibitem [{\citenamefont {Cottam}\ \emph {et~al.}(1983)\citenamefont {Cottam},
  \citenamefont {So}, \citenamefont {Lockwood}, \citenamefont {Katiyar},\ and\
  \citenamefont {Guggenheim}}]{cottam1983raman}%
  \BibitemOpen
  \bibfield  {author} {\bibinfo {author} {\bibfnamefont {M.}~\bibnamefont
  {Cottam}}, \bibinfo {author} {\bibfnamefont {V.}~\bibnamefont {So}}, \bibinfo
  {author} {\bibfnamefont {D.}~\bibnamefont {Lockwood}}, \bibinfo {author}
  {\bibfnamefont {R.}~\bibnamefont {Katiyar}}, \ and\ \bibinfo {author}
  {\bibfnamefont {H.}~\bibnamefont {Guggenheim}},\ }\href
  {https://iopscience.iop.org/article/10.1088/0022-3719/16/9/017/pdf?casa_token=lX9bT2YUM2QAAAAA:_dEixzpdMRx1LGh8r7S5SWv90eiskked_bsuZoc06M7-CCUhY39aM4nVFCgjzok8McJxo2Gw4BRhbj8}
  {\bibfield  {journal} {\bibinfo  {journal} {J. phys., C, Solid state phys.}\
  }\textbf {\bibinfo {volume} {16}},\ \bibinfo {pages} {1741} (\bibinfo {year}
  {1983})}\BibitemShut {NoStop}%
\bibitem [{\citenamefont {Lee}\ \emph {et~al.}(2006)\citenamefont {Lee},
  \citenamefont {Nagaosa},\ and\ \citenamefont {Wen}}]{Mott_Rev}%
  \BibitemOpen
  \bibfield  {author} {\bibinfo {author} {\bibfnamefont {P.~A.}\ \bibnamefont
  {Lee}}, \bibinfo {author} {\bibfnamefont {N.}~\bibnamefont {Nagaosa}}, \ and\
  \bibinfo {author} {\bibfnamefont {X.-G.}\ \bibnamefont {Wen}},\ }\href
  {\doibase 10.1103/RevModPhys.78.17} {\bibfield  {journal} {\bibinfo
  {journal} {Rev. Mod. Phys.}\ }\textbf {\bibinfo {volume} {78}},\ \bibinfo
  {pages} {17} (\bibinfo {year} {2006})}\BibitemShut {NoStop}%
\bibitem [{\citenamefont {Moriya}\ and\ \citenamefont
  {Ueda}(2003)}]{Moriya_2003}%
  \BibitemOpen
  \bibfield  {author} {\bibinfo {author} {\bibfnamefont {T.}~\bibnamefont
  {Moriya}}\ and\ \bibinfo {author} {\bibfnamefont {K.}~\bibnamefont {Ueda}},\
  }\href {\doibase 10.1088/0034-4885/66/8/202} {\bibfield  {journal} {\bibinfo
  {journal} {Rep. Prog. Phys.}\ }\textbf {\bibinfo {volume} {66}},\ \bibinfo
  {pages} {1299} (\bibinfo {year} {2003})}\BibitemShut {NoStop}%
\bibitem [{\citenamefont {Erlandsen}\ \emph {et~al.}(2019)\citenamefont
  {Erlandsen}, \citenamefont {Kamra}, \citenamefont {Brataas},\ and\
  \citenamefont {Sudb{\o}}}]{erlandsen2019enhancement}%
  \BibitemOpen
  \bibfield  {author} {\bibinfo {author} {\bibfnamefont {E.}~\bibnamefont
  {Erlandsen}}, \bibinfo {author} {\bibfnamefont {A.}~\bibnamefont {Kamra}},
  \bibinfo {author} {\bibfnamefont {A.}~\bibnamefont {Brataas}}, \ and\
  \bibinfo {author} {\bibfnamefont {A.}~\bibnamefont {Sudb{\o}}},\ }\href
  {\doibase https://doi.org/10.1103/PhysRevB.100.100503} {\bibfield  {journal}
  {\bibinfo  {journal} {Phys. Rev. B}\ }\textbf {\bibinfo {volume} {100}},\
  \bibinfo {pages} {100503} (\bibinfo {year} {2019})}\BibitemShut {NoStop}%
\bibitem [{\citenamefont {Canali}\ and\ \citenamefont
  {Girvin}(1992)}]{canali1992theory}%
  \BibitemOpen
  \bibfield  {author} {\bibinfo {author} {\bibfnamefont {C.}~\bibnamefont
  {Canali}}\ and\ \bibinfo {author} {\bibfnamefont {S.}~\bibnamefont
  {Girvin}},\ }\href {\doibase https://doi.org/10.1103/PhysRevB.45.7127}
  {\bibfield  {journal} {\bibinfo  {journal} {Phys. Rev. B}\ }\textbf {\bibinfo
  {volume} {45}},\ \bibinfo {pages} {7127} (\bibinfo {year}
  {1992})}\BibitemShut {NoStop}%
\bibitem [{\citenamefont {Kastner}\ \emph {et~al.}(1998)\citenamefont
  {Kastner}, \citenamefont {Birgeneau}, \citenamefont {Shirane},\ and\
  \citenamefont {Endoh}}]{kastner1998magnetic}%
  \BibitemOpen
  \bibfield  {author} {\bibinfo {author} {\bibfnamefont {M.}~\bibnamefont
  {Kastner}}, \bibinfo {author} {\bibfnamefont {R.}~\bibnamefont {Birgeneau}},
  \bibinfo {author} {\bibfnamefont {G.}~\bibnamefont {Shirane}}, \ and\
  \bibinfo {author} {\bibfnamefont {Y.}~\bibnamefont {Endoh}},\ }\href
  {\doibase https://doi.org/10.1103/RevModPhys.70.897} {\bibfield  {journal}
  {\bibinfo  {journal} {Rev. Mod. Phys.}\ }\textbf {\bibinfo {volume} {70}},\
  \bibinfo {pages} {897} (\bibinfo {year} {1998})}\BibitemShut {NoStop}%
\bibitem [{\citenamefont {Betto}\ \emph {et~al.}(2021)\citenamefont {Betto},
  \citenamefont {Fumagalli}, \citenamefont {Martinelli}, \citenamefont {Rossi},
  \citenamefont {Piombo}, \citenamefont {Yoshimi}, \citenamefont {Di~Castro},
  \citenamefont {Di~Gennaro}, \citenamefont {Sambri}, \citenamefont {Bonn}
  \emph {et~al.}}]{betto2021multiple}%
  \BibitemOpen
  \bibfield  {author} {\bibinfo {author} {\bibfnamefont {D.}~\bibnamefont
  {Betto}}, \bibinfo {author} {\bibfnamefont {R.}~\bibnamefont {Fumagalli}},
  \bibinfo {author} {\bibfnamefont {L.}~\bibnamefont {Martinelli}}, \bibinfo
  {author} {\bibfnamefont {M.}~\bibnamefont {Rossi}}, \bibinfo {author}
  {\bibfnamefont {R.}~\bibnamefont {Piombo}}, \bibinfo {author} {\bibfnamefont
  {K.}~\bibnamefont {Yoshimi}}, \bibinfo {author} {\bibfnamefont
  {D.}~\bibnamefont {Di~Castro}}, \bibinfo {author} {\bibfnamefont
  {E.}~\bibnamefont {Di~Gennaro}}, \bibinfo {author} {\bibfnamefont
  {A.}~\bibnamefont {Sambri}}, \bibinfo {author} {\bibfnamefont
  {D.}~\bibnamefont {Bonn}},  \emph {et~al.},\ }\href {\doibase
  https://doi.org/10.1103/PhysRevB.103.L140409} {\bibfield  {journal} {\bibinfo
   {journal} {Phys. Rev. B}\ }\textbf {\bibinfo {volume} {103}},\ \bibinfo
  {pages} {L140409} (\bibinfo {year} {2021})}\BibitemShut {NoStop}%
\bibitem [{\citenamefont {Scalapino}(2012)}]{scalapino2012common}%
  \BibitemOpen
  \bibfield  {author} {\bibinfo {author} {\bibfnamefont {D.~J.}\ \bibnamefont
  {Scalapino}},\ }\href {\doibase https://doi.org/10.1103/RevModPhys.84.1383}
  {\bibfield  {journal} {\bibinfo  {journal} {Rev. Mod. Phys.}\ }\textbf
  {\bibinfo {volume} {84}},\ \bibinfo {pages} {1383} (\bibinfo {year}
  {2012})}\BibitemShut {NoStop}%
\bibitem [{\citenamefont {Mitrano}\ \emph {et~al.}(2016)\citenamefont
  {Mitrano}, \citenamefont {Cantaluppi}, \citenamefont {Nicoletti},
  \citenamefont {Kaiser}, \citenamefont {Perucchi}, \citenamefont {Lupi},
  \citenamefont {Di~Pietro}, \citenamefont {Pontiroli}, \citenamefont
  {Ricc{\`o}}, \citenamefont {Clark} \emph {et~al.}}]{mitrano2016possible}%
  \BibitemOpen
  \bibfield  {author} {\bibinfo {author} {\bibfnamefont {M.}~\bibnamefont
  {Mitrano}}, \bibinfo {author} {\bibfnamefont {A.}~\bibnamefont {Cantaluppi}},
  \bibinfo {author} {\bibfnamefont {D.}~\bibnamefont {Nicoletti}}, \bibinfo
  {author} {\bibfnamefont {S.}~\bibnamefont {Kaiser}}, \bibinfo {author}
  {\bibfnamefont {A.}~\bibnamefont {Perucchi}}, \bibinfo {author}
  {\bibfnamefont {S.}~\bibnamefont {Lupi}}, \bibinfo {author} {\bibfnamefont
  {P.}~\bibnamefont {Di~Pietro}}, \bibinfo {author} {\bibfnamefont
  {D.}~\bibnamefont {Pontiroli}}, \bibinfo {author} {\bibfnamefont
  {M.}~\bibnamefont {Ricc{\`o}}}, \bibinfo {author} {\bibfnamefont {S.~R.}\
  \bibnamefont {Clark}},  \emph {et~al.},\ }\href {\doibase
  https://doi.org/10.1038/nature16522} {\bibfield  {journal} {\bibinfo
  {journal} {Nature}\ }\textbf {\bibinfo {volume} {530}},\ \bibinfo {pages}
  {461} (\bibinfo {year} {2016})}\BibitemShut {NoStop}%
\bibitem [{\citenamefont {Mankowsky}\ \emph {et~al.}(2014)\citenamefont
  {Mankowsky}, \citenamefont {Subedi}, \citenamefont {F{\"o}rst}, \citenamefont
  {Mariager}, \citenamefont {Chollet}, \citenamefont {Lemke}, \citenamefont
  {Robinson}, \citenamefont {Glownia}, \citenamefont {Minitti}, \citenamefont
  {Frano} \emph {et~al.}}]{mankowsky2014nonlinear}%
  \BibitemOpen
  \bibfield  {author} {\bibinfo {author} {\bibfnamefont {R.}~\bibnamefont
  {Mankowsky}}, \bibinfo {author} {\bibfnamefont {A.}~\bibnamefont {Subedi}},
  \bibinfo {author} {\bibfnamefont {M.}~\bibnamefont {F{\"o}rst}}, \bibinfo
  {author} {\bibfnamefont {S.~O.}\ \bibnamefont {Mariager}}, \bibinfo {author}
  {\bibfnamefont {M.}~\bibnamefont {Chollet}}, \bibinfo {author} {\bibfnamefont
  {H.}~\bibnamefont {Lemke}}, \bibinfo {author} {\bibfnamefont {J.~S.}\
  \bibnamefont {Robinson}}, \bibinfo {author} {\bibfnamefont {J.~M.}\
  \bibnamefont {Glownia}}, \bibinfo {author} {\bibfnamefont {M.~P.}\
  \bibnamefont {Minitti}}, \bibinfo {author} {\bibfnamefont {A.}~\bibnamefont
  {Frano}},  \emph {et~al.},\ }\href {\doibase
  https://doi.org/10.1038/nature13875} {\bibfield  {journal} {\bibinfo
  {journal} {Nature}\ }\textbf {\bibinfo {volume} {516}},\ \bibinfo {pages}
  {71} (\bibinfo {year} {2014})}\BibitemShut {NoStop}%
\bibitem [{\citenamefont {Odagaki}\ and\ \citenamefont
  {Tani}(1971)}]{Odagaki1971}%
  \BibitemOpen
  \bibfield  {author} {\bibinfo {author} {\bibfnamefont {T.}~\bibnamefont
  {Odagaki}}\ and\ \bibinfo {author} {\bibfnamefont {K.}~\bibnamefont {Tani}},\
  }\href {\doibase 10.1016/0375-9601(71)90278-7} {\bibfield  {journal}
  {\bibinfo  {journal} {Phys. Lett. A}\ }\textbf {\bibinfo {volume} {36}},\
  \bibinfo {pages} {399} (\bibinfo {year} {1971})}\BibitemShut {NoStop}%
\bibitem [{\citenamefont {Fedianin}\ \emph {et~al.}(2023)\citenamefont
  {Fedianin}, \citenamefont {Kalashnikova},\ and\ \citenamefont
  {Mentink}}]{fedianin2023selection}%
  \BibitemOpen
  \bibfield  {author} {\bibinfo {author} {\bibfnamefont {A.~E.}\ \bibnamefont
  {Fedianin}}, \bibinfo {author} {\bibfnamefont {A.~M.}\ \bibnamefont
  {Kalashnikova}}, \ and\ \bibinfo {author} {\bibfnamefont {J.~H.}\
  \bibnamefont {Mentink}},\ }\href {\doibase
  https://doi.org/10.1103/PhysRevB.107.144430} {\bibfield  {journal} {\bibinfo
  {journal} {Phys. Rev. B}\ }\textbf {\bibinfo {volume} {107}},\ \bibinfo
  {pages} {144430} (\bibinfo {year} {2023})}\BibitemShut {NoStop}%
\bibitem [{\citenamefont {Zhao}\ \emph {et~al.}(2004)\citenamefont {Zhao},
  \citenamefont {Bragas}, \citenamefont {Lockwood},\ and\ \citenamefont
  {Merlin}}]{zhao2004magnon}%
  \BibitemOpen
  \bibfield  {author} {\bibinfo {author} {\bibfnamefont {J.}~\bibnamefont
  {Zhao}}, \bibinfo {author} {\bibfnamefont {A.~V.}\ \bibnamefont {Bragas}},
  \bibinfo {author} {\bibfnamefont {D.~J.}\ \bibnamefont {Lockwood}}, \ and\
  \bibinfo {author} {\bibfnamefont {R.}~\bibnamefont {Merlin}},\ }\href
  {https://journals.aps.org/prl/pdf/10.1103/PhysRevLett.93.107203} {\bibfield
  {journal} {\bibinfo  {journal} {Phys. Rev. Lett.}\ }\textbf {\bibinfo
  {volume} {93}},\ \bibinfo {pages} {107203} (\bibinfo {year}
  {2004})}\BibitemShut {NoStop}%
\bibitem [{\citenamefont {Hu}\ and\ \citenamefont
  {Nori}(1996{\natexlab{a}})}]{hu1996quantum}%
  \BibitemOpen
  \bibfield  {author} {\bibinfo {author} {\bibfnamefont {X.}~\bibnamefont
  {Hu}}\ and\ \bibinfo {author} {\bibfnamefont {F.}~\bibnamefont {Nori}},\
  }\href {\doibase https://doi.org/10.1103/PhysRevB.53.2419} {\bibfield
  {journal} {\bibinfo  {journal} {Phys. Rev. B}\ }\textbf {\bibinfo {volume}
  {53}},\ \bibinfo {pages} {2419} (\bibinfo {year}
  {1996}{\natexlab{a}})}\BibitemShut {NoStop}%
\bibitem [{\citenamefont {Hu}\ and\ \citenamefont
  {Nori}(1996{\natexlab{b}})}]{hu1996squeezed}%
  \BibitemOpen
  \bibfield  {author} {\bibinfo {author} {\bibfnamefont {X.}~\bibnamefont
  {Hu}}\ and\ \bibinfo {author} {\bibfnamefont {F.}~\bibnamefont {Nori}},\
  }\href {\doibase https://doi.org/10.1103/PhysRevLett.76.2294} {\bibfield
  {journal} {\bibinfo  {journal} {Phys. Rev. Lett.}\ }\textbf {\bibinfo
  {volume} {76}},\ \bibinfo {pages} {2294} (\bibinfo {year}
  {1996}{\natexlab{b}})}\BibitemShut {NoStop}%
\bibitem [{\citenamefont {Aspelmeyer}\ \emph {et~al.}(2014)\citenamefont
  {Aspelmeyer}, \citenamefont {Kippenberg},\ and\ \citenamefont
  {Marquardt}}]{aspelmeyer2014cavity}%
  \BibitemOpen
  \bibfield  {author} {\bibinfo {author} {\bibfnamefont {M.}~\bibnamefont
  {Aspelmeyer}}, \bibinfo {author} {\bibfnamefont {T.~J.}\ \bibnamefont
  {Kippenberg}}, \ and\ \bibinfo {author} {\bibfnamefont {F.}~\bibnamefont
  {Marquardt}},\ }\href {\doibase 10.1103/RevModPhys.86.1391} {\bibfield
  {journal} {\bibinfo  {journal} {Rev. Mod. Phys.}\ }\textbf {\bibinfo {volume}
  {86}},\ \bibinfo {pages} {1391} (\bibinfo {year} {2014})}\BibitemShut
  {NoStop}%
\bibitem [{\citenamefont {Perelomov}(1975)}]{perelomov1975coherent}%
  \BibitemOpen
  \bibfield  {author} {\bibinfo {author} {\bibfnamefont {A.}~\bibnamefont
  {Perelomov}},\ }\href {\doibase 10.1007/BF01608832} {\bibfield  {journal}
  {\bibinfo  {journal} {Commun. Math. Phys.}\ }\textbf {\bibinfo {volume}
  {44}},\ \bibinfo {pages} {197} (\bibinfo {year} {1975})}\BibitemShut
  {NoStop}%
\bibitem [{\citenamefont {Kastrup}(2007)}]{Kastrup2007}%
  \BibitemOpen
  \bibfield  {author} {\bibinfo {author} {\bibfnamefont {H.}~\bibnamefont
  {Kastrup}},\ }\href {\doibase https://doi.org/10.1002/andp.200610245}
  {\bibfield  {journal} {\bibinfo  {journal} {Annalen der Physik}\ }\textbf
  {\bibinfo {volume} {16}},\ \bibinfo {pages} {439} (\bibinfo {year}
  {2007})}\BibitemShut {NoStop}%
\bibitem [{\citenamefont {Novaes}(2004)}]{novaes2004some}%
  \BibitemOpen
  \bibfield  {author} {\bibinfo {author} {\bibfnamefont {M.}~\bibnamefont
  {Novaes}},\ }\href {\doibase 10.1590/S1806-11172004000400008} {\bibfield
  {journal} {\bibinfo  {journal} {Revista Brasileira de Ensino de Fisica}\
  }\textbf {\bibinfo {volume} {26}},\ \bibinfo {pages} {351} (\bibinfo {year}
  {2004})}\BibitemShut {NoStop}%
\bibitem [{\citenamefont {Aravind}(1988)}]{Aravind:88}%
  \BibitemOpen
  \bibfield  {author} {\bibinfo {author} {\bibfnamefont {P.~K.}\ \bibnamefont
  {Aravind}},\ }\href {\doibase 10.1364/JOSAB.5.001545} {\bibfield  {journal}
  {\bibinfo  {journal} {J. Opt. Soc. Am. B}\ }\textbf {\bibinfo {volume} {5}},\
  \bibinfo {pages} {1545} (\bibinfo {year} {1988})}\BibitemShut {NoStop}%
\bibitem [{\citenamefont {Gerry}(1985)}]{gerry1985dynamics}%
  \BibitemOpen
  \bibfield  {author} {\bibinfo {author} {\bibfnamefont {C.~C.}\ \bibnamefont
  {Gerry}},\ }\href {\doibase 10.1103/PhysRevA.31.2721} {\bibfield  {journal}
  {\bibinfo  {journal} {Phys. Rev. A}\ }\textbf {\bibinfo {volume} {31}},\
  \bibinfo {pages} {2721} (\bibinfo {year} {1985})}\BibitemShut {NoStop}%
\bibitem [{\citenamefont {Mayergoyz}\ \emph {et~al.}(2009)\citenamefont
  {Mayergoyz}, \citenamefont {Bertotti},\ and\ \citenamefont
  {Serpico}}]{mayergoyz2009nonlinear}%
  \BibitemOpen
  \bibfield  {author} {\bibinfo {author} {\bibfnamefont {I.~D.}\ \bibnamefont
  {Mayergoyz}}, \bibinfo {author} {\bibfnamefont {G.}~\bibnamefont {Bertotti}},
  \ and\ \bibinfo {author} {\bibfnamefont {C.}~\bibnamefont {Serpico}},\
  }\href@noop {} {\emph {\bibinfo {title} {Nonlinear magnetization dynamics in
  nanosystems}}}\ (\bibinfo  {publisher} {Elsevier},\ \bibinfo {year}
  {2009})\BibitemShut {NoStop}%
\bibitem [{\citenamefont {Elyasi}\ \emph {et~al.}(2020)\citenamefont {Elyasi},
  \citenamefont {Blanter},\ and\ \citenamefont {Bauer}}]{elyasi2020resources}%
  \BibitemOpen
  \bibfield  {author} {\bibinfo {author} {\bibfnamefont {M.}~\bibnamefont
  {Elyasi}}, \bibinfo {author} {\bibfnamefont {Y.~M.}\ \bibnamefont {Blanter}},
  \ and\ \bibinfo {author} {\bibfnamefont {G.~E.}\ \bibnamefont {Bauer}},\
  }\href {\doibase 10.1103/PhysRevB.101.054402} {\bibfield  {journal} {\bibinfo
   {journal} {Phys. Rev. B}\ }\textbf {\bibinfo {volume} {101}},\ \bibinfo
  {pages} {054402} (\bibinfo {year} {2020})}\BibitemShut {NoStop}%
\bibitem [{\citenamefont {Rezende}\ and\ \citenamefont
  {de~Aguiar}(1990)}]{rezende1990spin}%
  \BibitemOpen
  \bibfield  {author} {\bibinfo {author} {\bibfnamefont {S.~M.}\ \bibnamefont
  {Rezende}}\ and\ \bibinfo {author} {\bibfnamefont {F.~M.}\ \bibnamefont
  {de~Aguiar}},\ }\href {\doibase 10.1109/5.56906} {\bibfield  {journal}
  {\bibinfo  {journal} {Proc. IEEE}\ }\textbf {\bibinfo {volume} {78}},\
  \bibinfo {pages} {893} (\bibinfo {year} {1990})}\BibitemShut {NoStop}%
\bibitem [{\citenamefont {Yuan}\ \emph {et~al.}(2022)\citenamefont {Yuan},
  \citenamefont {Cao}, \citenamefont {Kamra}, \citenamefont {Duine},\ and\
  \citenamefont {Yan}}]{yuan2022quantum}%
  \BibitemOpen
  \bibfield  {author} {\bibinfo {author} {\bibfnamefont {H.}~\bibnamefont
  {Yuan}}, \bibinfo {author} {\bibfnamefont {Y.}~\bibnamefont {Cao}}, \bibinfo
  {author} {\bibfnamefont {A.}~\bibnamefont {Kamra}}, \bibinfo {author}
  {\bibfnamefont {R.~A.}\ \bibnamefont {Duine}}, \ and\ \bibinfo {author}
  {\bibfnamefont {P.}~\bibnamefont {Yan}},\ }\href
  {https://www.sciencedirect.com/science/article/pii/S0370157322000977}
  {\bibfield  {journal} {\bibinfo  {journal} {Phys. Rep.}\ }\textbf {\bibinfo
  {volume} {965}},\ \bibinfo {pages} {1} (\bibinfo {year} {2022})}\BibitemShut
  {NoStop}%
\bibitem [{\citenamefont {Coldea}\ \emph {et~al.}(2001)\citenamefont {Coldea},
  \citenamefont {Hayden}, \citenamefont {Aeppli}, \citenamefont {Perring},
  \citenamefont {Frost}, \citenamefont {Mason}, \citenamefont {Cheong},\ and\
  \citenamefont {Fisk}}]{coldea2001spin}%
  \BibitemOpen
  \bibfield  {author} {\bibinfo {author} {\bibfnamefont {R.}~\bibnamefont
  {Coldea}}, \bibinfo {author} {\bibfnamefont {S.}~\bibnamefont {Hayden}},
  \bibinfo {author} {\bibfnamefont {G.}~\bibnamefont {Aeppli}}, \bibinfo
  {author} {\bibfnamefont {T.}~\bibnamefont {Perring}}, \bibinfo {author}
  {\bibfnamefont {C.}~\bibnamefont {Frost}}, \bibinfo {author} {\bibfnamefont
  {T.}~\bibnamefont {Mason}}, \bibinfo {author} {\bibfnamefont {S.-W.}\
  \bibnamefont {Cheong}}, \ and\ \bibinfo {author} {\bibfnamefont
  {Z.}~\bibnamefont {Fisk}},\ }\href {\doibase
  https://doi.org/10.1103/PhysRevLett.86.5377} {\bibfield  {journal} {\bibinfo
  {journal} {Phys. Rev. Lett.}\ }\textbf {\bibinfo {volume} {86}},\ \bibinfo
  {pages} {5377} (\bibinfo {year} {2001})}\BibitemShut {NoStop}%
\bibitem [{\citenamefont {Bonesteel}(1993)}]{bonesteel1993theory}%
  \BibitemOpen
  \bibfield  {author} {\bibinfo {author} {\bibfnamefont {N.}~\bibnamefont
  {Bonesteel}},\ }\href {\doibase https://doi.org/10.1103/PhysRevB.47.11302}
  {\bibfield  {journal} {\bibinfo  {journal} {Phys. Rev. B}\ }\textbf {\bibinfo
  {volume} {47}},\ \bibinfo {pages} {11302} (\bibinfo {year}
  {1993})}\BibitemShut {NoStop}%
\bibitem [{\citenamefont {Chen}\ \emph {et~al.}(2011)\citenamefont {Chen},
  \citenamefont {Sushkov},\ and\ \citenamefont {Tohyama}}]{chen2011angle}%
  \BibitemOpen
  \bibfield  {author} {\bibinfo {author} {\bibfnamefont {W.}~\bibnamefont
  {Chen}}, \bibinfo {author} {\bibfnamefont {O.~P.}\ \bibnamefont {Sushkov}}, \
  and\ \bibinfo {author} {\bibfnamefont {T.}~\bibnamefont {Tohyama}},\ }\href
  {\doibase https://doi.org/10.1103/PhysRevB.84.195125} {\bibfield  {journal}
  {\bibinfo  {journal} {Phys. Rev. B}\ }\textbf {\bibinfo {volume} {84}},\
  \bibinfo {pages} {195125} (\bibinfo {year} {2011})}\BibitemShut {NoStop}%
\bibitem [{\citenamefont {Govind}\ \emph {et~al.}(2001)\citenamefont {Govind},
  \citenamefont {Pratap}, \citenamefont {Ajay},\ and\ \citenamefont
  {Tripathi}}]{govind2001magnetic}%
  \BibitemOpen
  \bibfield  {author} {\bibinfo {author} {\bibnamefont {Govind}}, \bibinfo
  {author} {\bibfnamefont {A.}~\bibnamefont {Pratap}}, \bibinfo {author}
  {\bibnamefont {Ajay}}, \ and\ \bibinfo {author} {\bibfnamefont
  {R.}~\bibnamefont {Tripathi}},\ }\href {\doibase
  https://doi.org/10.1007/s100510170062} {\bibfield  {journal} {\bibinfo
  {journal} {Eur. Phys. J. B}\ }\textbf {\bibinfo {volume} {23}},\ \bibinfo
  {pages} {153} (\bibinfo {year} {2001})}\BibitemShut {NoStop}%
\bibitem [{\citenamefont {Dalla~Piazza}(2016)}]{dalla2016excitation}%
  \BibitemOpen
  \bibfield  {author} {\bibinfo {author} {\bibfnamefont {B.}~\bibnamefont
  {Dalla~Piazza}},\ }\href@noop {} {\emph {\bibinfo {title} {Excitation Spectra
  of Square Lattice Antiferromagnets: Theoretical Explanation of Experimental
  Observations}}}\ (\bibinfo  {publisher} {Springer},\ \bibinfo {year}
  {2016})\BibitemShut {NoStop}%
\bibitem [{\citenamefont {Wan}\ \emph {et~al.}(2009)\citenamefont {Wan},
  \citenamefont {Maier},\ and\ \citenamefont {Savrasov}}]{wan2009calculated}%
  \BibitemOpen
  \bibfield  {author} {\bibinfo {author} {\bibfnamefont {X.}~\bibnamefont
  {Wan}}, \bibinfo {author} {\bibfnamefont {T.~A.}\ \bibnamefont {Maier}}, \
  and\ \bibinfo {author} {\bibfnamefont {S.~Y.}\ \bibnamefont {Savrasov}},\
  }\href {\doibase 10.1103/PhysRevB.79.155114} {\bibfield  {journal} {\bibinfo
  {journal} {Phys. Rev. B}\ }\textbf {\bibinfo {volume} {79}},\ \bibinfo
  {pages} {155114} (\bibinfo {year} {2009})}\BibitemShut {NoStop}%
\bibitem [{\citenamefont {Sandvik}\ \emph {et~al.}(1998)\citenamefont
  {Sandvik}, \citenamefont {Capponi}, \citenamefont {Poilblanc},\ and\
  \citenamefont {Dagotto}}]{sandvik1998numerical}%
  \BibitemOpen
  \bibfield  {author} {\bibinfo {author} {\bibfnamefont {A.~W.}\ \bibnamefont
  {Sandvik}}, \bibinfo {author} {\bibfnamefont {S.}~\bibnamefont {Capponi}},
  \bibinfo {author} {\bibfnamefont {D.}~\bibnamefont {Poilblanc}}, \ and\
  \bibinfo {author} {\bibfnamefont {E.}~\bibnamefont {Dagotto}},\ }\href
  {\doibase 10.1103/PhysRevB.57.8478} {\bibfield  {journal} {\bibinfo
  {journal} {Phys. Rev. B}\ }\textbf {\bibinfo {volume} {57}},\ \bibinfo
  {pages} {8478} (\bibinfo {year} {1998})}\BibitemShut {NoStop}%
\bibitem [{\citenamefont {Manousakis}(1991)}]{manousakis1991spin}%
  \BibitemOpen
  \bibfield  {author} {\bibinfo {author} {\bibfnamefont {E.}~\bibnamefont
  {Manousakis}},\ }\href {\doibase https://doi.org/10.1103/RevModPhys.63.1}
  {\bibfield  {journal} {\bibinfo  {journal} {Rev. Mod. Phys.}\ }\textbf
  {\bibinfo {volume} {63}},\ \bibinfo {pages} {1} (\bibinfo {year}
  {1991})}\BibitemShut {NoStop}%
\bibitem [{\citenamefont {Chakravarty}\ \emph {et~al.}(1989)\citenamefont
  {Chakravarty}, \citenamefont {Halperin},\ and\ \citenamefont
  {Nelson}}]{chakravarty1989two}%
  \BibitemOpen
  \bibfield  {author} {\bibinfo {author} {\bibfnamefont {S.}~\bibnamefont
  {Chakravarty}}, \bibinfo {author} {\bibfnamefont {B.~I.}\ \bibnamefont
  {Halperin}}, \ and\ \bibinfo {author} {\bibfnamefont {D.~R.}\ \bibnamefont
  {Nelson}},\ }\href {\doibase https://doi.org/10.1103/PhysRevB.39.2344}
  {\bibfield  {journal} {\bibinfo  {journal} {Phys. Rev. B}\ }\textbf {\bibinfo
  {volume} {39}},\ \bibinfo {pages} {2344} (\bibinfo {year}
  {1989})}\BibitemShut {NoStop}%
\bibitem [{\citenamefont {Holstein}\ and\ \citenamefont
  {Primakoff}(1940)}]{holstein1940field}%
  \BibitemOpen
  \bibfield  {author} {\bibinfo {author} {\bibfnamefont {T.}~\bibnamefont
  {Holstein}}\ and\ \bibinfo {author} {\bibfnamefont {H.}~\bibnamefont
  {Primakoff}},\ }\href {\doibase https://doi.org/10.1103/PhysRev.58.1098}
  {\bibfield  {journal} {\bibinfo  {journal} {Phys. Rev.}\ }\textbf {\bibinfo
  {volume} {58}},\ \bibinfo {pages} {1098} (\bibinfo {year}
  {1940})}\BibitemShut {NoStop}%
\bibitem [{\citenamefont {Cottam}(1975)}]{cottam1975temperature}%
  \BibitemOpen
  \bibfield  {author} {\bibinfo {author} {\bibfnamefont {M.}~\bibnamefont
  {Cottam}},\ }\href {\doibase 10.1088/0022-3719/8/12/019} {\bibfield
  {journal} {\bibinfo  {journal} {J. Phys. C: Solid State Phys.}\ }\textbf
  {\bibinfo {volume} {8}},\ \bibinfo {pages} {1933} (\bibinfo {year}
  {1975})}\BibitemShut {NoStop}%
\bibitem [{\citenamefont {Cottam}\ and\ \citenamefont
  {Lockwood}(1986)}]{cottam1986light}%
  \BibitemOpen
  \bibfield  {author} {\bibinfo {author} {\bibfnamefont {M.~G.}\ \bibnamefont
  {Cottam}}\ and\ \bibinfo {author} {\bibfnamefont {D.~J.}\ \bibnamefont
  {Lockwood}},\ }\href@noop {} {\emph {\bibinfo {title} {Light scattering in
  magnetic solids}}}\ (\bibinfo  {publisher} {Wiley},\ \bibinfo {year}
  {1986})\BibitemShut {NoStop}%
\bibitem [{\citenamefont {Moriya}(1967)}]{moriya1967theory}%
  \BibitemOpen
  \bibfield  {author} {\bibinfo {author} {\bibfnamefont {T.}~\bibnamefont
  {Moriya}},\ }\href {\doibase 10.1143/JPSJ.23.490} {\bibfield  {journal}
  {\bibinfo  {journal} {J. Phys. Soc. Jpn.}\ }\textbf {\bibinfo {volume}
  {23}},\ \bibinfo {pages} {490} (\bibinfo {year} {1967})}\BibitemShut
  {NoStop}%
\bibitem [{\citenamefont {Bostr{\"o}m}\ \emph {et~al.}(2023)\citenamefont
  {Bostr{\"o}m}, \citenamefont {Parvini}, \citenamefont {McIver}, \citenamefont
  {Rubio}, \citenamefont {Viola~Kusminskiy},\ and\ \citenamefont
  {Sentef}}]{bostrom2023direct}%
  \BibitemOpen
  \bibfield  {author} {\bibinfo {author} {\bibfnamefont {E.~V.}\ \bibnamefont
  {Bostr{\"o}m}}, \bibinfo {author} {\bibfnamefont {T.~S.}\ \bibnamefont
  {Parvini}}, \bibinfo {author} {\bibfnamefont {J.~W.}\ \bibnamefont {McIver}},
  \bibinfo {author} {\bibfnamefont {A.}~\bibnamefont {Rubio}}, \bibinfo
  {author} {\bibfnamefont {S.}~\bibnamefont {Viola~Kusminskiy}}, \ and\
  \bibinfo {author} {\bibfnamefont {M.~A.}\ \bibnamefont {Sentef}},\ }\href
  {\doibase 10.1103/PhysRevLett.130.026701} {\bibfield  {journal} {\bibinfo
  {journal} {Phys. Rev. Lett.}\ }\textbf {\bibinfo {volume} {130}},\ \bibinfo
  {pages} {026701} (\bibinfo {year} {2023})}\BibitemShut {NoStop}%
\bibitem [{\citenamefont {Lockwood}\ and\ \citenamefont
  {Cottam}(2012)}]{lockwood2012magnetooptic}%
  \BibitemOpen
  \bibfield  {author} {\bibinfo {author} {\bibfnamefont {D.}~\bibnamefont
  {Lockwood}}\ and\ \bibinfo {author} {\bibfnamefont {M.}~\bibnamefont
  {Cottam}},\ }\href {\doibase https://doi.org/10.1063/1.4733682} {\bibfield
  {journal} {\bibinfo  {journal} {Low Temp. Phys.}\ }\textbf {\bibinfo {volume}
  {38}},\ \bibinfo {pages} {549} (\bibinfo {year} {2012})}\BibitemShut
  {NoStop}%
\bibitem [{\citenamefont {Odagaki}(1973{\natexlab{a}})}]{odagaki1973_part1}%
  \BibitemOpen
  \bibfield  {author} {\bibinfo {author} {\bibfnamefont {T.}~\bibnamefont
  {Odagaki}},\ }\href {\doibase 10.1143/JPSJ.35.40} {\bibfield  {journal}
  {\bibinfo  {journal} {J. Phys. Soc. Jpn.}\ }\textbf {\bibinfo {volume}
  {35}},\ \bibinfo {pages} {40} (\bibinfo {year}
  {1973}{\natexlab{a}})}\BibitemShut {NoStop}%
\bibitem [{\citenamefont {Odagaki}(1973{\natexlab{b}})}]{odagaki1973_part2}%
  \BibitemOpen
  \bibfield  {author} {\bibinfo {author} {\bibfnamefont {T.}~\bibnamefont
  {Odagaki}},\ }\href {\doibase 10.1143/JPSJ.35.1343} {\bibfield  {journal}
  {\bibinfo  {journal} {J. Phys. Soc. Jpn.}\ }\textbf {\bibinfo {volume}
  {35}},\ \bibinfo {pages} {1343} (\bibinfo {year}
  {1973}{\natexlab{b}})}\BibitemShut {NoStop}%
\bibitem [{\citenamefont {Formisano}\ \emph {et~al.}(2024)\citenamefont
  {Formisano}, \citenamefont {Gareev}, \citenamefont {Khusyainov},
  \citenamefont {Fedianin}, \citenamefont {Dubrovin}, \citenamefont {Syrnikov},
  \citenamefont {Afanasiev}, \citenamefont {Pisarev}, \citenamefont
  {Kalashnikova}, \citenamefont {Mentink} \emph
  {et~al.}}]{formisano2024coherent}%
  \BibitemOpen
  \bibfield  {author} {\bibinfo {author} {\bibfnamefont {F.}~\bibnamefont
  {Formisano}}, \bibinfo {author} {\bibfnamefont {T.}~\bibnamefont {Gareev}},
  \bibinfo {author} {\bibfnamefont {D.}~\bibnamefont {Khusyainov}}, \bibinfo
  {author} {\bibfnamefont {A.}~\bibnamefont {Fedianin}}, \bibinfo {author}
  {\bibfnamefont {R.}~\bibnamefont {Dubrovin}}, \bibinfo {author}
  {\bibfnamefont {P.}~\bibnamefont {Syrnikov}}, \bibinfo {author}
  {\bibfnamefont {D.}~\bibnamefont {Afanasiev}}, \bibinfo {author}
  {\bibfnamefont {R.}~\bibnamefont {Pisarev}}, \bibinfo {author} {\bibfnamefont
  {A.}~\bibnamefont {Kalashnikova}}, \bibinfo {author} {\bibfnamefont
  {J.}~\bibnamefont {Mentink}},  \emph {et~al.},\ }\href {\doibase
  10.1063/5.0180888} {\bibfield  {journal} {\bibinfo  {journal} {APL Mater.}\
  }\textbf {\bibinfo {volume} {12}} (\bibinfo {year} {2024}),\
  10.1063/5.0180888}\BibitemShut {NoStop}%
\bibitem [{\citenamefont {Dimmock}\ and\ \citenamefont
  {Wheeler}(1962)}]{dimmock1962symmetry}%
  \BibitemOpen
  \bibfield  {author} {\bibinfo {author} {\bibfnamefont {J.~O.}\ \bibnamefont
  {Dimmock}}\ and\ \bibinfo {author} {\bibfnamefont {R.}~\bibnamefont
  {Wheeler}},\ }\href {\doibase https://doi.org/10.1103/PhysRev.127.391}
  {\bibfield  {journal} {\bibinfo  {journal} {Phys. Rev.}\ }\textbf {\bibinfo
  {volume} {127}},\ \bibinfo {pages} {391} (\bibinfo {year}
  {1962})}\BibitemShut {NoStop}%
\bibitem [{\citenamefont {Sugano}\ and\ \citenamefont
  {Kojima}(2013)}]{sugano2013magneto}%
  \BibitemOpen
  \bibfield  {author} {\bibinfo {author} {\bibfnamefont {S.}~\bibnamefont
  {Sugano}}\ and\ \bibinfo {author} {\bibfnamefont {N.}~\bibnamefont
  {Kojima}},\ }\href@noop {} {\emph {\bibinfo {title} {Magneto-optics}}},\
  Vol.\ \bibinfo {volume} {128}\ (\bibinfo  {publisher} {Springer Science \&
  Business Media},\ \bibinfo {year} {2013})\BibitemShut {NoStop}%
\bibitem [{\citenamefont {Fleury}\ and\ \citenamefont
  {Guggenheim}(1970)}]{fleury1970magnon}%
  \BibitemOpen
  \bibfield  {author} {\bibinfo {author} {\bibfnamefont {P.}~\bibnamefont
  {Fleury}}\ and\ \bibinfo {author} {\bibfnamefont {H.}~\bibnamefont
  {Guggenheim}},\ }\href
  {https://journals.aps.org/prl/pdf/10.1103/PhysRevLett.24.1346} {\bibfield
  {journal} {\bibinfo  {journal} {Phys. Rev. Lett.}\ }\textbf {\bibinfo
  {volume} {24}},\ \bibinfo {pages} {1346} (\bibinfo {year}
  {1970})}\BibitemShut {NoStop}%
\bibitem [{\citenamefont {Amer}\ \emph {et~al.}(1975)\citenamefont {Amer},
  \citenamefont {Chiang},\ and\ \citenamefont {Shen}}]{amer1975two}%
  \BibitemOpen
  \bibfield  {author} {\bibinfo {author} {\bibfnamefont {N.~M.}\ \bibnamefont
  {Amer}}, \bibinfo {author} {\bibfnamefont {T.-c.}\ \bibnamefont {Chiang}}, \
  and\ \bibinfo {author} {\bibfnamefont {Y.}~\bibnamefont {Shen}},\ }\href
  {\doibase 10.1103/PhysRevLett.34.1454} {\bibfield  {journal} {\bibinfo
  {journal} {Phys. Rev. Lett.}\ }\textbf {\bibinfo {volume} {34}},\ \bibinfo
  {pages} {1454} (\bibinfo {year} {1975})}\BibitemShut {NoStop}%
\bibitem [{\citenamefont {Weber}\ and\ \citenamefont
  {Merlin}(2000)}]{weber2000raman}%
  \BibitemOpen
  \bibfield  {author} {\bibinfo {author} {\bibfnamefont {W.~H.}\ \bibnamefont
  {Weber}}\ and\ \bibinfo {author} {\bibfnamefont {R.}~\bibnamefont {Merlin}},\
  }\href@noop {} {\emph {\bibinfo {title} {Raman scattering in materials
  science}}},\ Vol.~\bibinfo {volume} {42}\ (\bibinfo  {publisher} {Springer
  Science \& Business Media},\ \bibinfo {year} {2000})\BibitemShut {NoStop}%
\bibitem [{\citenamefont {Devereaux}\ and\ \citenamefont
  {Hackl}(2007)}]{devereaux2007inelastic}%
  \BibitemOpen
  \bibfield  {author} {\bibinfo {author} {\bibfnamefont {T.~P.}\ \bibnamefont
  {Devereaux}}\ and\ \bibinfo {author} {\bibfnamefont {R.}~\bibnamefont
  {Hackl}},\ }\href {\doibase 10.1103/RevModPhys.79.175} {\bibfield  {journal}
  {\bibinfo  {journal} {Rev. Mod. Phys.}\ }\textbf {\bibinfo {volume} {79}},\
  \bibinfo {pages} {175} (\bibinfo {year} {2007})}\BibitemShut {NoStop}%
\bibitem [{\citenamefont {Poppinger}(1977)}]{poppinger1977temperature}%
  \BibitemOpen
  \bibfield  {author} {\bibinfo {author} {\bibfnamefont {M.}~\bibnamefont
  {Poppinger}},\ }\href {\doibase 10.1007/BF01315505} {\bibfield  {journal}
  {\bibinfo  {journal} {Zeitschrift f{\"u}r Physik B Condensed Matter}\
  }\textbf {\bibinfo {volume} {27}},\ \bibinfo {pages} {61} (\bibinfo {year}
  {1977})}\BibitemShut {NoStop}%
\bibitem [{\citenamefont {Davies}\ \emph {et~al.}(1971)\citenamefont {Davies},
  \citenamefont {Chinn},\ and\ \citenamefont {Zeiger}}]{davies1971spin}%
  \BibitemOpen
  \bibfield  {author} {\bibinfo {author} {\bibfnamefont {R.}~\bibnamefont
  {Davies}}, \bibinfo {author} {\bibfnamefont {S.}~\bibnamefont {Chinn}}, \
  and\ \bibinfo {author} {\bibfnamefont {H.}~\bibnamefont {Zeiger}},\ }\href
  {\doibase 10.1103/PhysRevB.4.992} {\bibfield  {journal} {\bibinfo  {journal}
  {Phys. Rev. B}\ }\textbf {\bibinfo {volume} {4}},\ \bibinfo {pages} {992}
  (\bibinfo {year} {1971})}\BibitemShut {NoStop}%
\bibitem [{\citenamefont {Vernay}\ \emph {et~al.}(2007)\citenamefont {Vernay},
  \citenamefont {Gingras},\ and\ \citenamefont
  {Devereaux}}]{vernay2007momentum}%
  \BibitemOpen
  \bibfield  {author} {\bibinfo {author} {\bibfnamefont {F.}~\bibnamefont
  {Vernay}}, \bibinfo {author} {\bibfnamefont {M.}~\bibnamefont {Gingras}}, \
  and\ \bibinfo {author} {\bibfnamefont {T.}~\bibnamefont {Devereaux}},\ }\href
  {\doibase 10.1103/PhysRevB.75.020403} {\bibfield  {journal} {\bibinfo
  {journal} {Phys. Rev. B}\ }\textbf {\bibinfo {volume} {75}},\ \bibinfo
  {pages} {020403} (\bibinfo {year} {2007})}\BibitemShut {NoStop}%
\bibitem [{\citenamefont {Chubukov}\ and\ \citenamefont
  {Frenkel}(1995)}]{chubukov1995resonant}%
  \BibitemOpen
  \bibfield  {author} {\bibinfo {author} {\bibfnamefont {A.~V.}\ \bibnamefont
  {Chubukov}}\ and\ \bibinfo {author} {\bibfnamefont {D.~M.}\ \bibnamefont
  {Frenkel}},\ }\href {\doibase 10.1103/PhysRevB.52.9760} {\bibfield  {journal}
  {\bibinfo  {journal} {Phys. Rev. B}\ }\textbf {\bibinfo {volume} {52}},\
  \bibinfo {pages} {9760} (\bibinfo {year} {1995})}\BibitemShut {NoStop}%
\bibitem [{\citenamefont {Shastry}\ and\ \citenamefont
  {Shraiman}(1990)}]{shastry1990theory}%
  \BibitemOpen
  \bibfield  {author} {\bibinfo {author} {\bibfnamefont {B.~S.}\ \bibnamefont
  {Shastry}}\ and\ \bibinfo {author} {\bibfnamefont {B.~I.}\ \bibnamefont
  {Shraiman}},\ }\href {\doibase 10.1103/PhysRevLett.65.1068} {\bibfield
  {journal} {\bibinfo  {journal} {Phys. Rev. Lett.}\ }\textbf {\bibinfo
  {volume} {65}},\ \bibinfo {pages} {1068} (\bibinfo {year}
  {1990})}\BibitemShut {NoStop}%
\bibitem [{\citenamefont {Zhao}\ \emph {et~al.}(2006)\citenamefont {Zhao},
  \citenamefont {Bragas}, \citenamefont {Merlin},\ and\ \citenamefont
  {Lockwood}}]{PhysRevB.73.184434}%
  \BibitemOpen
  \bibfield  {author} {\bibinfo {author} {\bibfnamefont {J.}~\bibnamefont
  {Zhao}}, \bibinfo {author} {\bibfnamefont {A.~V.}\ \bibnamefont {Bragas}},
  \bibinfo {author} {\bibfnamefont {R.}~\bibnamefont {Merlin}}, \ and\ \bibinfo
  {author} {\bibfnamefont {D.~J.}\ \bibnamefont {Lockwood}},\ }\href {\doibase
  10.1103/PhysRevB.73.184434} {\bibfield  {journal} {\bibinfo  {journal} {Phys.
  Rev. B}\ }\textbf {\bibinfo {volume} {73}},\ \bibinfo {pages} {184434}
  (\bibinfo {year} {2006})}\BibitemShut {NoStop}%
\bibitem [{\citenamefont {Sheshadri}\ \emph {et~al.}(2023)\citenamefont
  {Sheshadri}, \citenamefont {Malterre}, \citenamefont {Fujimori},\ and\
  \citenamefont {Chainani}}]{sheshadri2023connecting}%
  \BibitemOpen
  \bibfield  {author} {\bibinfo {author} {\bibfnamefont {K.}~\bibnamefont
  {Sheshadri}}, \bibinfo {author} {\bibfnamefont {D.}~\bibnamefont {Malterre}},
  \bibinfo {author} {\bibfnamefont {A.}~\bibnamefont {Fujimori}}, \ and\
  \bibinfo {author} {\bibfnamefont {A.}~\bibnamefont {Chainani}},\ }\href
  {\doibase https://doi.org/10.1103/PhysRevB.107.085125} {\bibfield  {journal}
  {\bibinfo  {journal} {Phys. Rev. B}\ }\textbf {\bibinfo {volume} {107}},\
  \bibinfo {pages} {085125} (\bibinfo {year} {2023})}\BibitemShut {NoStop}%
\bibitem [{\citenamefont {Weichselbaumer}\ \emph {et~al.}(2019)\citenamefont
  {Weichselbaumer}, \citenamefont {Natzkin}, \citenamefont {Zollitsch},
  \citenamefont {Weiler}, \citenamefont {Gross},\ and\ \citenamefont
  {Huebl}}]{weichselbaumer}%
  \BibitemOpen
  \bibfield  {author} {\bibinfo {author} {\bibfnamefont {S.}~\bibnamefont
  {Weichselbaumer}}, \bibinfo {author} {\bibfnamefont {P.}~\bibnamefont
  {Natzkin}}, \bibinfo {author} {\bibfnamefont {C.~W.}\ \bibnamefont
  {Zollitsch}}, \bibinfo {author} {\bibfnamefont {M.}~\bibnamefont {Weiler}},
  \bibinfo {author} {\bibfnamefont {R.}~\bibnamefont {Gross}}, \ and\ \bibinfo
  {author} {\bibfnamefont {H.}~\bibnamefont {Huebl}},\ }\href@noop {}
  {\bibfield  {journal} {\bibinfo  {journal} {Phys. Rev. Appl.}\ }\textbf
  {\bibinfo {volume} {12}},\ \bibinfo {pages} {024021} (\bibinfo {year}
  {2019})}\BibitemShut {NoStop}%
\bibitem [{\citenamefont {Graf}\ \emph {et~al.}(2021)\citenamefont {Graf},
  \citenamefont {Sharma}, \citenamefont {Huebl},\ and\ \citenamefont
  {Viola~Kusminskiy}}]{graf2021design}%
  \BibitemOpen
  \bibfield  {author} {\bibinfo {author} {\bibfnamefont {J.}~\bibnamefont
  {Graf}}, \bibinfo {author} {\bibfnamefont {S.}~\bibnamefont {Sharma}},
  \bibinfo {author} {\bibfnamefont {H.}~\bibnamefont {Huebl}}, \ and\ \bibinfo
  {author} {\bibfnamefont {S.}~\bibnamefont {Viola~Kusminskiy}},\ }\href@noop
  {} {\bibfield  {journal} {\bibinfo  {journal} {Phys. Rev. Res.}\ }\textbf
  {\bibinfo {volume} {3}},\ \bibinfo {pages} {013277} (\bibinfo {year}
  {2021})}\BibitemShut {NoStop}%
\bibitem [{\citenamefont {Gerry}\ \emph {et~al.}(1991)\citenamefont {Gerry},
  \citenamefont {Grobe},\ and\ \citenamefont {Vrscay}}]{gerry1991squeezed}%
  \BibitemOpen
  \bibfield  {author} {\bibinfo {author} {\bibfnamefont {C.~C.}\ \bibnamefont
  {Gerry}}, \bibinfo {author} {\bibfnamefont {R.}~\bibnamefont {Grobe}}, \ and\
  \bibinfo {author} {\bibfnamefont {E.~R.}\ \bibnamefont {Vrscay}},\ }\href
  {\doibase https://doi.org/10.1103/PhysRevA.43.361} {\bibfield  {journal}
  {\bibinfo  {journal} {Phys. Rev. A}\ }\textbf {\bibinfo {volume} {43}},\
  \bibinfo {pages} {361} (\bibinfo {year} {1991})}\BibitemShut {NoStop}%
\bibitem [{\citenamefont {Robert}\ and\ \citenamefont
  {Combescure}(2021)}]{robert2021coherent}%
  \BibitemOpen
  \bibfield  {author} {\bibinfo {author} {\bibfnamefont {D.}~\bibnamefont
  {Robert}}\ and\ \bibinfo {author} {\bibfnamefont {M.}~\bibnamefont
  {Combescure}},\ }\href@noop {} {\emph {\bibinfo {title} {Coherent states and
  applications in mathematical physics}}}\ (\bibinfo  {publisher} {Springer},\
  \bibinfo {year} {2021})\BibitemShut {NoStop}%
\bibitem [{\citenamefont {Perelomov}(1972)}]{perelomov1972coherent}%
  \BibitemOpen
  \bibfield  {author} {\bibinfo {author} {\bibfnamefont {A.~M.}\ \bibnamefont
  {Perelomov}},\ }\href {\doibase 10.1007/BF01645091} {\bibfield  {journal}
  {\bibinfo  {journal} {Commun. Math. Phys.}\ }\textbf {\bibinfo {volume}
  {26}},\ \bibinfo {pages} {222} (\bibinfo {year} {1972})}\BibitemShut
  {NoStop}%
\bibitem [{\citenamefont {Gerry}\ and\ \citenamefont
  {Kiefer}(1991)}]{gerry1991classical}%
  \BibitemOpen
  \bibfield  {author} {\bibinfo {author} {\bibfnamefont {C.}~\bibnamefont
  {Gerry}}\ and\ \bibinfo {author} {\bibfnamefont {J.}~\bibnamefont {Kiefer}},\
  }\href {\doibase 10.1088/0305-4470/24/15/020} {\bibfield  {journal} {\bibinfo
   {journal} {J. Phys. A}\ }\textbf {\bibinfo {volume} {24}},\ \bibinfo {pages}
  {3513} (\bibinfo {year} {1991})}\BibitemShut {NoStop}%
\bibitem [{\citenamefont {Gardiner}\ and\ \citenamefont
  {Collett}(1985)}]{GardColl}%
  \BibitemOpen
  \bibfield  {author} {\bibinfo {author} {\bibfnamefont {C.~W.}\ \bibnamefont
  {Gardiner}}\ and\ \bibinfo {author} {\bibfnamefont {M.~J.}\ \bibnamefont
  {Collett}},\ }\href {\doibase 10.1103/PhysRevA.31.3761} {\bibfield  {journal}
  {\bibinfo  {journal} {Phys. Rev. A}\ }\textbf {\bibinfo {volume} {31}},\
  \bibinfo {pages} {3761} (\bibinfo {year} {1985})}\BibitemShut {NoStop}%
\bibitem [{\citenamefont {Rezende}(2020)}]{Rezende2020}%
  \BibitemOpen
  \bibfield  {author} {\bibinfo {author} {\bibfnamefont {S.~M.}\ \bibnamefont
  {Rezende}},\ }\enquote {\bibinfo {title} {Magnons in antiferromagnets},}\ in\
  \href {\doibase 10.1007/978-3-030-41317-0_5} {\emph {\bibinfo {booktitle}
  {Fundamentals of Magnonics}}}\ (\bibinfo  {publisher} {Springer International
  Publishing},\ \bibinfo {address} {Cham},\ \bibinfo {year} {2020})\ pp.\
  \bibinfo {pages} {187--222}\BibitemShut {NoStop}%
\bibitem [{\citenamefont {Breuer}\ and\ \citenamefont
  {Petruccione}(2002)}]{BP_OQS}%
  \BibitemOpen
  \bibfield  {author} {\bibinfo {author} {\bibfnamefont {H.~P.}\ \bibnamefont
  {Breuer}}\ and\ \bibinfo {author} {\bibfnamefont {F.}~\bibnamefont
  {Petruccione}},\ }\href@noop {} {\emph {\bibinfo {title} {The theory of open
  quantum systems}}}\ (\bibinfo  {publisher} {Oxford University Press},\
  \bibinfo {year} {2002})\BibitemShut {NoStop}%
\bibitem [{\citenamefont {\'{A}ngel Rivas}\ and\ \citenamefont
  {Huelga}(2012)}]{RH_OQS}%
  \BibitemOpen
  \bibfield  {author} {\bibinfo {author} {\bibnamefont {\'{A}ngel Rivas}}\ and\
  \bibinfo {author} {\bibfnamefont {S.~F.}\ \bibnamefont {Huelga}},\ }\href
  {\doibase https://doi.org/10.1007/978-3-642-23354-8} {\emph {\bibinfo {title}
  {Open Quantum Systems: An Introduction}}}\ (\bibinfo  {publisher} {Springer
  Berlin, Heidelberg},\ \bibinfo {year} {2012})\BibitemShut {NoStop}%
\bibitem [{\citenamefont {Gilbert}(2004)}]{Gilbert2004}%
  \BibitemOpen
  \bibfield  {author} {\bibinfo {author} {\bibfnamefont {T.}~\bibnamefont
  {Gilbert}},\ }\href {\doibase 10.1109/TMAG.2004.836740} {\bibfield  {journal}
  {\bibinfo  {journal} {IEEE Trans. Magn.}\ }\textbf {\bibinfo {volume} {40}},\
  \bibinfo {pages} {3443} (\bibinfo {year} {2004})}\BibitemShut {NoStop}%
\bibitem [{\citenamefont {POPPINGER}(1977)}]{poppinger1977temperatureII}%
  \BibitemOpen
  \bibfield  {author} {\bibinfo {author} {\bibfnamefont {M.}~\bibnamefont
  {POPPINGER}},\ }\href {\doibase https://doi.org/10.1007/BF01315506}
  {\bibfield  {journal} {\bibinfo  {journal} {Zeitschrift für Physik B
  Condensed Matter}\ } (\bibinfo {year} {1977}),\
  https://doi.org/10.1007/BF01315506}\BibitemShut {NoStop}%
\bibitem [{\citenamefont {Ohlmann}\ and\ \citenamefont
  {Tinkham}(1961)}]{ohlmann1961antiferromagnetic}%
  \BibitemOpen
  \bibfield  {author} {\bibinfo {author} {\bibfnamefont {R.}~\bibnamefont
  {Ohlmann}}\ and\ \bibinfo {author} {\bibfnamefont {M.}~\bibnamefont
  {Tinkham}},\ }\href {\doibase 10.1103/PhysRev.123.425} {\bibfield  {journal}
  {\bibinfo  {journal} {Phys. Rev.}\ }\textbf {\bibinfo {volume} {123}},\
  \bibinfo {pages} {425} (\bibinfo {year} {1961})}\BibitemShut {NoStop}%
\bibitem [{\citenamefont {Barak}\ \emph {et~al.}(1980)\citenamefont {Barak},
  \citenamefont {Rezende}, \citenamefont {King},\ and\ \citenamefont
  {Jaccarino}}]{barak1980parallel}%
  \BibitemOpen
  \bibfield  {author} {\bibinfo {author} {\bibfnamefont {J.}~\bibnamefont
  {Barak}}, \bibinfo {author} {\bibfnamefont {S.}~\bibnamefont {Rezende}},
  \bibinfo {author} {\bibfnamefont {A.}~\bibnamefont {King}}, \ and\ \bibinfo
  {author} {\bibfnamefont {V.}~\bibnamefont {Jaccarino}},\ }\href {\doibase
  https://doi.org/10.1103/PhysRevB.21.3015} {\bibfield  {journal} {\bibinfo
  {journal} {Phys. Rev. B}\ }\textbf {\bibinfo {volume} {21}},\ \bibinfo
  {pages} {3015} (\bibinfo {year} {1980})}\BibitemShut {NoStop}%
\bibitem [{\citenamefont {Kotthaus}\ and\ \citenamefont
  {Jaccarino}(1972)}]{kotthaus1972}%
  \BibitemOpen
  \bibfield  {author} {\bibinfo {author} {\bibfnamefont {J.}~\bibnamefont
  {Kotthaus}}\ and\ \bibinfo {author} {\bibfnamefont {V.}~\bibnamefont
  {Jaccarino}},\ }\href {\doibase https://doi.org/10.1103/PhysRevLett.28.1649}
  {\bibfield  {journal} {\bibinfo  {journal} {Phys. Rev. Lett.}\ }\textbf
  {\bibinfo {volume} {28}},\ \bibinfo {pages} {1649} (\bibinfo {year}
  {1972})}\BibitemShut {NoStop}%
\bibitem [{\citenamefont {Vaidya}\ \emph {et~al.}(2020)\citenamefont {Vaidya},
  \citenamefont {Morley}, \citenamefont {van Tol}, \citenamefont {Liu},
  \citenamefont {Cheng}, \citenamefont {Brataas}, \citenamefont {Lederman},\
  and\ \citenamefont {Del~Barco}}]{vaidya2020subterahertz}%
  \BibitemOpen
  \bibfield  {author} {\bibinfo {author} {\bibfnamefont {P.}~\bibnamefont
  {Vaidya}}, \bibinfo {author} {\bibfnamefont {S.~A.}\ \bibnamefont {Morley}},
  \bibinfo {author} {\bibfnamefont {J.}~\bibnamefont {van Tol}}, \bibinfo
  {author} {\bibfnamefont {Y.}~\bibnamefont {Liu}}, \bibinfo {author}
  {\bibfnamefont {R.}~\bibnamefont {Cheng}}, \bibinfo {author} {\bibfnamefont
  {A.}~\bibnamefont {Brataas}}, \bibinfo {author} {\bibfnamefont
  {D.}~\bibnamefont {Lederman}}, \ and\ \bibinfo {author} {\bibfnamefont
  {E.}~\bibnamefont {Del~Barco}},\ }\href
  {https://www.science.org/doi/full/10.1126/science.aaz4247} {\bibfield
  {journal} {\bibinfo  {journal} {Science}\ }\textbf {\bibinfo {volume}
  {368}},\ \bibinfo {pages} {160} (\bibinfo {year} {2020})}\BibitemShut
  {NoStop}%
\bibitem [{\citenamefont {Peng}\ and\ \citenamefont {Hao}(2004)}]{PENG2004306}%
  \BibitemOpen
  \bibfield  {author} {\bibinfo {author} {\bibfnamefont {F.}~\bibnamefont
  {Peng}}\ and\ \bibinfo {author} {\bibfnamefont {B.}~\bibnamefont {Hao}},\
  }\href {\doibase https://doi.org/10.1016/j.physb.2003.12.014} {\bibfield
  {journal} {\bibinfo  {journal} {Physica B: Condensed Matter}\ }\textbf
  {\bibinfo {volume} {348}},\ \bibinfo {pages} {306} (\bibinfo {year}
  {2004})}\BibitemShut {NoStop}%
\bibitem [{\citenamefont {Jahn}(1973)}]{jahn1973linear}%
  \BibitemOpen
  \bibfield  {author} {\bibinfo {author} {\bibfnamefont {I.}~\bibnamefont
  {Jahn}},\ }\href
  {https://onlinelibrary.wiley.com/doi/abs/10.1002/pssb.2220570225} {\bibfield
  {journal} {\bibinfo  {journal} {physica status solidi (b)}\ }\textbf
  {\bibinfo {volume} {57}},\ \bibinfo {pages} {681} (\bibinfo {year}
  {1973})}\BibitemShut {NoStop}%
\bibitem [{\citenamefont {Jezek}\ and\ \citenamefont
  {Hernandez}(1990)}]{jezek1990nonlinear}%
  \BibitemOpen
  \bibfield  {author} {\bibinfo {author} {\bibfnamefont {D.}~\bibnamefont
  {Jezek}}\ and\ \bibinfo {author} {\bibfnamefont {E.}~\bibnamefont
  {Hernandez}},\ }\href {\doibase https://doi.org/10.1103/PhysRevA.42.96}
  {\bibfield  {journal} {\bibinfo  {journal} {Phys. Rev. A}\ }\textbf {\bibinfo
  {volume} {42}},\ \bibinfo {pages} {96} (\bibinfo {year} {1990})}\BibitemShut
  {NoStop}%
\bibitem [{\citenamefont {Chen}\ \emph {et~al.}(2000)\citenamefont {Chen},
  \citenamefont {Moiola},\ and\ \citenamefont {Wang}}]{chen2000bifurcation}%
  \BibitemOpen
  \bibfield  {author} {\bibinfo {author} {\bibfnamefont {G.}~\bibnamefont
  {Chen}}, \bibinfo {author} {\bibfnamefont {J.~L.}\ \bibnamefont {Moiola}}, \
  and\ \bibinfo {author} {\bibfnamefont {H.~O.}\ \bibnamefont {Wang}},\ }\href
  {\doibase https://doi.org/10.1142/S0218127400000360} {\bibfield  {journal}
  {\bibinfo  {journal} {Int. J. Bifurc. Chaos Appl. Sci. Eng.}\ }\textbf
  {\bibinfo {volume} {10}},\ \bibinfo {pages} {511} (\bibinfo {year}
  {2000})}\BibitemShut {NoStop}%
\bibitem [{\citenamefont {Collado}\ \emph {et~al.}(2018)\citenamefont
  {Collado}, \citenamefont {Lorenzana}, \citenamefont {Usaj},\ and\
  \citenamefont {Balseiro}}]{collado2018population}%
  \BibitemOpen
  \bibfield  {author} {\bibinfo {author} {\bibfnamefont {H.~O.}\ \bibnamefont
  {Collado}}, \bibinfo {author} {\bibfnamefont {J.}~\bibnamefont {Lorenzana}},
  \bibinfo {author} {\bibfnamefont {G.}~\bibnamefont {Usaj}}, \ and\ \bibinfo
  {author} {\bibfnamefont {C.~A.}\ \bibnamefont {Balseiro}},\ }\href {\doibase
  https://doi.org/10.1103/PhysRevB.98.214519} {\bibfield  {journal} {\bibinfo
  {journal} {Phys. Rev. B}\ }\textbf {\bibinfo {volume} {98}},\ \bibinfo
  {pages} {214519} (\bibinfo {year} {2018})}\BibitemShut {NoStop}%
\bibitem [{\citenamefont {Zhang}\ \emph
  {et~al.}(2021{\natexlab{b}})\citenamefont {Zhang}, \citenamefont {Zhang},
  \citenamefont {Hu}, \citenamefont {Niu},\ and\ \citenamefont
  {Liu}}]{zhang2021nonequilibrium}%
  \BibitemOpen
  \bibfield  {author} {\bibinfo {author} {\bibfnamefont {L.}~\bibnamefont
  {Zhang}}, \bibinfo {author} {\bibfnamefont {L.}~\bibnamefont {Zhang}},
  \bibinfo {author} {\bibfnamefont {Y.}~\bibnamefont {Hu}}, \bibinfo {author}
  {\bibfnamefont {S.}~\bibnamefont {Niu}}, \ and\ \bibinfo {author}
  {\bibfnamefont {X.-J.}\ \bibnamefont {Liu}},\ }\href {\doibase
  https://doi.org/10.1103/PhysRevB.103.224308} {\bibfield  {journal} {\bibinfo
  {journal} {Phys. Rev. B}\ }\textbf {\bibinfo {volume} {103}},\ \bibinfo
  {pages} {224308} (\bibinfo {year} {2021}{\natexlab{b}})}\BibitemShut
  {NoStop}%
\bibitem [{\citenamefont {Yuan}\ \emph {et~al.}(2021)\citenamefont {Yuan},
  \citenamefont {Yuan}, \citenamefont {Duine},\ and\ \citenamefont
  {Wang}}]{yuan2021recent}%
  \BibitemOpen
  \bibfield  {author} {\bibinfo {author} {\bibfnamefont {H.}~\bibnamefont
  {Yuan}}, \bibinfo {author} {\bibfnamefont {Z.}~\bibnamefont {Yuan}}, \bibinfo
  {author} {\bibfnamefont {R.~A.}\ \bibnamefont {Duine}}, \ and\ \bibinfo
  {author} {\bibfnamefont {X.}~\bibnamefont {Wang}},\ }\href {\doibase
  10.1209/0295-5075/132/57001} {\bibfield  {journal} {\bibinfo  {journal}
  {Europhysics Letters}\ }\textbf {\bibinfo {volume} {132}},\ \bibinfo {pages}
  {57001} (\bibinfo {year} {2021})}\BibitemShut {NoStop}%
\bibitem [{\citenamefont {R{\"o}mling}\ and\ \citenamefont
  {Kamra}(2024)}]{romling2024quantum}%
  \BibitemOpen
  \bibfield  {author} {\bibinfo {author} {\bibfnamefont {A.-L.~E.}\
  \bibnamefont {R{\"o}mling}}\ and\ \bibinfo {author} {\bibfnamefont
  {A.}~\bibnamefont {Kamra}},\ }\href {\doibase
  https://doi.org/10.1103/PhysRevB.109.174410} {\bibfield  {journal} {\bibinfo
  {journal} {Phys. Rev. B}\ }\textbf {\bibinfo {volume} {109}},\ \bibinfo
  {pages} {174410} (\bibinfo {year} {2024})}\BibitemShut {NoStop}%
\bibitem [{\citenamefont {Wang}\ \emph {et~al.}(2024)\citenamefont {Wang},
  \citenamefont {Zhang}, \citenamefont {Li}, \citenamefont {Wei}, \citenamefont
  {He}, \citenamefont {Xu}, \citenamefont {Xia}, \citenamefont {Luo},
  \citenamefont {Li}, \citenamefont {Dong} \emph
  {et~al.}}]{wang2024ultrastrong}%
  \BibitemOpen
  \bibfield  {author} {\bibinfo {author} {\bibfnamefont {Y.}~\bibnamefont
  {Wang}}, \bibinfo {author} {\bibfnamefont {Y.}~\bibnamefont {Zhang}},
  \bibinfo {author} {\bibfnamefont {C.}~\bibnamefont {Li}}, \bibinfo {author}
  {\bibfnamefont {J.}~\bibnamefont {Wei}}, \bibinfo {author} {\bibfnamefont
  {B.}~\bibnamefont {He}}, \bibinfo {author} {\bibfnamefont {H.}~\bibnamefont
  {Xu}}, \bibinfo {author} {\bibfnamefont {J.}~\bibnamefont {Xia}}, \bibinfo
  {author} {\bibfnamefont {X.}~\bibnamefont {Luo}}, \bibinfo {author}
  {\bibfnamefont {J.}~\bibnamefont {Li}}, \bibinfo {author} {\bibfnamefont
  {J.}~\bibnamefont {Dong}},  \emph {et~al.},\ }\href {\doibase
  https://doi.org/10.1038/s41467-024-46474-7} {\bibfield  {journal} {\bibinfo
  {journal} {Nat. Commun.}\ }\textbf {\bibinfo {volume} {15}},\ \bibinfo
  {pages} {2077} (\bibinfo {year} {2024})}\BibitemShut {NoStop}%
\bibitem [{\citenamefont {Makihara}\ \emph {et~al.}(2021)\citenamefont
  {Makihara}, \citenamefont {Hayashida}, \citenamefont {Noe~Ii}, \citenamefont
  {Li}, \citenamefont {Marquez~Peraca}, \citenamefont {Ma}, \citenamefont
  {Jin}, \citenamefont {Ren}, \citenamefont {Ma}, \citenamefont {Katayama}
  \emph {et~al.}}]{makihara2021ultrastrong}%
  \BibitemOpen
  \bibfield  {author} {\bibinfo {author} {\bibfnamefont {T.}~\bibnamefont
  {Makihara}}, \bibinfo {author} {\bibfnamefont {K.}~\bibnamefont {Hayashida}},
  \bibinfo {author} {\bibfnamefont {G.~T.}\ \bibnamefont {Noe~Ii}}, \bibinfo
  {author} {\bibfnamefont {X.}~\bibnamefont {Li}}, \bibinfo {author}
  {\bibfnamefont {N.}~\bibnamefont {Marquez~Peraca}}, \bibinfo {author}
  {\bibfnamefont {X.}~\bibnamefont {Ma}}, \bibinfo {author} {\bibfnamefont
  {Z.}~\bibnamefont {Jin}}, \bibinfo {author} {\bibfnamefont {W.}~\bibnamefont
  {Ren}}, \bibinfo {author} {\bibfnamefont {G.}~\bibnamefont {Ma}}, \bibinfo
  {author} {\bibfnamefont {I.}~\bibnamefont {Katayama}},  \emph {et~al.},\
  }\href {\doibase https://doi.org/10.1038/s41467-021-23159-z} {\bibfield
  {journal} {\bibinfo  {journal} {Nat. Commun.}\ }\textbf {\bibinfo {volume}
  {12}},\ \bibinfo {pages} {3115} (\bibinfo {year} {2021})}\BibitemShut
  {NoStop}%
\bibitem [{\citenamefont {Azimi~Mousolou}\ \emph {et~al.}(2020)\citenamefont
  {Azimi~Mousolou}, \citenamefont {Bagrov}, \citenamefont {Bergman},
  \citenamefont {Delin}, \citenamefont {Eriksson}, \citenamefont {Liu},
  \citenamefont {Pereiro}, \citenamefont {Thonig},\ and\ \citenamefont
  {Sj{\"o}qvist}}]{azimi2020hierarchy}%
  \BibitemOpen
  \bibfield  {author} {\bibinfo {author} {\bibfnamefont {V.}~\bibnamefont
  {Azimi~Mousolou}}, \bibinfo {author} {\bibfnamefont {A.}~\bibnamefont
  {Bagrov}}, \bibinfo {author} {\bibfnamefont {A.}~\bibnamefont {Bergman}},
  \bibinfo {author} {\bibfnamefont {A.}~\bibnamefont {Delin}}, \bibinfo
  {author} {\bibfnamefont {O.}~\bibnamefont {Eriksson}}, \bibinfo {author}
  {\bibfnamefont {Y.}~\bibnamefont {Liu}}, \bibinfo {author} {\bibfnamefont
  {M.}~\bibnamefont {Pereiro}}, \bibinfo {author} {\bibfnamefont
  {D.}~\bibnamefont {Thonig}}, \ and\ \bibinfo {author} {\bibfnamefont
  {E.}~\bibnamefont {Sj{\"o}qvist}},\ }\href {\doibase
  https://doi.org/10.1103/PhysRevB.102.224418} {\bibfield  {journal} {\bibinfo
  {journal} {Phys. Rev. B}\ }\textbf {\bibinfo {volume} {102}},\ \bibinfo
  {pages} {224418} (\bibinfo {year} {2020})}\BibitemShut {NoStop}%
\bibitem [{\citenamefont {Kubo}(1966)}]{kubo1966fluctuation}%
  \BibitemOpen
  \bibfield  {author} {\bibinfo {author} {\bibfnamefont {R.}~\bibnamefont
  {Kubo}},\ }\href {\doibase 10.1088/0034-4885/29/1/306} {\bibfield  {journal}
  {\bibinfo  {journal} {Rep. Prog. Phys.}\ }\textbf {\bibinfo {volume} {29}},\
  \bibinfo {pages} {255} (\bibinfo {year} {1966})}\BibitemShut {NoStop}%
\bibitem [{\citenamefont {Abaimov}(2015)}]{abaimov2015statistical}%
  \BibitemOpen
  \bibfield  {author} {\bibinfo {author} {\bibfnamefont {S.~G.}\ \bibnamefont
  {Abaimov}},\ }\href@noop {} {\emph {\bibinfo {title} {Statistical physics of
  non-thermal phase transitions: from foundations to applications}}}\ (\bibinfo
   {publisher} {Springer},\ \bibinfo {year} {2015})\BibitemShut {NoStop}%
\bibitem [{\citenamefont {Hutchings}\ \emph {et~al.}(1970)\citenamefont
  {Hutchings}, \citenamefont {Rainford},\ and\ \citenamefont
  {Guggenheim}}]{hutchings1970spin}%
  \BibitemOpen
  \bibfield  {author} {\bibinfo {author} {\bibfnamefont {M.}~\bibnamefont
  {Hutchings}}, \bibinfo {author} {\bibfnamefont {B.}~\bibnamefont {Rainford}},
  \ and\ \bibinfo {author} {\bibfnamefont {H.}~\bibnamefont {Guggenheim}},\
  }\href {\doibase 10.1088/0022-3719/3/2/013} {\bibfield  {journal} {\bibinfo
  {journal} {J. Phys. C: Solid State Phys.}\ }\textbf {\bibinfo {volume} {3}},\
  \bibinfo {pages} {307} (\bibinfo {year} {1970})}\BibitemShut {NoStop}%
\bibitem [{\citenamefont {Rezende}(1978)}]{rezende1978antiferromagnetic}%
  \BibitemOpen
  \bibfield  {author} {\bibinfo {author} {\bibfnamefont {S.}~\bibnamefont
  {Rezende}},\ }\href {\doibase 10.1088/0022-3719/11/16/005} {\bibfield
  {journal} {\bibinfo  {journal} {J. Phys. C: Solid State Phys.}\ }\textbf
  {\bibinfo {volume} {11}},\ \bibinfo {pages} {L701} (\bibinfo {year}
  {1978})}\BibitemShut {NoStop}%
\bibitem [{\citenamefont {Rezende}\ \emph {et~al.}(2019)\citenamefont
  {Rezende}, \citenamefont {Azevedo},\ and\ \citenamefont
  {Rodr{\'\i}guez-Su{\'a}rez}}]{rezende2019}%
  \BibitemOpen
  \bibfield  {author} {\bibinfo {author} {\bibfnamefont {S.~M.}\ \bibnamefont
  {Rezende}}, \bibinfo {author} {\bibfnamefont {A.}~\bibnamefont {Azevedo}}, \
  and\ \bibinfo {author} {\bibfnamefont {R.~L.}\ \bibnamefont
  {Rodr{\'\i}guez-Su{\'a}rez}},\ }\href {\doibase
  https://doi.org/10.1063/1.5109132} {\bibfield  {journal} {\bibinfo  {journal}
  {J. Appl. Phys.}\ }\textbf {\bibinfo {volume} {126}},\ \bibinfo {pages}
  {151101} (\bibinfo {year} {2019})}\BibitemShut {NoStop}%
\bibitem [{\citenamefont {Bostr{\"o}m}\ \emph {et~al.}(2021)\citenamefont
  {Bostr{\"o}m}, \citenamefont {Parvini}, \citenamefont {McIver}, \citenamefont
  {Rubio}, \citenamefont {Viola~Kusminskiy},\ and\ \citenamefont
  {Sentef}}]{bostrom2021all}%
  \BibitemOpen
  \bibfield  {author} {\bibinfo {author} {\bibfnamefont {E.~V.}\ \bibnamefont
  {Bostr{\"o}m}}, \bibinfo {author} {\bibfnamefont {T.~S.}\ \bibnamefont
  {Parvini}}, \bibinfo {author} {\bibfnamefont {J.~W.}\ \bibnamefont {McIver}},
  \bibinfo {author} {\bibfnamefont {A.}~\bibnamefont {Rubio}}, \bibinfo
  {author} {\bibfnamefont {S.}~\bibnamefont {Viola~Kusminskiy}}, \ and\
  \bibinfo {author} {\bibfnamefont {M.~A.}\ \bibnamefont {Sentef}},\ }\href
  {\doibase 10.1103/PhysRevB.104.L100404} {\bibfield  {journal} {\bibinfo
  {journal} {Phys. Rev. B}\ }\textbf {\bibinfo {volume} {104}},\ \bibinfo
  {pages} {L100404} (\bibinfo {year} {2021})}\BibitemShut {NoStop}%
\bibitem [{\citenamefont {Nikotin}\ \emph {et~al.}(1969)\citenamefont
  {Nikotin}, \citenamefont {Lindg{\aa}rd},\ and\ \citenamefont
  {Dietrich}}]{nikotin1969magnon}%
  \BibitemOpen
  \bibfield  {author} {\bibinfo {author} {\bibfnamefont {O.}~\bibnamefont
  {Nikotin}}, \bibinfo {author} {\bibfnamefont {P.-A.}\ \bibnamefont
  {Lindg{\aa}rd}}, \ and\ \bibinfo {author} {\bibfnamefont {O.}~\bibnamefont
  {Dietrich}},\ }\href
  {https://iopscience.iop.org/article/10.1088/0022-3719/2/7/309/meta?casa_token=rwj512wuqn8AAAAA:9xA6hzYQ4-PlY5shJxpJJ_jTcotG4h2u9gQ4HTk2EHzATaqh9euWJZboYnVs3f55wxJWsUn4dETT9Kc}
  {\bibfield  {journal} {\bibinfo  {journal} {J. Phys. Condens. Matter}\
  }\textbf {\bibinfo {volume} {2}},\ \bibinfo {pages} {1168} (\bibinfo {year}
  {1969})}\BibitemShut {NoStop}%
\bibitem [{\citenamefont {Tam}\ \emph {et~al.}(2022)\citenamefont {Tam},
  \citenamefont {Zhu}, \citenamefont {Ayres}, \citenamefont {Kummer},
  \citenamefont {Yakhou-Harris}, \citenamefont {Cooper}, \citenamefont
  {Carrington},\ and\ \citenamefont {Hayden}}]{tam2022charge}%
  \BibitemOpen
  \bibfield  {author} {\bibinfo {author} {\bibfnamefont {C.}~\bibnamefont
  {Tam}}, \bibinfo {author} {\bibfnamefont {M.}~\bibnamefont {Zhu}}, \bibinfo
  {author} {\bibfnamefont {J.}~\bibnamefont {Ayres}}, \bibinfo {author}
  {\bibfnamefont {K.}~\bibnamefont {Kummer}}, \bibinfo {author} {\bibfnamefont
  {F.}~\bibnamefont {Yakhou-Harris}}, \bibinfo {author} {\bibfnamefont
  {J.}~\bibnamefont {Cooper}}, \bibinfo {author} {\bibfnamefont
  {A.}~\bibnamefont {Carrington}}, \ and\ \bibinfo {author} {\bibfnamefont
  {S.}~\bibnamefont {Hayden}},\ }\href {\doibase
  https://doi.org/10.1038/s41467-022-28124-y} {\bibfield  {journal} {\bibinfo
  {journal} {Nat. Commun.}\ }\textbf {\bibinfo {volume} {13}},\ \bibinfo
  {pages} {570} (\bibinfo {year} {2022})}\BibitemShut {NoStop}%
\bibitem [{\citenamefont {Keimer}\ \emph {et~al.}(1992)\citenamefont {Keimer},
  \citenamefont {Aharony}, \citenamefont {Auerbach}, \citenamefont {Birgeneau},
  \citenamefont {Cassanho}, \citenamefont {Endoh}, \citenamefont {Erwin},
  \citenamefont {Kastner},\ and\ \citenamefont {Shirane}}]{keimer1992neel}%
  \BibitemOpen
  \bibfield  {author} {\bibinfo {author} {\bibfnamefont {B.}~\bibnamefont
  {Keimer}}, \bibinfo {author} {\bibfnamefont {A.}~\bibnamefont {Aharony}},
  \bibinfo {author} {\bibfnamefont {A.}~\bibnamefont {Auerbach}}, \bibinfo
  {author} {\bibfnamefont {R.}~\bibnamefont {Birgeneau}}, \bibinfo {author}
  {\bibfnamefont {A.}~\bibnamefont {Cassanho}}, \bibinfo {author}
  {\bibfnamefont {Y.}~\bibnamefont {Endoh}}, \bibinfo {author} {\bibfnamefont
  {R.}~\bibnamefont {Erwin}}, \bibinfo {author} {\bibfnamefont
  {M.}~\bibnamefont {Kastner}}, \ and\ \bibinfo {author} {\bibfnamefont
  {G.}~\bibnamefont {Shirane}},\ }\href {\doibase 10.1103/PhysRevB.45.7430}
  {\bibfield  {journal} {\bibinfo  {journal} {Phys. Rev. B}\ }\textbf {\bibinfo
  {volume} {45}},\ \bibinfo {pages} {7430} (\bibinfo {year}
  {1992})}\BibitemShut {NoStop}%
\bibitem [{\citenamefont {Armstrong}(2010)}]{armstrong2010variable}%
  \BibitemOpen
  \bibfield  {author} {\bibinfo {author} {\bibfnamefont {H.}~\bibnamefont
  {Armstrong}},\ }\emph {\bibinfo {title} {Variable-temperature
  photoluminescence emission instrumentation and measurements on low yield
  metals}},\ \href@noop {} {Ph.D. thesis},\ \bibinfo  {school} {Durham
  University} (\bibinfo {year} {2010})\BibitemShut {NoStop}%
\bibitem [{\citenamefont {Sadat~Parvini}\ \emph {et~al.}(2023)\citenamefont
  {Sadat~Parvini}, \citenamefont {Paz}, \citenamefont {B{\"o}hnert},
  \citenamefont {Schulman}, \citenamefont {Benetti}, \citenamefont {Oberbauer},
  \citenamefont {Walowski}, \citenamefont {Moradi}, \citenamefont {Ferreira},\
  and\ \citenamefont {M{\"u}nzenberg}}]{2023enhancing}%
  \BibitemOpen
  \bibfield  {author} {\bibinfo {author} {\bibfnamefont {T.}~\bibnamefont
  {Sadat~Parvini}}, \bibinfo {author} {\bibfnamefont {E.}~\bibnamefont {Paz}},
  \bibinfo {author} {\bibfnamefont {T.}~\bibnamefont {B{\"o}hnert}}, \bibinfo
  {author} {\bibfnamefont {A.}~\bibnamefont {Schulman}}, \bibinfo {author}
  {\bibfnamefont {L.}~\bibnamefont {Benetti}}, \bibinfo {author} {\bibfnamefont
  {F.}~\bibnamefont {Oberbauer}}, \bibinfo {author} {\bibfnamefont
  {J.}~\bibnamefont {Walowski}}, \bibinfo {author} {\bibfnamefont
  {F.}~\bibnamefont {Moradi}}, \bibinfo {author} {\bibfnamefont
  {R.}~\bibnamefont {Ferreira}}, \ and\ \bibinfo {author} {\bibfnamefont
  {M.}~\bibnamefont {M{\"u}nzenberg}},\ }\href {\doibase 10.1063/5.0151480}
  {\bibfield  {journal} {\bibinfo  {journal} {J. Appl. Phys.}\ }\textbf
  {\bibinfo {volume} {133}} (\bibinfo {year} {2023}),\
  10.1063/5.0151480}\BibitemShut {NoStop}%
\bibitem [{\citenamefont {Urazhdin}\ \emph {et~al.}(2010)\citenamefont
  {Urazhdin}, \citenamefont {Tabor}, \citenamefont {Tiberkevich},\ and\
  \citenamefont {Slavin}}]{2010fractional}%
  \BibitemOpen
  \bibfield  {author} {\bibinfo {author} {\bibfnamefont {S.}~\bibnamefont
  {Urazhdin}}, \bibinfo {author} {\bibfnamefont {P.}~\bibnamefont {Tabor}},
  \bibinfo {author} {\bibfnamefont {V.}~\bibnamefont {Tiberkevich}}, \ and\
  \bibinfo {author} {\bibfnamefont {A.}~\bibnamefont {Slavin}},\ }\href
  {\doibase 10.1103/PhysRevLett.105.104101} {\bibfield  {journal} {\bibinfo
  {journal} {Phys. Rev. Lett.}\ }\textbf {\bibinfo {volume} {105}},\ \bibinfo
  {pages} {104101} (\bibinfo {year} {2010})}\BibitemShut {NoStop}%
\end{thebibliography}%
\end{document}